\documentclass[aps,prl,twocolumn,superscriptaddress,showpacs,floatfix]{revtex4-2}

\usepackage{amsmath,amsthm,amssymb,amsfonts,float,graphics,epsfig,epstopdf,color,verbatim,tabularx,bm,multirow,appendix}
\usepackage[utf8]{inputenc}
\usepackage[T1]{fontenc}
\usepackage{xcolor}
\usepackage{dsfont}
\usepackage{textcomp}
\usepackage{yfonts}
\usepackage{footnote}
\usepackage{bm}
\usepackage{subfigure}
\usepackage{mathrsfs}
\usepackage{MnSymbol}
\usepackage{graphicx}
\usepackage{verbatim}
\usepackage[colorlinks=true, citecolor=blue, linkcolor=blue, urlcolor=blue]{hyperref}
\usepackage{multirow}
\usepackage{braket}
\usepackage[normalem]{ulem}
\usepackage{tikz}
\usetikzlibrary{calc}
\usetikzlibrary{shapes.multipart}
\usepackage{orcidlink}
\usepackage{xr} 

\graphicspath{ {images/} }

\newtheorem*{prop}{Abelian separability}

\newcommand{\comments}[1]{}


\begin{document}

\title{Multiparty entanglement loops in quantum spin liquids}

\author{Liuke Lyu}
\affiliation{D\'epartement de Physique, Universit\'e de Montr\'eal, Montr\'eal, QC H3C 3J7, Canada}
\affiliation{
 Institut Courtois, Universit\'e de Montr\'eal, Montr\'eal (Qu\'ebec), H2V 0B3, Canada
}
\affiliation{
 Centre de Recherches Math\'ematiques, Universit\'e de Montr\'eal, Montr\'eal, QC, Canada, HC3 3J7
}

\author{Deeksha Chandorkar}
\author{Samarth Kapoor}
\author{So Takei\,\orcidlink{0000-0001-9177-1895}}
\email{stakei@qc.cuny.edu}
\affiliation{Department of Physics, Queens College of the City University of New York, Queens, New York 11367, USA}
\affiliation{Physics Doctoral Program, Graduate Center of the City University of New York, New York, NY 10016, USA}


\author{Erik S. S{\o}rensen\,\orcidlink{0000-0002-5956-1190}}
\email{sorensen@mcmaster.ca}
\affiliation{Department of Physics and Astronomy, McMaster University, Hamilton, Ontario L8S 4M1, Canada}

\author{William Witczak-Krempa}
\email{w.witczak-krempa@umontreal.ca}
\affiliation{D\'epartement de Physique, Universit\'e de Montr\'eal, Montr\'eal, QC H3C 3J7, Canada}
\affiliation{
 Institut Courtois, Universit\'e de Montr\'eal, Montr\'eal (Qu\'ebec), H2V 0B3, Canada
}
\affiliation{
 Centre de Recherches Math\'ematiques, Universit\'e de Montr\'eal, Montr\'eal, QC, Canada, HC3 3J7
}

\begin{abstract}
Quantum spin liquids (QSLs) give rise to exotic emergent particles by weaving intricate entanglement patterns in the underlying electrons. 
Bipartite measures between subregions can detect the presence of anyons, but little is known about the full entanglement structure of QSLs. Here, we study the multiparty entanglement of QSLs via entanglement microscopy. We find that in contrast to conventional matter, the genuine multiparty entanglement (GME) between spins is absent in the smallest subregions, a phenomenon we call ``entanglement frustration.'' Instead, GME is more collective, and arises solely in loops. By exploiting exact results and large-scale numerics, we confirm these properties in various gapped and gapless QSLs realized in physically motivated Hamiltonians, as well as with string-net wavefunctions hosting abelian or non-abelian anyons. Our results shed new light on the phase diagram of Kitaev’s honeycomb model in a Zeeman field, and the Kagome Heisenberg model under various perturbations. Going beyond QSLs, we provide evidence that entanglement loops are a universal property of quantum gauge theories. This leads to a new understanding of fractionalization, and the means by which gauge bosons encode quantum information. 
\end{abstract}

\date{\today}
\maketitle


\section{Introduction}

\begin{figure*}[hbt!]
\centering
\includegraphics[width=0.8\textwidth]{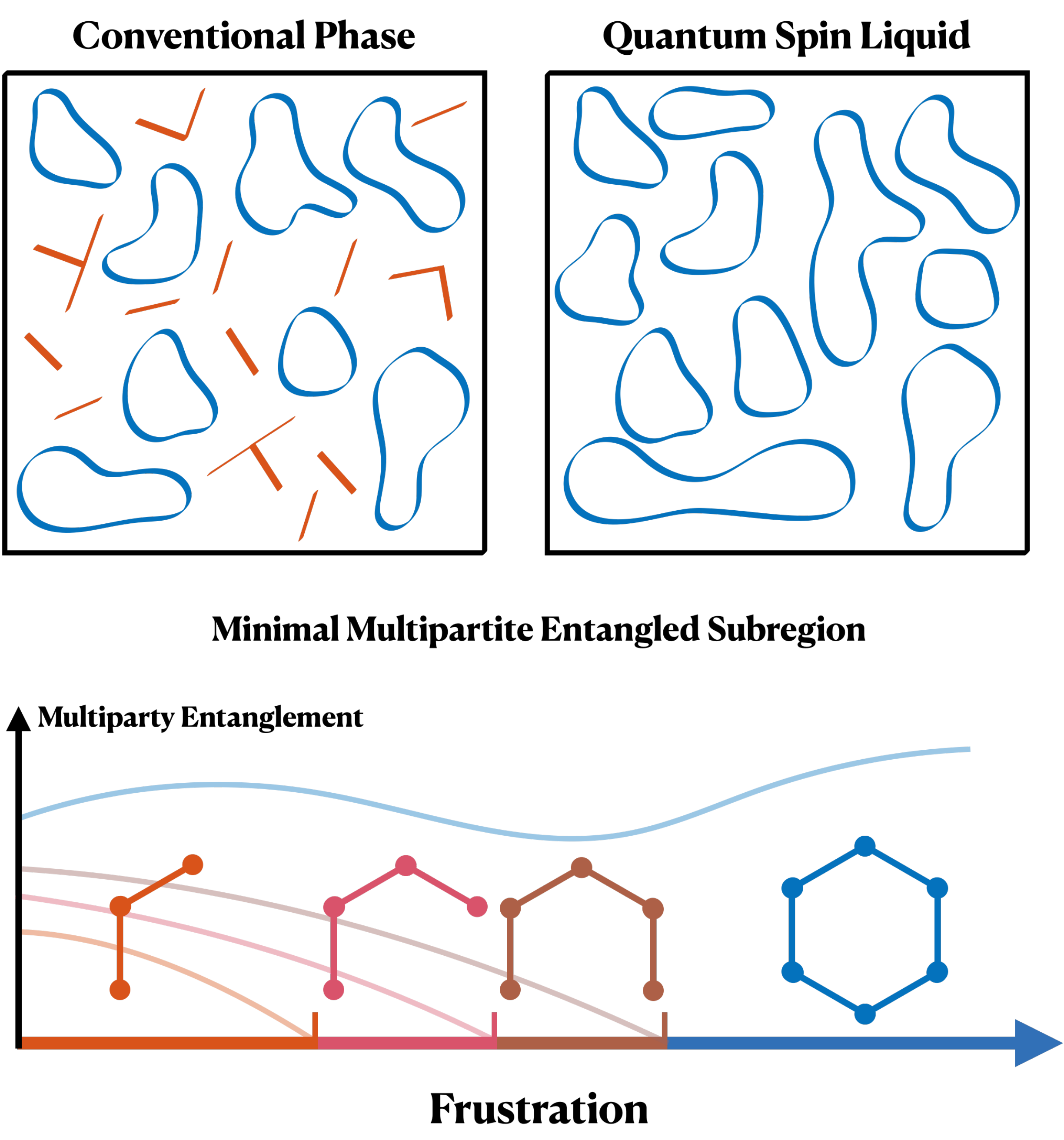}
\caption{\textbf{Multiparty Entanglement Signature of Quantum Spin Liquids.} 
Upper panel: In conventional phases of matter (left), genuine multipartite entanglement is broadly distributed, occurring in both small non-loopy clusters and larger loops. In quantum spin liquids (right),  ``entanglement frustration'' occurs: GME vanishes in small, non-loopy regions, becoming concentrated in loopy regions. 
Lower panel: schematic evolution of the minimal multipartite entangled subregion (MMES) as frustration increases. Three-spin entanglement vanishes first, then four- and five-spin, leaving only the six-spin hexagon in the highly frustrated spin-liquid regime. See Fig.~\ref{fig:kit-phasediagram} for a detailed GME evolution in the Kitaev model in a magnetic field.
}
\label{fig:QSL_vs_conventional}
\end{figure*}


Certain quantum materials do not order at low temperature due to strong quantum fluctuations. 
A key case arises in frustrated spin models where no order can satisfy the local interactions, preventing the spins from ``solidifying''. 
Such quantum spin liquids (QSLs) host emergent fractionalized particles that are forbidden in conventional phases of matter~\cite{Savary2017,Sachdev_2023}. A striking example is the Kitaev honeycomb model that leads to the emergence of Majorana fermions, which are their own antiparticle~\cite {Kitaev2006}. Furthermore, the addition of a Zeeman magnetic field gaps these fermions, leaving behind an exotic topologically ordered phase with non-abelian anyons. Experimental observation of a topological thermal Hall effect in $\alpha$-RuCl$_3$ provided evidence for such a phase~\cite{Yokoi2021}, although alternative explanations have been proposed~\cite{Czajka2023,Grissonnanche2022,PALee2021,Kee2023}. Another important realization of QSL states with non-abelian anyons was recently established in quantum computers with a small number of spins~\cite{google,minev2025,Xu2024}. The flexibility of such architectures allowed the creation and braiding of anyons, which can lead to fault-tolerant quantum computation. 


Unlike magnetically ordered states, QSLs lack local order parameters,
but it was shown that for 2d gapped QSL, a linear combination of entanglement entropies known as the topological entanglement entropy, detects the presence of anyons~\cite{Levin2006TEE,TEEKitaev2006}. These entropies measure the bipartite entanglement between one subregion and its complement when the system is pure, namely at zero temperature. But bipartite entanglement is the simplest form: many other forms of entanglement exist between multiple subregions that cannot be captured by bipartite measures. 
Of particular interest is genuine multiparty entanglement, meaning that all parties contribute to the entanglement. A precise definition and computable measures are described below, and in the Methods. We can consider an informative example inspired by Einstein-Podolsky-Rosen: the cat state for 3 spins-$1/2$: $\tfrac{1}{\sqrt2}(\ket{\uparrow \uparrow \uparrow}+\ket{\downarrow \downarrow \downarrow})$. 
There is no entanglement between any pair of spins 
once the third is traced out,
yet the three spins are genuinely multiparty entangled.
A hint of this can be seen by performing a measurement on a single spin. If the outcome is $S_z=+ 1/2$, it collapses the state of other 2 spins to be up as well, whereas they were entirely undetermined prior to the measurement (we use natural units where $\hbar=1$).

This leads to important questions: What is the multiparty entanglement structure in QSLs? Are the features distinctive compared to conventional phases? If so, are they robust in realistic settings such as generic interactions, finite temperature, or coupling to a more general environment?
We shall see that the answer is positive to all three questions. However, there exists an important hurdle: evaluating multiparty entanglement measures for generic quantum states is difficult, even for a small number of spins. To make progress, we shall use entanglement microscopy~\cite{wangEntanglement2024}: we will obtain the full quantum state of a microscopic subregion, and then extract its multiparty entanglement structure. This offers a substantial generalisation of the study of bipartite entanglement involving skeletal regions~\cite{berthiere}.
Our first main result is that strong frustration leads to missing entanglement between nearby spins, a phenomenon that we call ``entanglement frustration.''  Entanglement becomes more collective, but how is it distributed? The second main result is that GME in QSLs robustly arises in closed loops, and is entirely absent in subregions that do not contain a loop. We will see how this is borne out in a variety of models that realize both gapped and gapless QSLs, including the Kitaev honeycomb model with a Zeeman field, and the paradigmatic antiferromagnetic Heisenberg model on the Kagome lattice. Our large-scale numerical results on these models will shed new light on the parameter regimes where a QSL can emerge.
More generally, we will see how the loopy entanglement arises in string-net wavefunctions for gapped 2D QSLs; these can be seen as the coarse-grained universal description of topologically ordered states. Going beyond the study of QSL, we will argue that loopy entanglement arises in generic quantum gauge theories, bringing new insight into how gauge bosons encode quantum information. 

The past decades have seen a sustained effort in entanglement detection, for an early review see~\cite{Guhne2009}. 
Recently, Hauke et al.~\cite{Hauke2016} showed that the quantum Fisher information (QFI) can be obtained from the dynamical structure factor, $S(k,\omega$. Since $S(k,\omega)$ is experimentally measurable, this initiated a significant ongoing experimental and theoretical effort in characterizing entanglement and more generally quantum correlations, directly targeting materials and many body systems. For recent reviews, see~\cite{Frerot_2023,Laurell2025}.
If the QFI density, $f_Q$, satisfy, $f_Q>k$ it follows that the state is at least ($k$+1)-particle entangled~\cite{Toth2012,Hyllus2012,Hauke2016}. Importantly, $f_Q<k$ does not imply less than ($k$+1)-partite entanglement. The QFI therefore only yields an incomplete measure of the multipartite entanglement, and we hope that our results will spur the development of more complete experimental probes of multipartite entanglement.

\section{Entanglement frustration and loops}

\begin{figure}[hbt!]
\centering
\hfill
\subfigure[]{\includegraphics[width=0.48\columnwidth]{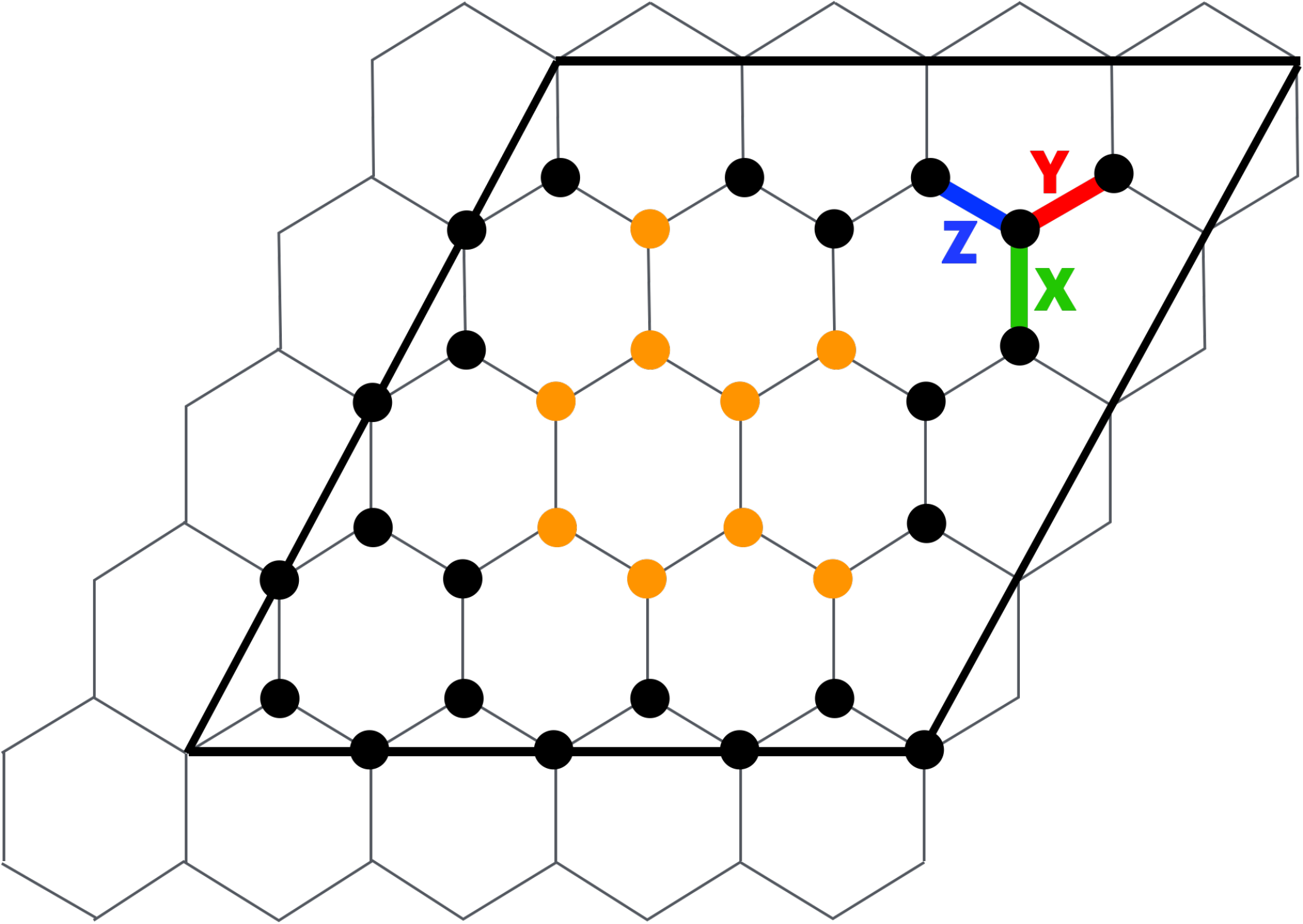}}
\hfill
\subfigure[]{\includegraphics[width=0.50\columnwidth]{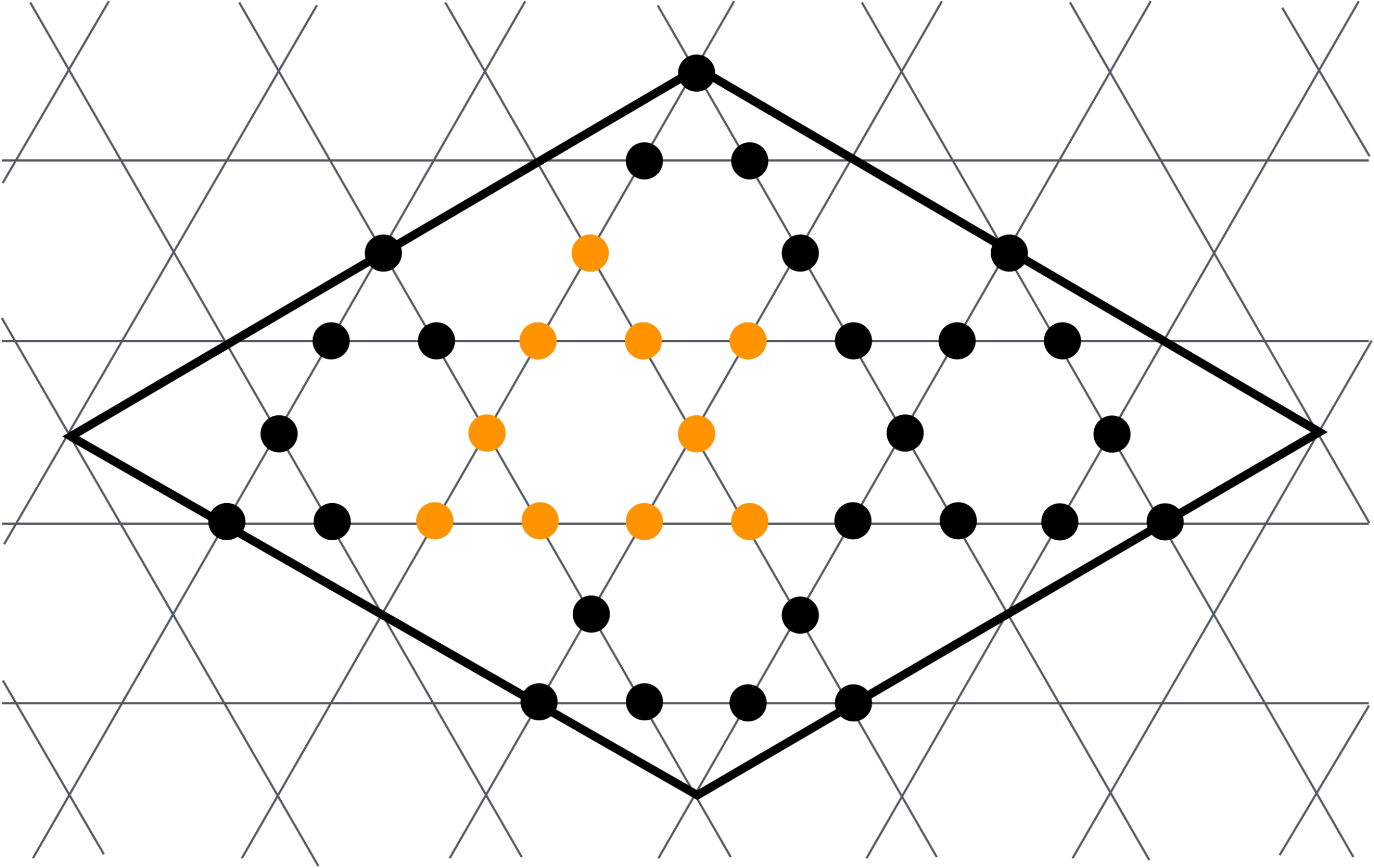}}
\hfill
\caption{\textbf{Exact Diagonalization Clusters} 
(a) A 32‑site honeycomb cluster (Dyck graph) for the Kitaev model, with bonds coloured to distinguish the three orientation‑dependent couplings. 
(b) A 36‑site Kagome cluster. 
In both lattices, the orange sites indicate the compact region, chosen to minimize finite‑size effects, within which we study the GME of various subregions.}
\label{fig:geometries}
\end{figure}

\noindent\textbf{Kitaev honeycomb model}\\
We begin with a model of Kitaev on the honeycomb lattice~\cite{Kitaev2006}, 
\begin{equation}
H = \sum_{\langle i, j \rangle_\gamma} K_\gamma S_i^\gamma S_j^\gamma - \sum_i \mathbf{h}\cdot \mathbf{S}_i, \label{eq:Kitaev111}
\end{equation}
with anisotropic interactions and a magnetic field along the [111] direction, $\mathbf{h}=h /\sqrt{3}(\hat{x}+\hat{y}+\hat{z})$.
This model possesses gapless, gapped abelian and non-abelian QSL phases.
The first two types exist at vanishing Zeeman field $h=0$, as can be shown by an exact solution in terms of Majorana fermion operators. The non-abelian QSL arises at finite but small $h$. The intermediate $h$ regime has been argued to potentially realize an exotic QSL, but no consensus exists~\cite{Hickey2019, Zhu2018, Pollmann2018, Holdhusen2024, Zhang2022}. Large values of the Zeeman field stabilize a conventional paramagnet. We shall work both with the $h=0$ exact solution in the thermodynamic limit, and with large-scale exact diagonalization for $h>0$ on a 32-site lattice with high symmetry (Fig.~\ref{fig:geometries}a). We will also employ an exactly solvable effective Hamiltonian obtained by perturbation theory at small $h$, which includes a 3-spin interaction in addition to the Kitaev Hamiltonian.
We will study the GME of microscopic subregions by evaluating the genuine multipartite negativity (GMN) between $m$ parties~\cite{Guhne2011}, where each party is a group of spins and $m\leq 6$. This powerful entanglement monotone leverages the positive partial transpose (PPT) criterion that was formulated to detect bipartite entanglement via the well-known entanglement negativity~\cite{Peres1996}, to which the GMN reduces for 2 parties. 
For a pure state, the $m$-party GMN is given by the minimal negativity among all bipartitions, $1|23\cdots m,12|3\cdots m$, etc. A convex roof extension lifts this definition to a general density matrix by minimizing over all mixtures realizing the state.
Although convex roof measures are generally not tractable, the GMN can be formulated in terms of a semidefinite program, making it numerically efficient and robust. Further details are given in the Methods. 

\begin{figure*}[h!tbp]
\centering
\includegraphics[width=0.99\textwidth]{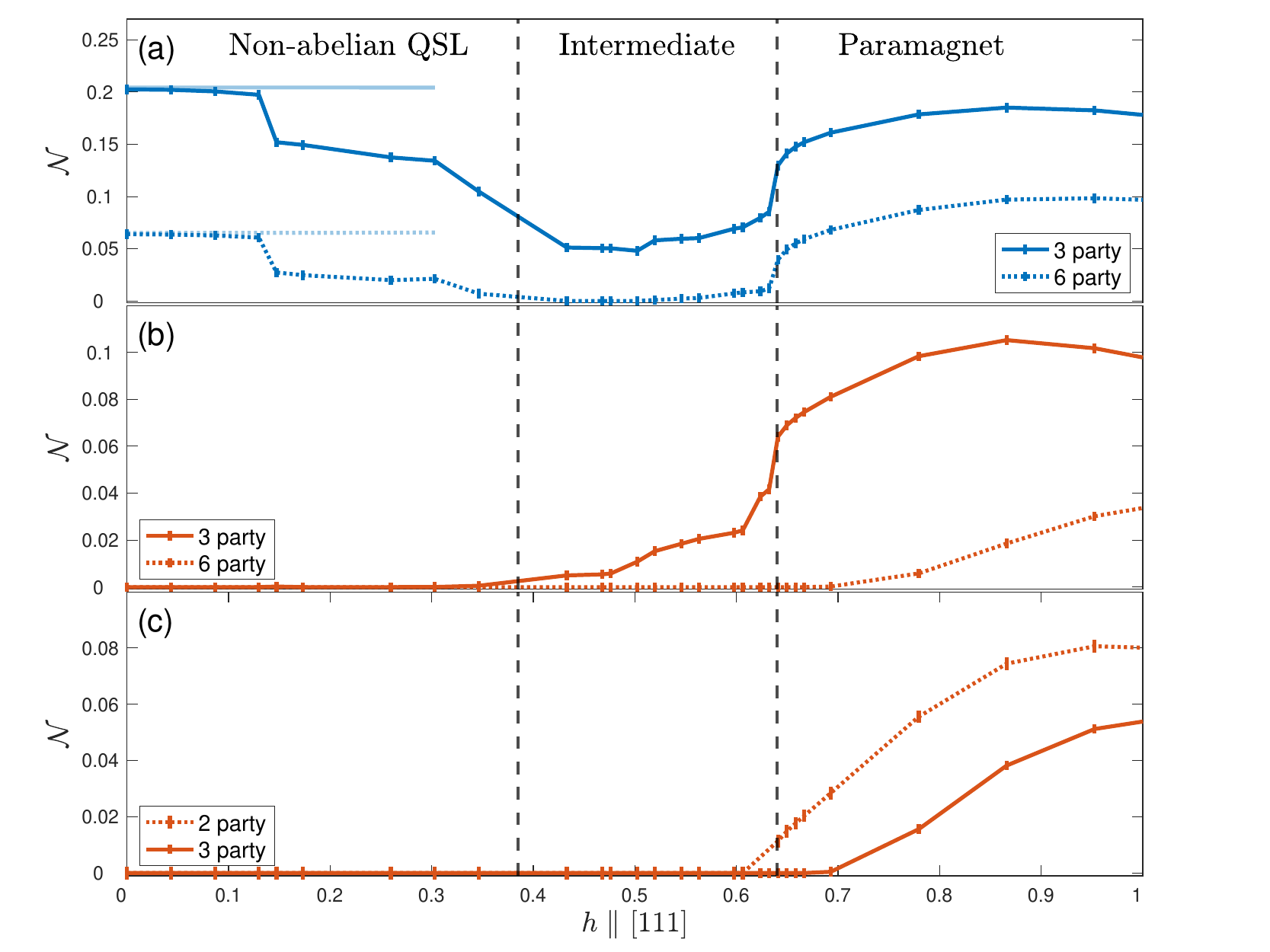}\llap{\shortstack{%
        \includegraphics[scale=.15]{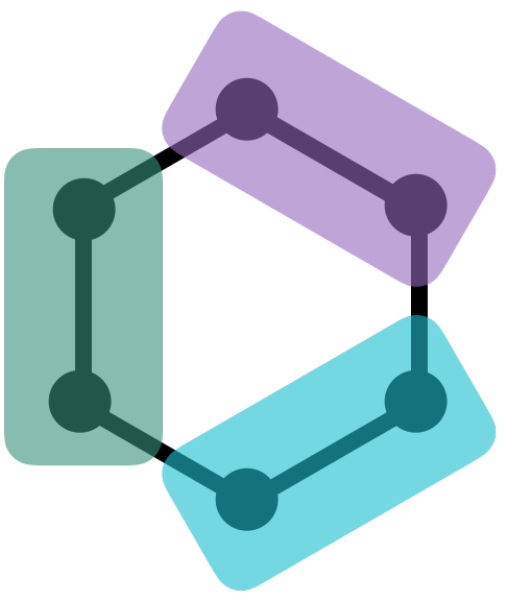}\\
        \rule{0ex}{3.9in}%
      }
  \rule{3.1in}{0ex}}\llap{\shortstack{%
        \includegraphics[scale=.15]{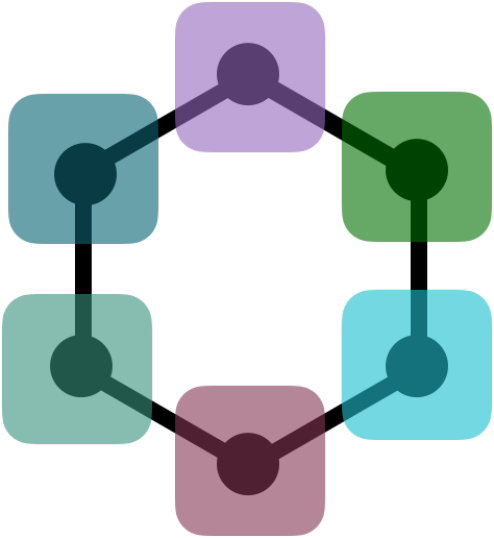}\\
        \rule{0ex}{3.95in}%
      }
  \rule{5.6in}{0ex}}\llap{\shortstack{%
        \includegraphics[scale=.15]{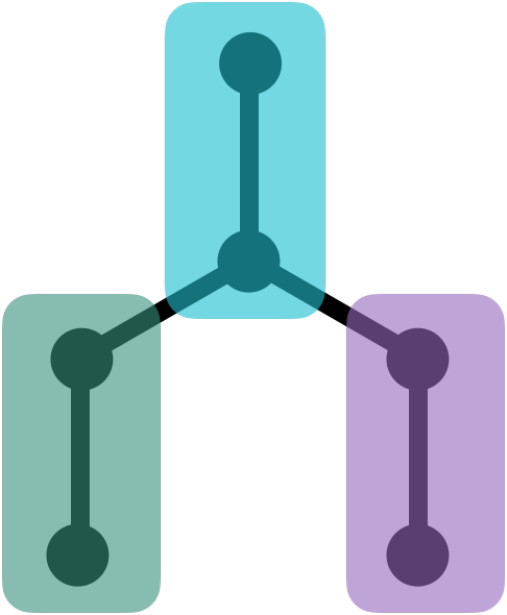}\\
        \rule{0ex}{2.35in}%
      }
  \rule{2.85in}{0ex}}\llap{\shortstack{%
        \includegraphics[scale=.15]{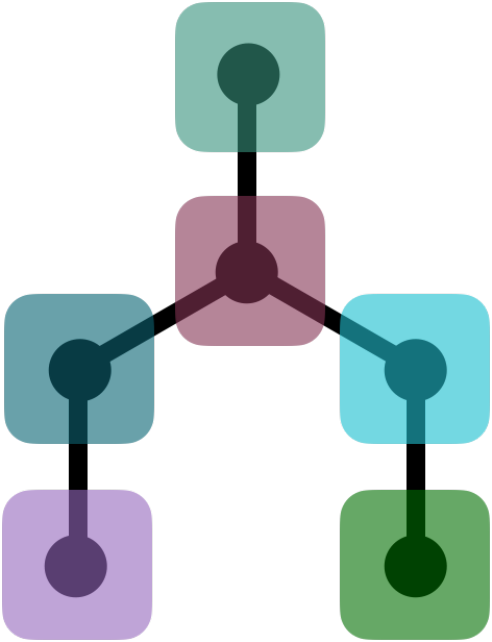}\\
        \rule{0ex}{2.2in}%
      }
  \rule{1.5in}{0ex}}\llap{\shortstack{%
        \includegraphics[scale=.15]{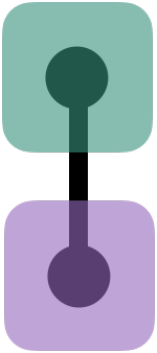}\\
        \rule{0ex}{1.0in}%
      }
  \rule{2.2in}{0ex}}\llap{\shortstack{%
        \includegraphics[scale=.15]{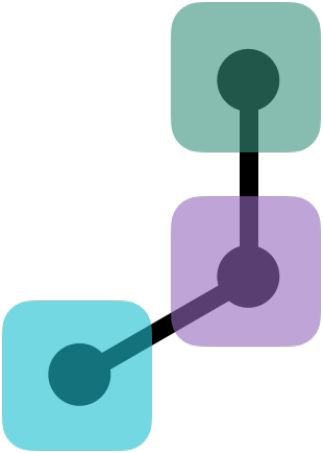}\\
        \rule{0ex}{0.45in}%
      }
  \rule{1.0in}{0ex}}
\caption{\textbf{GMN vs $h$ for the Kitaev Honeycomb Model.} 
GMN is plotted as a function of the [111] field $h$ for (a) the hexagonal plaquette, (b) a non-loopy “fork” subregion, and (c) two- and three-spin subregions.  In each panel, solid (dashed) lines show the tripartite GMN $\mathcal{N}_3$ (six‑partite GMN $\mathcal{N}_6$ or bipartite negativity $\mathcal{N}_2$) from exact diagonalization on the 32‑site cluster. In (a), lighter curves show the same quantities obtained exactly in the thermodynamic limit using the effective Hamiltonian obtained by perturbation theory at small $h$ (see SM section~\ref{sec:effHamiltonian-kitaev}), in excellent agreement within the non‑Abelian QSL at small fields.  In panels (b) and (c), the perturbative predictions vanish and are omitted.  Vertical dashed lines at $h_{c1}=0.38$ and $h_{c2}=0.64$ denote the entry and exit of the intermediate phase based on the energy susceptibility~\cite{inprep}. Data for other subregions is shown in the SM Fig.~\ref{fig:kitaev_hex_phase_diagram}. Results for a 24-site cluster agree closely with those shown here (SM Fig.~\ref{fig:kitaev_nauru}).
}
\label{fig:kit-phasediagram}
\end{figure*}
Figure~\ref{fig:kit-phasediagram} shows the results for the GMN vs $h$ for different subregions and partitions. In the simplest case, the negativity between 2 spins vanishes until $h\approx0.6$, well beyond the region of stability of the non-abelian QSL. Next, we consider the 3-party GMN between 3 adjacent spins, which vanishes in an even larger window and starts to rise in the trivial paramagnetic phase, see Fig.~\ref{fig:kit-phasediagram}. Both these results are striking since adjacent 2-site entanglement and 3-site GME should be present in generic phases, as is the case in the 1d, 2d, and 3d ferromagnetic quantum Ising models for all values of the transverse field~\cite{wangEntanglement2024,Lyu2024}. We call this phenomenon {\bf entanglement frustration}: the frustrated interactions lead to quantum fluctuations that prevent nearby spins from sharing entanglement. Going to 6 spins, Fig.~\ref{fig:kit-phasediagram}b shows that entanglement frustration occurs in a large window for the GME in fork subregions divided into either 3 or 6 parties. 
In order to find GME in the QSL phases at small $h$, we need to consider the minimal subregion with a loop: the hexagonal plaquette. Fig.~\ref{fig:kit-phasediagram}a shows that the hexagon possesses strong 3- and 6-party GME, and the 3-party case is maximal at zero field compared to any $h>0$. 
Larger loops made of two, three, or four adjacent hexagons likewise retain finite GMN at small fields, further supporting the picture that entanglement becomes increasingly collective in a QSL.

One can make the above observations more systematic. We define the {\bf minimal multipartite entangled subregion (MMES)}, which is the smallest subregion of spins that hosts GME. 
The evolution of the MMES is shown in Fig.~\ref{fig:QSL_vs_conventional}.
In the high-field paramagnet, every near-neighbour cluster is entangled.  Increasing frustration extinguishes GME in the smallest clusters first, so the MMES grows step-by-step: from three adjacent spins, to four, then five, and finally to the full six-spin hexagon, which remains the only multiparty-entangled subregion with 6 spins or fewer in the QSLs at zero and small $h$. In contrast, the intermediate phase shows the simultaneous suppression of plaquette GME, and the appearance of non-loopy GME as shown in the middle panel of Fig.~\ref{fig:kit-phasediagram} for the ``fork'' subregion. In fact, the smaller non-loopy 5-spin GMN shown in Fig.~\ref{fig:QSL_vs_conventional} starts rising at $h=0.5$ (see SM Fig.~\ref{fig:kitaev_hex_phase_diagram} for details). As we shall see, this behaviour suggests a departure from a QSL in the intermediate phase.

\begin{figure}[hbt!]
\centering
\includegraphics[width=1\columnwidth]{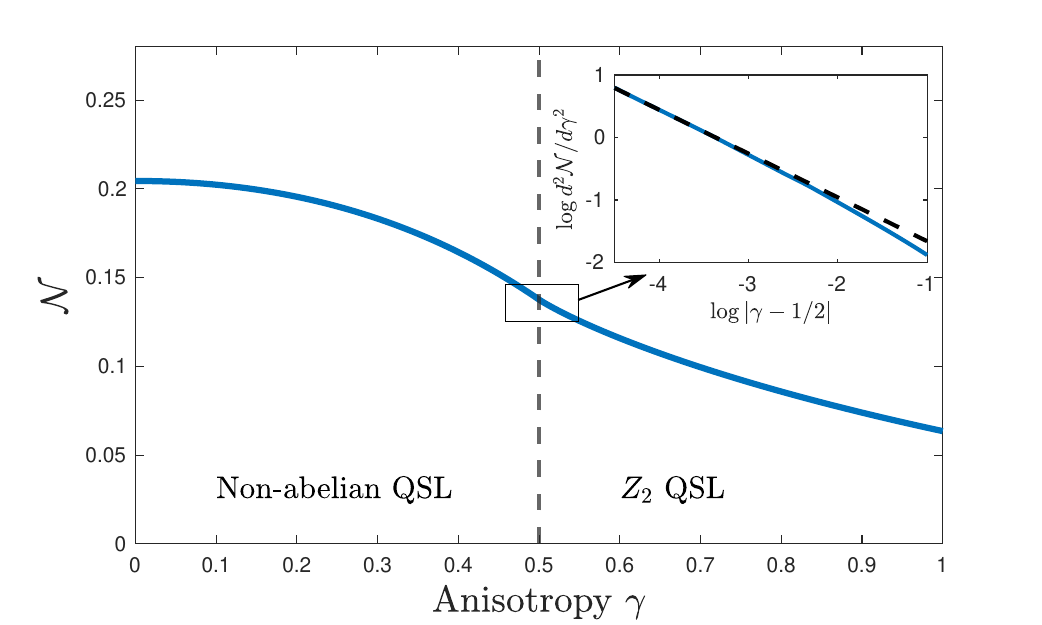}\llap{\shortstack{%
        \includegraphics[scale=.13]{Figures/hexagon-222.pdf}\\
        \rule{0ex}{0.75in}%
      }
  \rule{2.1in}{0ex}}
\caption{\textbf{GMN vs anisotropy for the Kitaev Honeycomb model.} 
Tripartite GMN for the hexagonal plaquette as a function of anisotropy \( \gamma \). The Kitaev interactions are defined by $K_x=K_y=1 - \gamma/2$ and $K_z = 1 + \gamma$. 
At $\gamma = 1/2$, a phase transition occurs from a non-Abelian spin liquid to a \(\mathbb{Z}_2\) spin liquid, and GMN exhibits a singularity. 
The inset shows a power-law divergence for the second derivative of GMN near the phase transition: $d^2\mathcal{N}/d\gamma^2 \sim |\gamma-1/2|^{-0.70}$.
The RDMs are computed using the exact Majorana fermion solution (see SM section \ref{appendix:RDM_exact_methods}), corresponding to the thermodynamic limit of the Kitaev Honeycomb model.
}
\label{fig:kit-anisotropy}
\end{figure}
{\emph{Anisotropy}---We consider a cut in the phase diagram by setting $h=0$ and changing $K_z$ compared to $K_x=K_y$, which allows us to use the exact solution in the thermodynamic limit. At large anisotropy, the system becomes a gapped QSL described by a $\mathbb{Z}_2$ gauge theory (toric code universality class). We find that for all values of $K_z$, the GME is loopy, and the MMES is the hexagon.
Interestingly, at the topological phase transition between gapless and gapped QSLs, the second derivative of the GMN diverges, echoing similar behaviour seen in the quantum Fisher information but with a distinct exponent~\cite{Lambert2020}.
\\

\begin{figure*}[hbt!]
\centering
\includegraphics[width=0.95\textwidth]{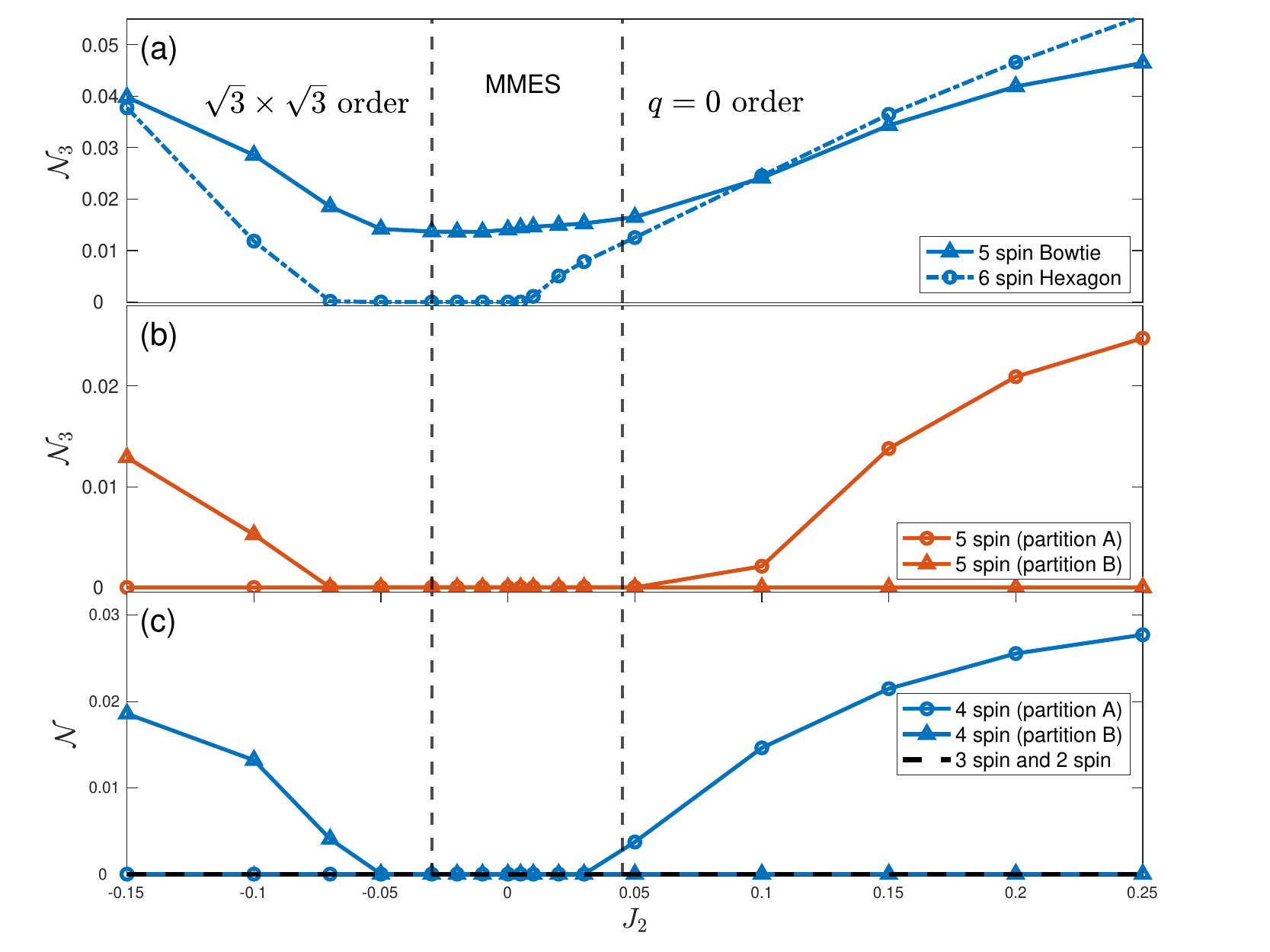}\llap{\shortstack{%
        \includegraphics[scale=.15]{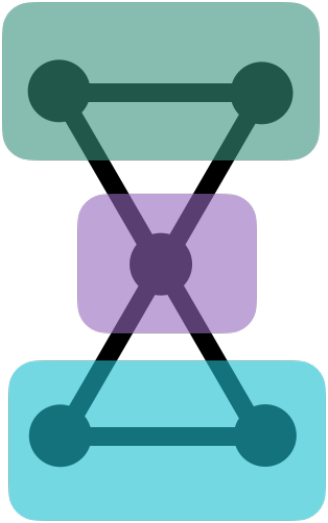}\\
        \rule{0ex}{3.9in}%
      }
  \rule{3.70in}{0ex}}\llap{\shortstack{%
        \includegraphics[scale=.15]{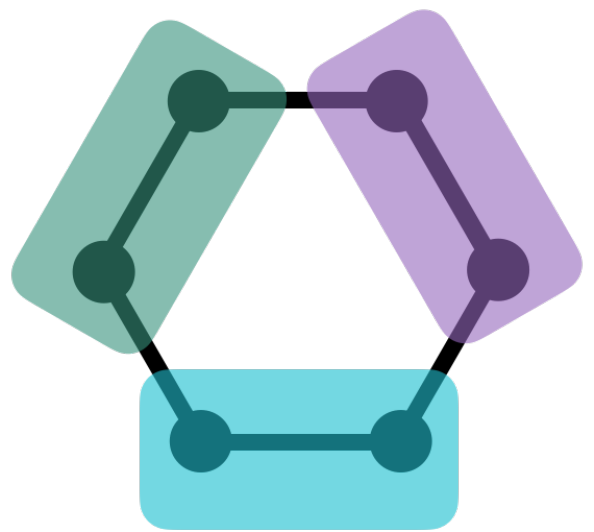}\\
        \rule{0ex}{3.4in}%
      }
  \rule{2.4in}{0ex}}\llap{\shortstack{%
        \includegraphics[scale=.15]{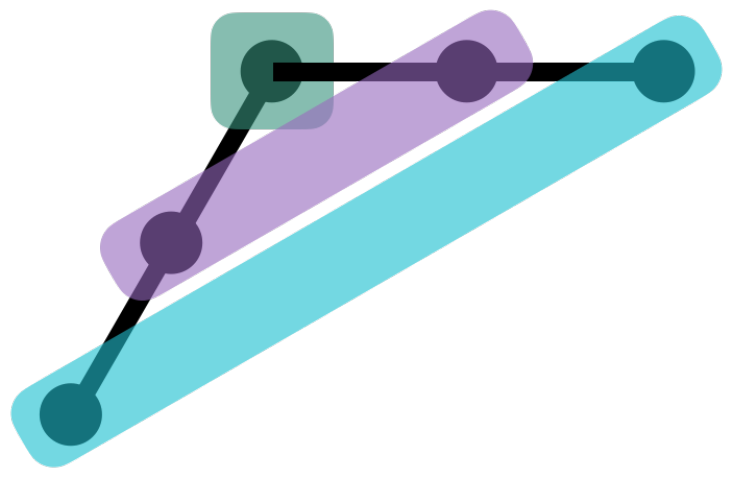}\\
        \rule{0ex}{2.4in}%
      }
  \rule{1.8in}{0ex}}\llap{\shortstack{%
        \includegraphics[scale=.15]{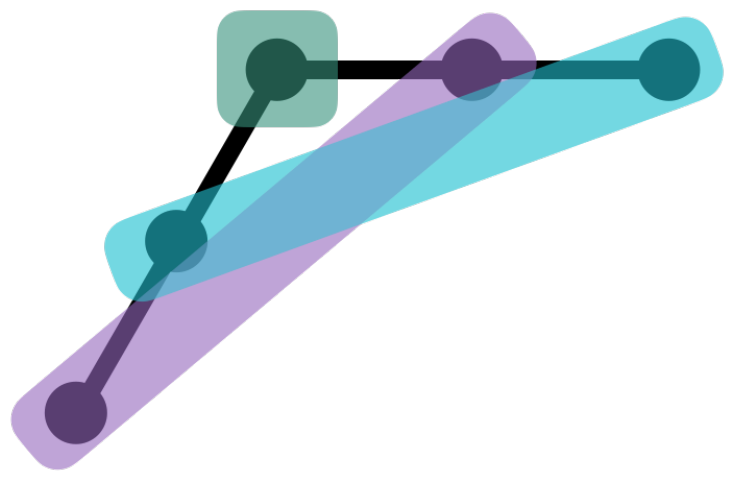}\\
        \rule{0ex}{2.35in}%
      }
  \rule{4.8in}{0ex}}\llap{\shortstack{%
        \includegraphics[scale=.15]{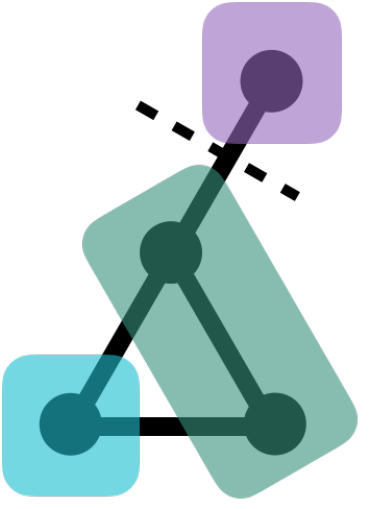}\\
        \rule{0ex}{1.0in}%
      }
  \rule{2.7in}{0ex}}\llap{\shortstack{%
        \includegraphics[scale=.15]{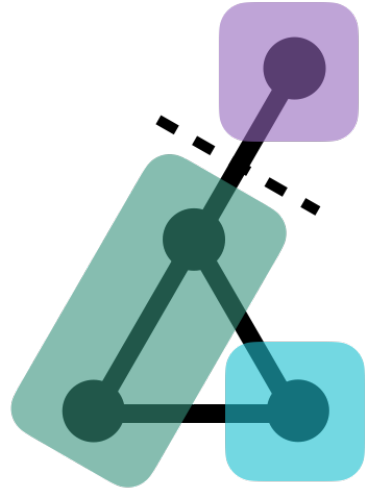}\\
        \rule{0ex}{1.0in}%
      }
  \rule{5.0in}{0ex}}\llap{\shortstack{%
        \includegraphics[scale=.15]{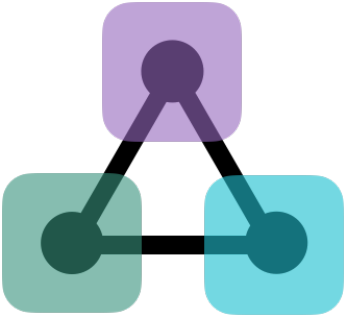}\\
        \rule{0ex}{0.4in}%
      }
  \rule{2.0in}{0ex}}\llap{\shortstack{%
        \includegraphics[scale=.15]{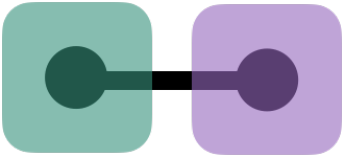}\\
        \rule{0ex}{0.5in}%
      }
  \rule{1.2in}{0ex}}
\caption{\textbf{GMN vs \( J_2 \) for the Kagome Heisenberg Model.} 
Panels show tripartite GMN \(\mathcal{N}_3\) for (a) the loopy hexagon and two‑loop bowtie, (b) two partitions of a five‑spin non‑loopy cluster, and (c) four‑ and three‑spin subregions, alongside the bipartite negativity \(\mathcal{N}_2\) for two spins.  
The vertical dashed lines (\(J_2\!\simeq\!-0.03\) and \(0.045\)) indicate approximate boundaries from tensor-network techniques~\cite{Liao2017}: a central region consistent with a QSL, bounded on the left by a $\sqrt3\times\sqrt3$ ordered phase, and on the right by a $q=0$ ordered phase.  
Within this window, the bowtie subregion—the MMES—retains GME, whereas the hexagon shows GME only for \(J_2>0\).  Non-loopy and ``less-loopy'' clusters acquire GME only outside the QSL window. In (c), the four-spin cluster is deemed less loopy because a single bond cut (dashed line) isolates the purple party, and both two- and three-spin GMN remain zero for all \(J_2\), underscoring strong entanglement frustration.
}
\label{fig:kag-J2}
\end{figure*}

\noindent\textbf{Kagome Heisenberg model}\\
We next turn to one of the most important Hamiltonians in quantum magnetism: the spin-1/2 Kagome~\cite{Mekata2003} antiferromagnetic Heisenberg model, 
\begin{equation}
H_{\rm AF} =  J_1 \sum_{\langle i, j \rangle} \mathbf{S}_i\cdot \mathbf{S}_j + J_2 \sum_{\langle\langle i, j \rangle\rangle} \mathbf{S}_i\cdot \mathbf{S}_j\,.  \label{eq:kagome-J1J2}
\end{equation}
In our analysis, the nearest-neighbour anti-ferromagnetic coupling $J_1$ will be set to unity, and we will vary the next-nearest neighbour coupling, taken to be either ferro- or anti-ferromagnetic.
It is believed to host a QSL in a small window near $J_2=0$. Recent evidence has pointed towards a gapless Dirac QSL~\cite{He2017, Iqbal2011,Clark2013}, which is an intricate state of matter with massless Dirac fermions coupled to an emergent photon. We have obtained the ground state as a function of $J_2$ with ED for a 36-site lattice (Fig.~\ref{fig:geometries}b), in agreement with previous results~\cite{Leung1993,Lecheminant1997,Lauchli2011,Andreas2019}.
The data shows strong entanglement frustration:
both two- and three-spin GMN vanish for $-0.15<J_2<0.25$ (Fig.~\ref{fig:kag-J2}c). 
Instead, within the estimated QSL window marked by the dashed lines (based on tensor-network results~\cite{Liao2017}), GME is entirely loopy, with the MMES given by the 5-spin bowtie (two triangles sharing a vertex) shown in Fig.~\ref{fig:kag-J2}a. Outside this window, non-loopy GME starts to appear in larger five-spin clusters (e.g. Fig.~\ref{fig:kag-J2}b). We even observe GME in certain four-spin subregions containing a single triangle, but this entanglement is far less robust: removing a single bond can isolate one party and destroy the GME. By contrast, the bowtie's two-loop structure admits no single-edge cut that disconnects all three parties. 
We thus uncover a hierarchy of “loopiness.”
Our results provide new support for the existence of a stable QSL in a finite window around $J_2=0$. 
Moreover, the hexagon's tripartite GMN distinguishes the weak-AFM and weak-FM sides: it becomes finite only for antiferromagnetic $J_2$, which also corresponds to the regime where the groundstate shows strong overlap with a Dirac QSL variational wavefunction~\cite{Iqbal2021}.

\begin{figure*}[hbt!]
\centering
\includegraphics[width=0.9\textwidth]{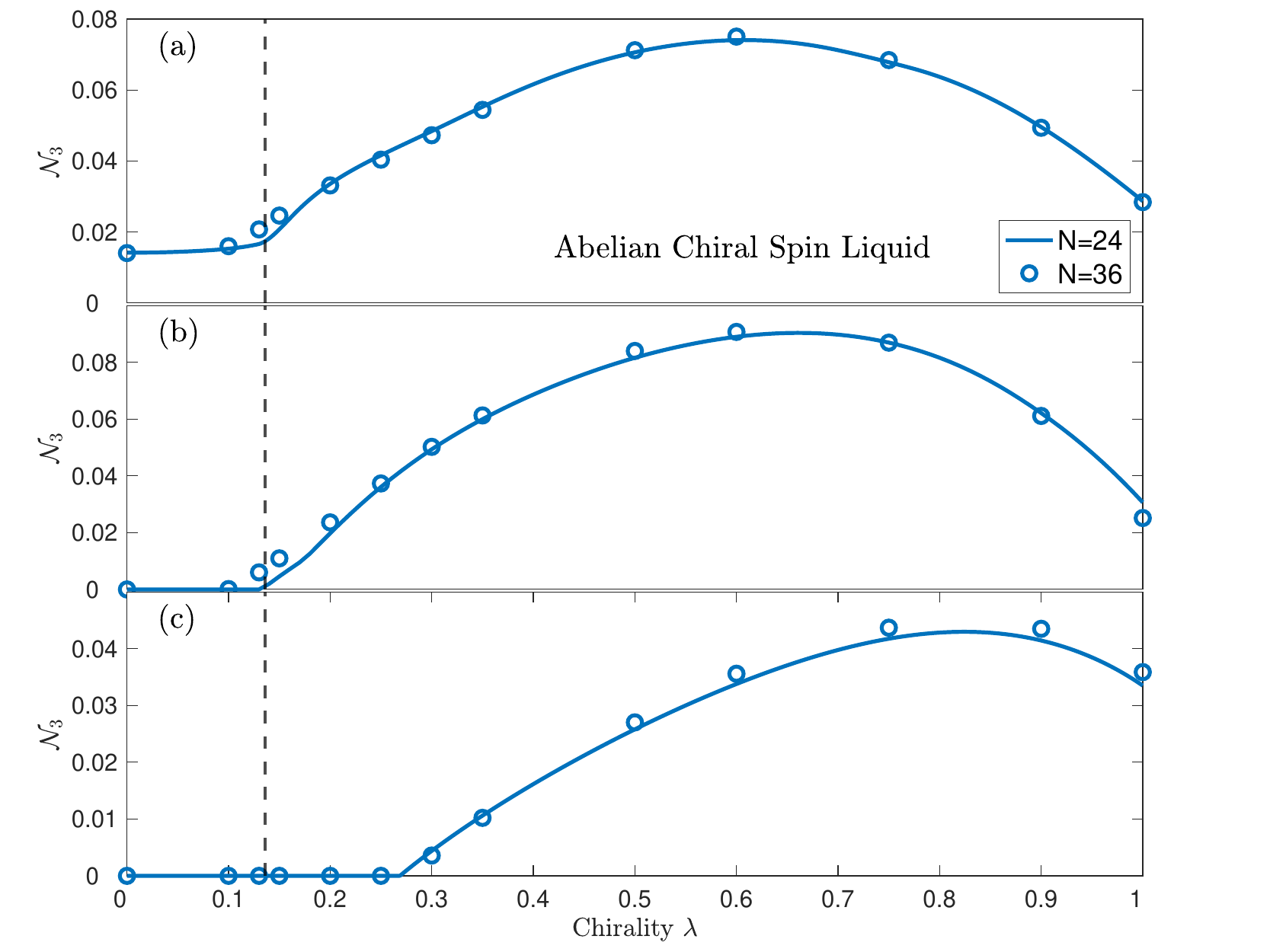}\llap{\shortstack{%
        \includegraphics[scale=.2]{Figures/bowtie-212.pdf}\\
        \rule{0ex}{3.7in}%
      }
  \rule{2.2in}{0ex}}\llap{\shortstack{%
        \includegraphics[scale=.2]{Figures/hexagon-222-horizontal.pdf}\\
        \rule{0ex}{2.1in}%
      }
  \rule{2.1in}{0ex}}\llap{\shortstack{%
        \includegraphics[scale=.2]{Figures/triangle.pdf}\\
        \rule{0ex}{0.7in}%
      }
  \rule{2.2in}{0ex}}
\caption{\textbf{GMN vs Chirality 
for the Kagome Heisenberg Model.} 
Tripartite GMN \(\mathcal{N}_3\) is shown as a function of the scalar‑chirality coupling \(\lambda\) for (a) the bowtie, (b) the hexagon, and (c) the triangle subregions. 
The vertical dashed line at \(\lambda_c\simeq0.136\) marks the onset of the Abelian chiral spin liquid with emergent semions~\cite{Bauer2014}. Beyond \(\lambda_c\), the hexagon gains GME, and at higher $\lambda$ the triangle becomes the MMES, highlighting the enhanced looped entanglement of the chiral phase.
Solid blue lines (24‑site cluster) and circles (36‑site cluster) are close, demonstrating minimal finite‑size effects.
 }
\label{fig:kag-lam}
\end{figure*}
We next consider a different perturbation to the model by including a spin-chirality interaction at $J_2=0$:
\begin{align}
   H=(1-\lambda) H_{\rm AF} + \lambda J_\chi \sum_{\Delta, \nabla} \mathbf{S}_i\cdot\Bigl(\mathbf{S}_j\times\mathbf{S}_k\Bigr)
   \label{eq:kagome-chiral}
\end{align}
A sufficiently large $\lambda$ has been argued to lead to an abelian chiral QSL with an emergent semion~\cite{Bauer2014}: an exotic excitation with exchange statistics midway between those of a boson and a fermion. We observe a qualitative change in the GME structure near the transition reported in previous works~\cite{Bauer2014} $\lambda_c\approx 0.136$: the ground state becomes more loopy because the hexagon acquires GME, and the bowtie GME shows a sharp increase, see Fig.~\ref{fig:kag-lam}a and b. At larger values of $\lambda$, the MMES changes to the smallest subregion with a loop: a triangle made of nearest neighbours. This shows that the chiral QSL is more loopy than the putative parent QSL at $\lambda=0$, and further provides evidence that these are 
distinct types of spin liquids. \\

\noindent\textbf{Resonating Valence Bond states}\\ 
We now turn to an important family of wavefunctions: Resonating Valence Bond (RVB) states, built as equal-weight superpositions of nearest neighbour singlet coverings—including every topological sector—on a given lattice. These have played an important role in various contexts, including frustrated magnetism and high-temperature superconductivity~\cite{rvb}. We begin with the Kagome lattice, where the RVB state has been shown to be a $\mathbb{Z}_2$ QSL~\cite{Schuch2012, Yang2012}. Working on the same 36-site lattice, we find that the GME is entirely loopy, and the MMES is the bowtie, just as for the Kagome Heisenberg model at small $|J_2|$. 
However, the GMN values are markedly reduced compared with the Heisenberg model: $4\times10^{-6}$ for the partition used in Fig.~\ref{fig:kag-J2} and $10^{-2}$ for an alternative split, which indicates an even stronger entanglement frustration in the Kagome RVB state. Our findings complement the general result that the multiparty entanglement of $m\geq 2$ disjoint subregions in RVB states is bounded by a function that decreases exponentially with separation~\cite{Gilles2023}. 


Now, we study the square lattice RVB for $6\times6$ sites. Contrary to the Kagome case, the bipartite nature of the lattice leads to a state with power-law correlations so that the parent Hamiltonian is gapless~\cite{RK1988,Tang2011,Albuquerque2010}. It can be understood as a quantum phase transition point.
We nevertheless find vanishing GMN for all 3-spin, and non-loopy 4-spin subregions. The MMES is a square plaquette, which is the smallest subregion containing at least one loop. Interestingly, this state can also be understood as a quantum gauge theory with a gapless $U(1)$ gauge field. This suggests that the loopy GME structure extends beyond QSLs phases, and is a broader feature of deconfined quantum gauge theories. 

\section{Robustness and universality}
\noindent\textbf{Robust loops}\\
We first ask whether the loopy entanglement structure found above remains robust against perturbations such as varying interactions or adding a finite temperature. 
To answer the question, we choose a non-loopy subregion from one of the systems discussed above, and consider the fate of entanglement under the evolution induced by the perturbation~\cite{parez2024fateentanglement}. 
We can go beyond GMN, and show (a non-trivial task) that the reduced density matrix is biseparable so that \emph{any} measure of GME vanishes, see Methods for details. 
Because the biseparable set is convex and of full measure, a generic biseparable density matrix lies inside it rather than on its boundary; only fine-tuned Hamiltonians would place the state exactly at the edge. Any sufficiently small perturbation therefore keeps the state within the biseparable set, and its GME remains strictly zero. 

This is indeed the case in the above examples, where we found finite ranges of purely loopy entanglement (see SM, section~\ref{sec:certify_SEP}}).
We further found that the loopy GME persists even under strong local perturbations: performing a projective measurement on any single spin—inside or outside the region—fails to break GME in loopy subregions.
For example, measuring one site of the hexagon in the Kitaev model leaves the remaining five spins with finite GME, which establishes that the quantum information is shared collectively around the loop, and is robust to local measurements. 

\noindent\textbf{String Nets}\\
Second, we extend our analysis to a large class of gapped 2d QSLs described by string-net wavefunctions~\cite{Levin2005}. These wavefunctions represent fixed points under coarse-graining (renormalization group) transformations, and we consider them on a honeycomb lattice with spins ($S = 1/2, 1, \dots$) residing on the bonds. For Abelian string-net models, we prove in SM section~\ref{sec:GME_string-net} a general result:
\begin{prop}
The reduced density matrix for any non-loopy subregion in an abelian string-net is fully separable, implying the complete absence of entanglement.
\end{prop}
\noindent The proof exploits the fact that knowing the configuration of the spins adjacent to $A$ fixes the state of $A$.
Furthermore, even for the smallest loopy subregion—the hexagonal plaquette—no GME emerges in the Abelian models we studied. Specifically, the simplest $\mathbb{Z}_2$ gauge theory ($S=1/2$) yields a fully separable plaquette state, while for the double-semion model ($S=1/2$) we show that the tripartite state composed of the 3 pairs around the plaquette is biseparable, implying no GME, a condition in agreement but stronger than the vanishing of GMN. Similarly, for the $\mathbb{Z}_3$ gauge theory ($S=1$), no GME was detected for the plaquette density matrix. 

Turning to non-Abelian cases, in SM section~\ref{sec:GME_string-net} we also examine the Fibonacci string-net ($S=1/2$), $S_3$ gauge theory ($S=1$) and Ising anyon string-net ($S=1$); the last is a doubled version of the theory realized at small fields in the Kitaev model discussed above. We again find an absence of GME in non-loopy subregions. However, in contrast to the abelian nets, bipartite entanglement does arise in non-loopy regions.
Tripartite GME emerges for the hexagonal plaquette in the Ising string-net, and on an extended 12-site region (hexagon and the 6 edges that touch it) in the Fibonacci string-net, while no GME is found in the $S_3$ gauge theory.
This strongly suggests that the absence of non-loopy GME holds across a wide class of QSLs. Moreover, we observe that in string-net states—owing to their strict local gauge constraints—entanglement frustration is particularly severe, often fully suppressing GME even in the smallest loopy subregion. Interestingly, some QSLs have a more loopy entanglement structure than others, exhibiting GME already for the minimal loop. In fact, our results indicate that the most loopy QSLs are the chiral ones. We now examine this point further.

\noindent\textbf{Chiral loops}\\
A chiral QSL hosts chiral edge modes on its boundaries. These propagating degrees of freedom cannot be gapped due to the lack of backscattering channels. In other words, they possess a gravitational anomaly related to a non-zero thermal Hall conductance~\cite{Cappelli2002}. 
The entanglement cut that defines the partial trace can be viewed as introducing a physical boundary whose low-energy spectrum matches the edge CFT of the phase~\cite{TEEKitaev2006, Li_Haldane_2008}.
We can use this relation to obtain the GMN of a macroscopic region, such as a disk. We partition the disk into three equal pie slices. In order to obtain the universal contribution to the tripartite GMN, we trade the mixed reduced density matrix with the pure one of the disk in the presence of a real boundary (with the vacuum). The GMN is then given by the minimal negativity among all bipartitions, which by virtue of rotational symmetry is the negativity between one slice and the other two. The first contribution in the thermodynamic limit is the boundary law, which arises due to the short-ranged entanglement along the slice’s two edges. Second, there is a longer-range term arising from the propagating 1d chiral mode(s), and can be determined from the associated chiral conformal field theory (CFT): 
the logarithm of the GMN scales as
$\log\mathcal{N} \sim (c/4) \log(R/a)$,
which is half the value for a non-chiral CFT~\cite{Calabrese2013}.
Here $R$ is the disk’s radius, $a$ a short-distance cutoff scale, and $c$ is the chiral central charge of the QSL. 
In particular, this universal CFT term can be isolated by dividing the disk into three and four equal slices and taking the difference $\log\mathcal N_3 - \log\mathcal N_4$. The boundary-law terms cancel out, leaving only the chiral edge-mode contribution. We have thus established chiral-mode enhanced GME in large subregions. Interestingly, our results from the previous section on the Kitaev and chiral Kagome models, suggest that such physics percolates down to the smallest scales.

\section{Discussion}
We have established that a robust loop structure for GME emerges in a variety of QSLs irrespective of the gap, chirality, anyon structure, etc. Entanglement frustration leads to a transfer of entanglement from non-loopy to loopy subregions, thus making it more collective. One can then use this property as a diagnostic to detect QSLs. For instance, the strong loopy structure in the Kagome antiferromagnetic Heisenberg model points towards a finite window with a stable QSL. In contrast, in the Kitaev model with a magnetic field along the [111] direction, the intermediate phase displays non-loopy GME, which provides evidence against a QSL in the entire intermediate regime. In addition, we have also observed the same properties in an RVB model that hosts a quantum critical point with an emergent U(1) gauge field.
The common trait of these systems is that they are deconfined quantum gauge theories. Our results point towards a fundamental property of systems described by quantum gauge theories: states with loopy GME. This provides a more basic description that does not require one to derive an effective quantum field theory. 

We now turn to future directions and applications of our results.
First, it will be interesting to map multiparty entanglement in a wide class of exotic phases—ranging from additional spin‑liquid models to symmetry‑protected topological states (SPT), fracton orders, and other unconventional states of quantum matter—to determine the scope of the properties we identified in quantum gauge theories. For example, let us briefly discuss a 2d SPT realized in the CZX bosonic/fermionic model~\cite{Chen_2011} where each site of the square lattice contains 4 spins/fermionic modes. The ground state is obtained by covering the lattice with cat states involving one spin/mode from each site around a square plaquette. As such, we readily conclude that 1) the MMES is the square plaquette (the smallest loopy subregion); 2) any non-loopy region has no GME. The result invites the investigation of further SPTs, especially in experimentally realistic models. 
Second, an advantage of using
the GMN is that it yields a GME witness, namely an observable $W$ defined on the subregion, and whose expectation value detects the presence of GME. As such, $W$ can be used for entanglement detection in the implementation of QSL states in actual quantum computers, or quantum simulation platforms such as ultra-cold atomic gases. This would circumvent the need to obtain the full reduced density matrix (tomography). 
Third, by understanding how multiparty entanglement is organized in fractionalized phases, our results lay the groundwork for harnessing these robust, non-local correlations as a resource for quantum information processing.
Finally, it would be of great interest to exploit the entanglement frustration and loop structure as a novel means to construct quantum states and Hamiltonians for QSLs, and deconfined gauge theories more generally. 

\section{Methods}
\noindent\textbf{Multiparty Entanglement Measures}\\
Entanglement in a multipartite quantum system comes in different forms. Consider a pure quantum state $\ket{\psi}$ defined on a $n$-party Hilbert space $\mathcal{H}_1 \otimes \cdots \otimes \mathcal{H}_n$. The state is \textit{fully separable} if it factorizes as $\ket{\psi}=\ket{\psi^{(1)}}\otimes\ket{\psi^{(2)}}\otimes \cdots \otimes \ket{\psi^{(n)}}$ and otherwise hosts some form of entanglement. For any bipartition $A|B$, it is biseparable if $\ket{\psi^{\mathrm{bs}}_{A|B}} = \ket{\alpha}_A \otimes \ket{\beta}_B$, signifying no entanglement across the cut. Finally, if it fails to be biseparable under every bipartition, it is genuinely multipartite entangled (GME), the strongest form of entanglement. 
These definitions can be extended to mixed quantum states, described by density matrices, such as the reduced states of subregions studied in this work. A mixed state is fully separable if it admits a convex decomposition $\rho^{\mathrm{sep}}=\sum_k p_k\,\ket{\psi_k^{\mathrm{sep}}}\bra{\psi_k^{\mathrm{sep}}}$, where each $\ket{\psi_k^{\mathrm{sep}}}$ is fully separable. It is biseparable if $\rho^{\mathrm{bs}}=\sum_k p_k\,\ket{\psi_k^{\mathrm{bs}}}\bra{\psi_k^{\mathrm{bs}}}$, where each $\ket{\psi_k^{\mathrm{bs}}}$ is biseparable along some partition. Finally, if no such decomposition exists, $\rho$ is GME.
Mixed‑state GME is generally difficult to quantify, but for small systems (a few spins) we can employ the genuine multipartite negativity (GMN)~\cite{Guhne2011}, which builds on the bipartite quantum negativity $N(\rho)=(\left\| \rho^{T_A} \right\|_1 -1)/2$, where $\|X\|_1$ is the trace norm and $T_A$ denotes partial transpose on subsystem $A$~\cite{Peres1996, Horodecki1996PPT}. A naive extension to multiparty entanglement is the minimum negativity $N^{\min }(\rho)=\min _M N_M(\rho)$ which is minimized over all bipartitions $M$, but $N^{\mathrm min}>0$ does not guarantee GME. Instead, GMN is defined by the mixed convex‑roof extension of the minimum negativity~\cite{Hofmann_2014}
\begin{equation}
    \mathcal{N}(\varrho)=\min _{\{p_k,\rho_k\}} \sum_k p_k N^{\text{min}}\left(\rho_k\right),
    \label{eq:CRE_GMN}
\end{equation}
where each ensemble $\{p_k,\rho_k\}$ satisfies $\rho=\sum_k p_k \rho_k$. GMN can be computed efficiently via a semidefinite program, which guarantees an optimal solution of the minimization;
implementation details can be found in SM section~\ref{sec:GMN_SDP}.
A positive value, $\mathcal{N}(\varrho)>0$, certifies GME.  The converse, however, is not true: 
$\mathcal{N}(\varrho)=0$ does not exclude the presence of GME. 
To close this gap, we employ the adaptive‑polytope algorithm~\cite{Ties2024}, which recursively approximates the biseparable set with polytopes and reformulates the GME robustness problem as a series of semidefinite programs. Concretely, one finds the largest $t$ for which $\rho(t)=t\rho+(1-t)I/D$ is biseparable. Whenever this critical $t^*>1$, the state is guaranteed to be biseparable and thus has no GME. 
In practice, for the majority of density matrices examined in this work, a vanishing GMN coincides with biseparability.

\noindent\textbf{Exact Diagonalization}\\
The ground state wavefunctions, from which the density matrices forming the basis of the GME calculations are derived, are obtained using fully parallelized~\cite{Lauchli2011} Lanczos-based exact diagonalizations of finite clusters with periodic boundary conditions.
The clusters used, mainly of size $N=32$ for the honeycomb Kitaev model and
$N=36$ for the Kagome model, are shown in Fig.~\ref{fig:geometries}.
In the 32-site Kitaev system, we find 3- and 6-fold ground-state degeneracies within both the small-field non-abelian spin liquid and intermediate regimes; the specific field values will be reported in a forthcoming work~\cite{inprep}. In each case, we pick a single, translationally symmetric ground state for our reduced-density-matrix calculations. This choice guarantees that every equivalent local subregion yields the same RDM. 
Equivalently, one may take the $T=0$ limit of the thermal (equal-weight) mixture over the degenerate manifold to obtain the RDM, which yields almost identical GMN to any single symmetric ground state.
The ground-state degeneracy reflects the symmetry of the Hamiltonian: a 60° ($C_6$) rotation of the lattice about a hexagon center combined with a 120° ($C_3$) spin rotation about the [111] axis
—which cyclically permutes the spin operators $(S^x,S^y,S^z)$—
leaves the Hamiltonian invariant.


\section*{Acknowledgements}
We thank Andreas M.~L\"auchli for valuable discussions and for sharing Kitaev-honeycomb ED data, and Ziyang Meng for collaboration on entanglement microscopy of relevant systems.
W.W.-K.\/ and L.L.\/ are supported by a grant from the Fondation Courtois, a Chair of the Institut Courtois, a Discovery Grant from NSERC, and a Canada Research Chair.
This research was enabled in part by support provided by SHARCNET (sharcnet.ca) and the Digital Research Alliance of Canada (alliancecan.ca). E.S.S. acknowledges the support of
the Natural Sciences and Engineering Research Council of Canada (NSERC) through Discovery
Grant RGPIN-2024-06711. S.T., D.C., and S.K. acknowledge support by NSF CAREER Award No. DMR-2238135.


\bibliographystyle{longapsrev4-2}
\bibliography{bibtex}

\begin{thebibliography}{67}%
\makeatletter
\providecommand \@ifxundefined [1]{%
 \@ifx{#1\undefined}
}%
\providecommand \@ifnum [1]{%
 \ifnum #1\expandafter \@firstoftwo
 \else \expandafter \@secondoftwo
 \fi
}%
\providecommand \@ifx [1]{%
 \ifx #1\expandafter \@firstoftwo
 \else \expandafter \@secondoftwo
 \fi
}%
\providecommand \natexlab [1]{#1}%
\providecommand \enquote  [1]{``#1''}%
\providecommand \bibnamefont  [1]{#1}%
\providecommand \bibfnamefont [1]{#1}%
\providecommand \citenamefont [1]{#1}%
\providecommand \href@noop [0]{\@secondoftwo}%
\providecommand \href [0]{\begingroup \@sanitize@url \@href}%
\providecommand \@href[1]{\@@startlink{#1}\@@href}%
\providecommand \@@href[1]{\endgroup#1\@@endlink}%
\providecommand \@sanitize@url [0]{\catcode `\\12\catcode `\$12\catcode `\&12\catcode `\#12\catcode `\^12\catcode `\_12\catcode `\%12\relax}%
\providecommand \@@startlink[1]{}%
\providecommand \@@endlink[0]{}%
\providecommand \url  [0]{\begingroup\@sanitize@url \@url }%
\providecommand \@url [1]{\endgroup\@href {#1}{\urlprefix }}%
\providecommand \urlprefix  [0]{URL }%
\providecommand \Eprint [0]{\href }%
\providecommand \doibase [0]{https://doi.org/}%
\providecommand \selectlanguage [0]{\@gobble}%
\providecommand \bibinfo  [0]{\@secondoftwo}%
\providecommand \bibfield  [0]{\@secondoftwo}%
\providecommand \translation [1]{[#1]}%
\providecommand \BibitemOpen [0]{}%
\providecommand \bibitemStop [0]{}%
\providecommand \bibitemNoStop [0]{.\EOS\space}%
\providecommand \EOS [0]{\spacefactor3000\relax}%
\providecommand \BibitemShut  [1]{\csname bibitem#1\endcsname}%
\let\auto@bib@innerbib\@empty
\bibitem [{\citenamefont {Savary}\ and\ \citenamefont {Balents}(2016)}]{Savary2017}%
  \BibitemOpen
  \bibfield  {author} {\bibinfo {author} {\bibfnamefont {L.}~\bibnamefont {Savary}}\ and\ \bibinfo {author} {\bibfnamefont {L.}~\bibnamefont {Balents}},\ }\bibfield  {title} {\bibinfo {title} {Quantum spin liquids: a review},\ }\href {https://doi.org/10.1088/0034-4885/80/1/016502} {\bibfield  {journal} {\bibinfo  {journal} {Reports on Progress in Physics}\ }\textbf {\bibinfo {volume} {80}},\ \bibinfo {pages} {016502} (\bibinfo {year} {2016})}\BibitemShut {NoStop}%
\bibitem [{\citenamefont {Sachdev}(2023)}]{Sachdev_2023}%
  \BibitemOpen
  \bibfield  {author} {\bibinfo {author} {\bibfnamefont {S.}~\bibnamefont {Sachdev}},\ }\href@noop {} {\emph {\bibinfo {title} {Quantum Phases of Matter}}}\ (\bibinfo  {publisher} {Cambridge University Press},\ \bibinfo {year} {2023})\BibitemShut {NoStop}%
\bibitem [{\citenamefont {Kitaev}(2006)}]{Kitaev2006}%
  \BibitemOpen
  \bibfield  {author} {\bibinfo {author} {\bibfnamefont {A.}~\bibnamefont {Kitaev}},\ }\bibfield  {title} {\bibinfo {title} {Anyons in an exactly solved model and beyond},\ }\href {https://doi.org/https://doi.org/10.1016/j.aop.2005.10.005} {\bibfield  {journal} {\bibinfo  {journal} {Annals of Physics}\ }\textbf {\bibinfo {volume} {321}},\ \bibinfo {pages} {2} (\bibinfo {year} {2006})}\BibitemShut {NoStop}%
\bibitem [{\citenamefont {Yokoi}\ \emph {et~al.}(2021)\citenamefont {Yokoi}, \citenamefont {Ma}, \citenamefont {Kasahara}, \citenamefont {Kasahara}, \citenamefont {Shibauchi}, \citenamefont {Kurita}, \citenamefont {Tanaka}, \citenamefont {Nasu}, \citenamefont {Motome}, \citenamefont {Hickey}, \citenamefont {Trebst},\ and\ \citenamefont {Matsuda}}]{Yokoi2021}%
  \BibitemOpen
  \bibfield  {author} {\bibinfo {author} {\bibfnamefont {T.}~\bibnamefont {Yokoi}}, \bibinfo {author} {\bibfnamefont {S.}~\bibnamefont {Ma}}, \bibinfo {author} {\bibfnamefont {Y.}~\bibnamefont {Kasahara}}, \bibinfo {author} {\bibfnamefont {S.}~\bibnamefont {Kasahara}}, \bibinfo {author} {\bibfnamefont {T.}~\bibnamefont {Shibauchi}}, \bibinfo {author} {\bibfnamefont {N.}~\bibnamefont {Kurita}}, \bibinfo {author} {\bibfnamefont {H.}~\bibnamefont {Tanaka}}, \bibinfo {author} {\bibfnamefont {J.}~\bibnamefont {Nasu}}, \bibinfo {author} {\bibfnamefont {Y.}~\bibnamefont {Motome}}, \bibinfo {author} {\bibfnamefont {C.}~\bibnamefont {Hickey}}, \bibinfo {author} {\bibfnamefont {S.}~\bibnamefont {Trebst}},\ and\ \bibinfo {author} {\bibfnamefont {Y.}~\bibnamefont {Matsuda}},\ }\bibfield  {title} {\bibinfo {title} {Half-integer quantized anomalous thermal Hall effect in the Kitaev material candidate $\alpha-RuCl_3$},\ }\href {https://doi.org/10.1126/science.aay5551} {\bibfield  {journal} {\bibinfo  {journal} {Science}\
  }\textbf {\bibinfo {volume} {373}},\ \bibinfo {pages} {568} (\bibinfo {year} {2021})}\BibitemShut {NoStop}%
\bibitem [{\citenamefont {Czajka}\ \emph {et~al.}(2023)\citenamefont {Czajka}, \citenamefont {Gao}, \citenamefont {Hirschberger}, \citenamefont {Lampen-Kelley}, \citenamefont {Banerjee}, \citenamefont {Quirk}, \citenamefont {Mandrus}, \citenamefont {Nagler},\ and\ \citenamefont {Ong}}]{Czajka2023}%
  \BibitemOpen
  \bibfield  {author} {\bibinfo {author} {\bibfnamefont {P.}~\bibnamefont {Czajka}}, \bibinfo {author} {\bibfnamefont {T.}~\bibnamefont {Gao}}, \bibinfo {author} {\bibfnamefont {M.}~\bibnamefont {Hirschberger}}, \bibinfo {author} {\bibfnamefont {P.}~\bibnamefont {Lampen-Kelley}}, \bibinfo {author} {\bibfnamefont {A.}~\bibnamefont {Banerjee}}, \bibinfo {author} {\bibfnamefont {N.}~\bibnamefont {Quirk}}, \bibinfo {author} {\bibfnamefont {D.~G.}\ \bibnamefont {Mandrus}}, \bibinfo {author} {\bibfnamefont {S.~E.}\ \bibnamefont {Nagler}},\ and\ \bibinfo {author} {\bibfnamefont {N.~P.}\ \bibnamefont {Ong}},\ }\bibfield  {title} {\bibinfo {title} {Planar thermal Hall effect of topological bosons in the Kitaev magnet $\alpha$-RuCl3},\ }\href {https://doi.org/10.1038/s41563-022-01397-w} {\bibfield  {journal} {\bibinfo  {journal} {Nature Materials}\ }\textbf {\bibinfo {volume} {22}},\ \bibinfo {pages} {36} (\bibinfo {year} {2023})}\BibitemShut {NoStop}%
\bibitem [{\citenamefont {Lefran\ifmmode~\mbox{\c{c}}\else \c{c}\fi{}ois}\ \emph {et~al.}(2022)\citenamefont {Lefran\ifmmode~\mbox{\c{c}}\else \c{c}\fi{}ois}, \citenamefont {Grissonnanche}, \citenamefont {Baglo}, \citenamefont {Lampen-Kelley}, \citenamefont {Yan}, \citenamefont {Balz}, \citenamefont {Mandrus}, \citenamefont {Nagler}, \citenamefont {Kim}, \citenamefont {Kim}, \citenamefont {Doiron-Leyraud},\ and\ \citenamefont {Taillefer}}]{Grissonnanche2022}%
  \BibitemOpen
  \bibfield  {author} {\bibinfo {author} {\bibfnamefont {E.}~\bibnamefont {Lefran\ifmmode~\mbox{\c{c}}\else \c{c}\fi{}ois}}, \bibinfo {author} {\bibfnamefont {G.}~\bibnamefont {Grissonnanche}}, \bibinfo {author} {\bibfnamefont {J.}~\bibnamefont {Baglo}}, \bibinfo {author} {\bibfnamefont {P.}~\bibnamefont {Lampen-Kelley}}, \bibinfo {author} {\bibfnamefont {J.-Q.}\ \bibnamefont {Yan}}, \bibinfo {author} {\bibfnamefont {C.}~\bibnamefont {Balz}}, \bibinfo {author} {\bibfnamefont {D.}~\bibnamefont {Mandrus}}, \bibinfo {author} {\bibfnamefont {S.~E.}\ \bibnamefont {Nagler}}, \bibinfo {author} {\bibfnamefont {S.}~\bibnamefont {Kim}}, \bibinfo {author} {\bibfnamefont {Y.-J.}\ \bibnamefont {Kim}}, \bibinfo {author} {\bibfnamefont {N.}~\bibnamefont {Doiron-Leyraud}},\ and\ \bibinfo {author} {\bibfnamefont {L.}~\bibnamefont {Taillefer}},\ }\bibfield  {title} {\bibinfo {title} {Evidence of a Phonon Hall Effect in the Kitaev Spin Liquid Candidate $\ensuremath{\alpha}\text{\ensuremath{-}}{\mathrm{RuCl}}_{3}$},\ }\href
  {https://doi.org/10.1103/PhysRevX.12.021025} {\bibfield  {journal} {\bibinfo  {journal} {Phys. Rev. X}\ }\textbf {\bibinfo {volume} {12}},\ \bibinfo {pages} {021025} (\bibinfo {year} {2022})}\BibitemShut {NoStop}%
\bibitem [{\citenamefont {Lee}(2021)}]{PALee2021}%
  \BibitemOpen
  \bibfield  {author} {\bibinfo {author} {\bibfnamefont {P.~A.}\ \bibnamefont {Lee}},\ }\bibfield  {title} {\bibinfo {title} {Quantized (or not quantized) thermal Hall effect and oscillations in the thermal conductivity in the Kitaev spin liquid candidate $\alpha$-RuCl$_3$},\ }\bibfield  {journal} {\bibinfo  {journal} {Condensed matter Journal Club}\ }\href {https://doi.org/10.36471/JCCM\_November\_2021\_02} {10.36471/JCCM\_November\_2021\_02} (\bibinfo {year} {2021})\BibitemShut {NoStop}%
\bibitem [{\citenamefont {Kee}(2023)}]{Kee2023}%
  \BibitemOpen
  \bibfield  {author} {\bibinfo {author} {\bibfnamefont {H.-Y.}\ \bibnamefont {Kee}},\ }\bibfield  {title} {\bibinfo {title} {Thermal Hall conductivity of $\alpha$-RuCl$_3$},\ }\href {https://doi.org/10.1038/s41563-022-01444-6} {\bibfield  {journal} {\bibinfo  {journal} {Nature Materials}\ }\textbf {\bibinfo {volume} {22}},\ \bibinfo {pages} {6} (\bibinfo {year} {2023})}\BibitemShut {NoStop}%
\bibitem [{goo(2023)}]{google}%
  \BibitemOpen
  \bibfield  {title} {\bibinfo {title} {Non-Abelian braiding of graph vertices in a superconducting processor},\ }\href@noop {} {\bibfield  {journal} {\bibinfo  {journal} {Nature}\ }\textbf {\bibinfo {volume} {618}},\ \bibinfo {pages} {264} (\bibinfo {year} {2023})}\BibitemShut {NoStop}%
\bibitem [{\citenamefont {Minev}\ \emph {et~al.}(2025)\citenamefont {Minev}, \citenamefont {Najafi}, \citenamefont {Majumder}, \citenamefont {Wang}, \citenamefont {Stern}, \citenamefont {Kim}, \citenamefont {Jian},\ and\ \citenamefont {Zhu}}]{minev2025}%
  \BibitemOpen
  \bibfield  {author} {\bibinfo {author} {\bibfnamefont {Z.~K.}\ \bibnamefont {Minev}}, \bibinfo {author} {\bibfnamefont {K.}~\bibnamefont {Najafi}}, \bibinfo {author} {\bibfnamefont {S.}~\bibnamefont {Majumder}}, \bibinfo {author} {\bibfnamefont {J.}~\bibnamefont {Wang}}, \bibinfo {author} {\bibfnamefont {A.}~\bibnamefont {Stern}}, \bibinfo {author} {\bibfnamefont {E.-A.}\ \bibnamefont {Kim}}, \bibinfo {author} {\bibfnamefont {C.-M.}\ \bibnamefont {Jian}},\ and\ \bibinfo {author} {\bibfnamefont {G.}~\bibnamefont {Zhu}},\ }\bibinfo {title} {Realizing string-net condensation: Fibonacci anyon braiding for universal gates and sampling chromatic polynomials},\ \Eprint {https://arxiv.org/abs/2406.12820} {arXiv:2406.12820} \BibitemShut {NoStop}%
\bibitem [{\citenamefont {Xu}\ \emph {et~al.}(2024)\citenamefont {Xu}, \citenamefont {Sun}, \citenamefont {Wang}, \citenamefont {Li}, \citenamefont {Zhu}, \citenamefont {Dong}, \citenamefont {Deng}, \citenamefont {Zhang}, \citenamefont {Chen}, \citenamefont {Wu}, \citenamefont {Zhang}, \citenamefont {Jin}, \citenamefont {Zhu}, \citenamefont {Gao}, \citenamefont {Zhang}, \citenamefont {Wang}, \citenamefont {Zou}, \citenamefont {Tan}, \citenamefont {Shen}, \citenamefont {Zhong}, \citenamefont {Bao}, \citenamefont {Li}, \citenamefont {Jiang}, \citenamefont {Yu}, \citenamefont {Song}, \citenamefont {Zhang}, \citenamefont {Xiang}, \citenamefont {Guo}, \citenamefont {Wang}, \citenamefont {Song}, \citenamefont {Wang},\ and\ \citenamefont {Deng}}]{Xu2024}%
  \BibitemOpen
  \bibfield  {author} {\bibinfo {author} {\bibfnamefont {S.}~\bibnamefont {Xu}}, \bibinfo {author} {\bibfnamefont {Z.-Z.}\ \bibnamefont {Sun}}, \bibinfo {author} {\bibfnamefont {K.}~\bibnamefont {Wang}}, \bibinfo {author} {\bibfnamefont {H.}~\bibnamefont {Li}}, \bibinfo {author} {\bibfnamefont {Z.}~\bibnamefont {Zhu}}, \bibinfo {author} {\bibfnamefont {H.}~\bibnamefont {Dong}}, \bibinfo {author} {\bibfnamefont {J.}~\bibnamefont {Deng}}, \bibinfo {author} {\bibfnamefont {X.}~\bibnamefont {Zhang}}, \bibinfo {author} {\bibfnamefont {J.}~\bibnamefont {Chen}}, \bibinfo {author} {\bibfnamefont {Y.}~\bibnamefont {Wu}}, \bibinfo {author} {\bibfnamefont {C.}~\bibnamefont {Zhang}}, \bibinfo {author} {\bibfnamefont {F.}~\bibnamefont {Jin}}, \bibinfo {author} {\bibfnamefont {X.}~\bibnamefont {Zhu}}, \bibinfo {author} {\bibfnamefont {Y.}~\bibnamefont {Gao}}, \bibinfo {author} {\bibfnamefont {A.}~\bibnamefont {Zhang}}, \bibinfo {author} {\bibfnamefont {N.}~\bibnamefont {Wang}}, \bibinfo {author} {\bibfnamefont
  {Y.}~\bibnamefont {Zou}}, \bibinfo {author} {\bibfnamefont {Z.}~\bibnamefont {Tan}}, \bibinfo {author} {\bibfnamefont {F.}~\bibnamefont {Shen}}, \bibinfo {author} {\bibfnamefont {J.}~\bibnamefont {Zhong}}, \bibinfo {author} {\bibfnamefont {Z.}~\bibnamefont {Bao}}, \bibinfo {author} {\bibfnamefont {W.}~\bibnamefont {Li}}, \bibinfo {author} {\bibfnamefont {W.}~\bibnamefont {Jiang}}, \bibinfo {author} {\bibfnamefont {L.-W.}\ \bibnamefont {Yu}}, \bibinfo {author} {\bibfnamefont {Z.}~\bibnamefont {Song}}, \bibinfo {author} {\bibfnamefont {P.}~\bibnamefont {Zhang}}, \bibinfo {author} {\bibfnamefont {L.}~\bibnamefont {Xiang}}, \bibinfo {author} {\bibfnamefont {Q.}~\bibnamefont {Guo}}, \bibinfo {author} {\bibfnamefont {Z.}~\bibnamefont {Wang}}, \bibinfo {author} {\bibfnamefont {C.}~\bibnamefont {Song}}, \bibinfo {author} {\bibfnamefont {H.}~\bibnamefont {Wang}},\ and\ \bibinfo {author} {\bibfnamefont {D.-L.}\ \bibnamefont {Deng}},\ }\bibfield  {title} {\bibinfo {title} {Non-Abelian braiding of Fibonacci anyons with
  a superconducting processor},\ }\href {https://doi.org/10.1038/s41567-024-02529-6} {\bibfield  {journal} {\bibinfo  {journal} {Nature Physics}\ }\textbf {\bibinfo {volume} {20}},\ \bibinfo {pages} {1469} (\bibinfo {year} {2024})}\BibitemShut {NoStop}%
\bibitem [{\citenamefont {Levin}\ and\ \citenamefont {Wen}(2006)}]{Levin2006TEE}%
  \BibitemOpen
  \bibfield  {author} {\bibinfo {author} {\bibfnamefont {M.}~\bibnamefont {Levin}}\ and\ \bibinfo {author} {\bibfnamefont {X.-G.}\ \bibnamefont {Wen}},\ }\bibfield  {title} {\bibinfo {title} {Detecting Topological Order in a Ground State Wave Function},\ }\href {https://doi.org/10.1103/PhysRevLett.96.110405} {\bibfield  {journal} {\bibinfo  {journal} {Phys. Rev. Lett.}\ }\textbf {\bibinfo {volume} {96}},\ \bibinfo {pages} {110405} (\bibinfo {year} {2006})}\BibitemShut {NoStop}%
\bibitem [{\citenamefont {Kitaev}\ and\ \citenamefont {Preskill}(2006)}]{TEEKitaev2006}%
  \BibitemOpen
  \bibfield  {author} {\bibinfo {author} {\bibfnamefont {A.}~\bibnamefont {Kitaev}}\ and\ \bibinfo {author} {\bibfnamefont {J.}~\bibnamefont {Preskill}},\ }\bibfield  {title} {\bibinfo {title} {Topological Entanglement Entropy},\ }\href {https://doi.org/10.1103/PhysRevLett.96.110404} {\bibfield  {journal} {\bibinfo  {journal} {Phys. Rev. Lett.}\ }\textbf {\bibinfo {volume} {96}},\ \bibinfo {pages} {110404} (\bibinfo {year} {2006})}\BibitemShut {NoStop}%
\bibitem [{\citenamefont {Wang}\ \emph {et~al.}(2025)\citenamefont {Wang}, \citenamefont {Song}, \citenamefont {Lyu}, \citenamefont {Witczak-Krempa},\ and\ \citenamefont {Meng}}]{wangEntanglement2024}%
  \BibitemOpen
  \bibfield  {author} {\bibinfo {author} {\bibfnamefont {T.-T.}\ \bibnamefont {Wang}}, \bibinfo {author} {\bibfnamefont {M.}~\bibnamefont {Song}}, \bibinfo {author} {\bibfnamefont {L.}~\bibnamefont {Lyu}}, \bibinfo {author} {\bibfnamefont {W.}~\bibnamefont {Witczak-Krempa}},\ and\ \bibinfo {author} {\bibfnamefont {Z.~Y.}\ \bibnamefont {Meng}},\ }\bibfield  {title} {\bibinfo {title} {Entanglement microscopy and tomography in many-body systems},\ }\href {https://doi.org/10.1038/s41467-024-55354-z} {\bibfield  {journal} {\bibinfo  {journal} {Nature Communications}\ }\textbf {\bibinfo {volume} {16}},\ \bibinfo {pages} {96} (\bibinfo {year} {2025})}\BibitemShut {NoStop}%
\bibitem [{\citenamefont {Berthiere}\ and\ \citenamefont {Witczak-Krempa}(2022)}]{berthiere}%
  \BibitemOpen
  \bibfield  {author} {\bibinfo {author} {\bibfnamefont {C.}~\bibnamefont {Berthiere}}\ and\ \bibinfo {author} {\bibfnamefont {W.}~\bibnamefont {Witczak-Krempa}},\ }\bibfield  {title} {\bibinfo {title} {Entanglement of Skeletal Regions},\ }\href {https://doi.org/10.1103/PhysRevLett.128.240502} {\bibfield  {journal} {\bibinfo  {journal} {Phys. Rev. Lett.}\ }\textbf {\bibinfo {volume} {128}},\ \bibinfo {pages} {240502} (\bibinfo {year} {2022})}\BibitemShut {NoStop}%
\bibitem [{\citenamefont {Gühne}\ and\ \citenamefont {Tóth}(2009)}]{Guhne2009}%
  \BibitemOpen
  \bibfield  {author} {\bibinfo {author} {\bibfnamefont {O.}~\bibnamefont {Gühne}}\ and\ \bibinfo {author} {\bibfnamefont {G.}~\bibnamefont {Tóth}},\ }\bibfield  {title} {\bibinfo {title} {Entanglement detection},\ }\href {https://doi.org/https://doi.org/10.1016/j.physrep.2009.02.004} {\bibfield  {journal} {\bibinfo  {journal} {Physics Reports}\ }\textbf {\bibinfo {volume} {474}},\ \bibinfo {pages} {1} (\bibinfo {year} {2009})}\BibitemShut {NoStop}%
\bibitem [{\citenamefont {Hauke}\ \emph {et~al.}(2016)\citenamefont {Hauke}, \citenamefont {Heyl}, \citenamefont {Tagliacozzo},\ and\ \citenamefont {Zoller}}]{Hauke2016}%
  \BibitemOpen
  \bibfield  {author} {\bibinfo {author} {\bibfnamefont {P.}~\bibnamefont {Hauke}}, \bibinfo {author} {\bibfnamefont {M.}~\bibnamefont {Heyl}}, \bibinfo {author} {\bibfnamefont {L.}~\bibnamefont {Tagliacozzo}},\ and\ \bibinfo {author} {\bibfnamefont {P.}~\bibnamefont {Zoller}},\ }\bibfield  {title} {\bibinfo {title} {Measuring multipartite entanglement through dynamic susceptibilities},\ }\href@noop {} {\bibfield  {journal} {\bibinfo  {journal} {Nature Physics}\ }\textbf {\bibinfo {volume} {12}},\ \bibinfo {pages} {778} (\bibinfo {year} {2016})}\BibitemShut {NoStop}%
\bibitem [{\citenamefont {Frérot}\ \emph {et~al.}(2023)\citenamefont {Frérot}, \citenamefont {Fadel},\ and\ \citenamefont {Lewenstein}}]{Frerot_2023}%
  \BibitemOpen
  \bibfield  {author} {\bibinfo {author} {\bibfnamefont {I.}~\bibnamefont {Frérot}}, \bibinfo {author} {\bibfnamefont {M.}~\bibnamefont {Fadel}},\ and\ \bibinfo {author} {\bibfnamefont {M.}~\bibnamefont {Lewenstein}},\ }\bibfield  {title} {\bibinfo {title} {Probing quantum correlations in many-body systems: a review of scalable methods},\ }\href {https://doi.org/10.1088/1361-6633/acf8d7} {\bibfield  {journal} {\bibinfo  {journal} {Reports on Progress in Physics}\ }\textbf {\bibinfo {volume} {86}},\ \bibinfo {pages} {114001} (\bibinfo {year} {2023})}\BibitemShut {NoStop}%
\bibitem [{\citenamefont {Laurell}\ \emph {et~al.}(2025)\citenamefont {Laurell}, \citenamefont {Scheie}, \citenamefont {Dagotto},\ and\ \citenamefont {Tennant}}]{Laurell2025}%
  \BibitemOpen
  \bibfield  {author} {\bibinfo {author} {\bibfnamefont {P.}~\bibnamefont {Laurell}}, \bibinfo {author} {\bibfnamefont {A.}~\bibnamefont {Scheie}}, \bibinfo {author} {\bibfnamefont {E.}~\bibnamefont {Dagotto}},\ and\ \bibinfo {author} {\bibfnamefont {D.~A.}\ \bibnamefont {Tennant}},\ }\bibfield  {title} {\bibinfo {title} {Witnessing Entanglement and Quantum Correlations in Condensed Matter: A Review},\ }\href {https://doi.org/https://doi.org/10.1002/qute.202400196} {\bibfield  {journal} {\bibinfo  {journal} {Advanced Quantum Technologies}\ }\textbf {\bibinfo {volume} {8}},\ \bibinfo {pages} {2400196} (\bibinfo {year} {2025})}\BibitemShut {NoStop}%
\bibitem [{\citenamefont {T\'oth}(2012)}]{Toth2012}%
  \BibitemOpen
  \bibfield  {author} {\bibinfo {author} {\bibfnamefont {G.}~\bibnamefont {T\'oth}},\ }\bibfield  {title} {\bibinfo {title} {Multipartite entanglement and high-precision metrology},\ }\href {https://doi.org/10.1103/PhysRevA.85.022322} {\bibfield  {journal} {\bibinfo  {journal} {Phys. Rev. A}\ }\textbf {\bibinfo {volume} {85}},\ \bibinfo {pages} {022322} (\bibinfo {year} {2012})}\BibitemShut {NoStop}%
\bibitem [{\citenamefont {Hyllus}\ \emph {et~al.}(2012)\citenamefont {Hyllus}, \citenamefont {Laskowski}, \citenamefont {Krischek}, \citenamefont {Schwemmer}, \citenamefont {Wieczorek}, \citenamefont {Weinfurter}, \citenamefont {Pezz\'e},\ and\ \citenamefont {Smerzi}}]{Hyllus2012}%
  \BibitemOpen
  \bibfield  {author} {\bibinfo {author} {\bibfnamefont {P.}~\bibnamefont {Hyllus}}, \bibinfo {author} {\bibfnamefont {W.}~\bibnamefont {Laskowski}}, \bibinfo {author} {\bibfnamefont {R.}~\bibnamefont {Krischek}}, \bibinfo {author} {\bibfnamefont {C.}~\bibnamefont {Schwemmer}}, \bibinfo {author} {\bibfnamefont {W.}~\bibnamefont {Wieczorek}}, \bibinfo {author} {\bibfnamefont {H.}~\bibnamefont {Weinfurter}}, \bibinfo {author} {\bibfnamefont {L.}~\bibnamefont {Pezz\'e}},\ and\ \bibinfo {author} {\bibfnamefont {A.}~\bibnamefont {Smerzi}},\ }\bibfield  {title} {\bibinfo {title} {Fisher information and multiparticle entanglement},\ }\href {https://doi.org/10.1103/PhysRevA.85.022321} {\bibfield  {journal} {\bibinfo  {journal} {Phys. Rev. A}\ }\textbf {\bibinfo {volume} {85}},\ \bibinfo {pages} {022321} (\bibinfo {year} {2012})}\BibitemShut {NoStop}%
\bibitem [{\citenamefont {Hickey}\ and\ \citenamefont {Trebst}(2019)}]{Hickey2019}%
  \BibitemOpen
  \bibfield  {author} {\bibinfo {author} {\bibfnamefont {C.}~\bibnamefont {Hickey}}\ and\ \bibinfo {author} {\bibfnamefont {S.}~\bibnamefont {Trebst}},\ }\bibfield  {title} {\bibinfo {title} {Emergence of a field-driven U(1) spin liquid in the Kitaev honeycomb model},\ }\href@noop {} {\bibfield  {journal} {\bibinfo  {journal} {Nature Communications}\ }\textbf {\bibinfo {volume} {10}},\ \bibinfo {pages} {530} (\bibinfo {year} {2019})}\BibitemShut {NoStop}%
\bibitem [{\citenamefont {Zhu}\ \emph {et~al.}(2018)\citenamefont {Zhu}, \citenamefont {Kimchi}, \citenamefont {Sheng},\ and\ \citenamefont {Fu}}]{Zhu2018}%
  \BibitemOpen
  \bibfield  {author} {\bibinfo {author} {\bibfnamefont {Z.}~\bibnamefont {Zhu}}, \bibinfo {author} {\bibfnamefont {I.}~\bibnamefont {Kimchi}}, \bibinfo {author} {\bibfnamefont {D.~N.}\ \bibnamefont {Sheng}},\ and\ \bibinfo {author} {\bibfnamefont {L.}~\bibnamefont {Fu}},\ }\bibfield  {title} {\bibinfo {title} {Robust non-Abelian spin liquid and a possible intermediate phase in the antiferromagnetic Kitaev model with magnetic field},\ }\href {https://doi.org/10.1103/PhysRevB.97.241110} {\bibfield  {journal} {\bibinfo  {journal} {Phys. Rev. B}\ }\textbf {\bibinfo {volume} {97}},\ \bibinfo {pages} {241110} (\bibinfo {year} {2018})}\BibitemShut {NoStop}%
\bibitem [{\citenamefont {Gohlke}\ \emph {et~al.}(2018)\citenamefont {Gohlke}, \citenamefont {Moessner},\ and\ \citenamefont {Pollmann}}]{Pollmann2018}%
  \BibitemOpen
  \bibfield  {author} {\bibinfo {author} {\bibfnamefont {M.}~\bibnamefont {Gohlke}}, \bibinfo {author} {\bibfnamefont {R.}~\bibnamefont {Moessner}},\ and\ \bibinfo {author} {\bibfnamefont {F.}~\bibnamefont {Pollmann}},\ }\bibfield  {title} {\bibinfo {title} {Dynamical and topological properties of the Kitaev model in a [111] magnetic field},\ }\href {https://doi.org/10.1103/PhysRevB.98.014418} {\bibfield  {journal} {\bibinfo  {journal} {Physical Review B}\ }\textbf {\bibinfo {volume} {98}},\ \bibinfo {pages} {014418} (\bibinfo {year} {2018})},\ \bibinfo {note} {pRB}\BibitemShut {NoStop}%
\bibitem [{\citenamefont {Holdhusen}\ \emph {et~al.}(2024)\citenamefont {Holdhusen}, \citenamefont {Huerga},\ and\ \citenamefont {Ortiz}}]{Holdhusen2024}%
  \BibitemOpen
  \bibfield  {author} {\bibinfo {author} {\bibfnamefont {W.}~\bibnamefont {Holdhusen}}, \bibinfo {author} {\bibfnamefont {D.}~\bibnamefont {Huerga}},\ and\ \bibinfo {author} {\bibfnamefont {G.}~\bibnamefont {Ortiz}},\ }\bibfield  {title} {\bibinfo {title} {Emergent magnetic order in the antiferromagnetic Kitaev model in a [111] field},\ }\href {https://doi.org/10.1103/PhysRevB.109.174411} {\bibfield  {journal} {\bibinfo  {journal} {Phys. Rev. B}\ }\textbf {\bibinfo {volume} {109}},\ \bibinfo {pages} {174411} (\bibinfo {year} {2024})}\BibitemShut {NoStop}%
\bibitem [{\citenamefont {Zhang}\ \emph {et~al.}(2022)\citenamefont {Zhang}, \citenamefont {Hal{\'a}sz},\ and\ \citenamefont {Batista}}]{Zhang2022}%
  \BibitemOpen
  \bibfield  {author} {\bibinfo {author} {\bibfnamefont {S.-S.}\ \bibnamefont {Zhang}}, \bibinfo {author} {\bibfnamefont {G.~B.}\ \bibnamefont {Hal{\'a}sz}},\ and\ \bibinfo {author} {\bibfnamefont {C.~D.}\ \bibnamefont {Batista}},\ }\bibfield  {title} {\bibinfo {title} {Theory of the Kitaev model in a [111] magnetic field},\ }\href {https://doi.org/10.1038/s41467-022-28014-3} {\bibfield  {journal} {\bibinfo  {journal} {Nature Communications}\ }\textbf {\bibinfo {volume} {13}},\ \bibinfo {pages} {399} (\bibinfo {year} {2022})}\BibitemShut {NoStop}%
\bibitem [{\citenamefont {Jungnitsch}\ \emph {et~al.}(2011)\citenamefont {Jungnitsch}, \citenamefont {Moroder},\ and\ \citenamefont {Gühne}}]{Guhne2011}%
  \BibitemOpen
  \bibfield  {author} {\bibinfo {author} {\bibfnamefont {B.}~\bibnamefont {Jungnitsch}}, \bibinfo {author} {\bibfnamefont {T.}~\bibnamefont {Moroder}},\ and\ \bibinfo {author} {\bibfnamefont {O.}~\bibnamefont {Gühne}},\ }\bibfield  {title} {\bibinfo {title} {Taming Multiparticle Entanglement},\ }\href {https://doi.org/10.1103/PhysRevLett.106.190502} {\bibfield  {journal} {\bibinfo  {journal} {Physical Review Letters}\ }\textbf {\bibinfo {volume} {106}},\ \bibinfo {pages} {190502} (\bibinfo {year} {2011})}\BibitemShut {NoStop}%
\bibitem [{\citenamefont {Peres}(1996)}]{Peres1996}%
  \BibitemOpen
  \bibfield  {author} {\bibinfo {author} {\bibfnamefont {A.}~\bibnamefont {Peres}},\ }\bibfield  {title} {\bibinfo {title} {Separability Criterion for Density Matrices},\ }\href {https://doi.org/10.1103/PhysRevLett.77.1413} {\bibfield  {journal} {\bibinfo  {journal} {Phys. Rev. Lett.}\ }\textbf {\bibinfo {volume} {77}},\ \bibinfo {pages} {1413} (\bibinfo {year} {1996})}\BibitemShut {NoStop}%
\bibitem [{\citenamefont {Lyu}\ \emph {et~al.}()\citenamefont {Lyu}, \citenamefont {Witczak‑Krempa}, \citenamefont {Läuchli},\ and\ \citenamefont {Sørensen}}]{inprep}%
  \BibitemOpen
  \bibfield  {author} {\bibinfo {author} {\bibfnamefont {L.}~\bibnamefont {Lyu}}, \bibinfo {author} {\bibfnamefont {W.}~\bibnamefont {Witczak‑Krempa}}, \bibinfo {author} {\bibfnamefont {A.}~\bibnamefont {Läuchli}},\ and\ \bibinfo {author} {\bibfnamefont {E.~S.}\ \bibnamefont {Sørensen}},\ }\bibinfo {note} {in preparation}\BibitemShut {NoStop}%
\bibitem [{\citenamefont {Lyu}\ \emph {et~al.}(2025)\citenamefont {Lyu}, \citenamefont {Song}, \citenamefont {Wang}, \citenamefont {Meng},\ and\ \citenamefont {Witczak-Krempa}}]{Lyu2024}%
  \BibitemOpen
  \bibfield  {author} {\bibinfo {author} {\bibfnamefont {L.}~\bibnamefont {Lyu}}, \bibinfo {author} {\bibfnamefont {M.}~\bibnamefont {Song}}, \bibinfo {author} {\bibfnamefont {T.-T.}\ \bibnamefont {Wang}}, \bibinfo {author} {\bibfnamefont {Z.~Y.}\ \bibnamefont {Meng}},\ and\ \bibinfo {author} {\bibfnamefont {W.}~\bibnamefont {Witczak-Krempa}},\ }\bibfield  {title} {\bibinfo {title} {Multiparty entanglement microscopy of quantum Ising models in one, two, and three dimensions},\ }\href {https://doi.org/10.1103/PhysRevB.111.245108} {\bibfield  {journal} {\bibinfo  {journal} {Phys. Rev. B}\ }\textbf {\bibinfo {volume} {111}},\ \bibinfo {pages} {245108} (\bibinfo {year} {2025})}\BibitemShut {NoStop}%
\bibitem [{\citenamefont {Lambert}\ and\ \citenamefont {S\o{}rensen}(2020)}]{Lambert2020}%
  \BibitemOpen
  \bibfield  {author} {\bibinfo {author} {\bibfnamefont {J.}~\bibnamefont {Lambert}}\ and\ \bibinfo {author} {\bibfnamefont {E.~S.}\ \bibnamefont {S\o{}rensen}},\ }\bibfield  {title} {\bibinfo {title} {Revealing divergent length scales using quantum Fisher information in the Kitaev honeycomb model},\ }\href {https://doi.org/10.1103/PhysRevB.102.224401} {\bibfield  {journal} {\bibinfo  {journal} {Phys. Rev. B}\ }\textbf {\bibinfo {volume} {102}},\ \bibinfo {pages} {224401} (\bibinfo {year} {2020})}\BibitemShut {NoStop}%
\bibitem [{\citenamefont {Liao}\ \emph {et~al.}(2017)\citenamefont {Liao}, \citenamefont {Xie}, \citenamefont {Chen}, \citenamefont {Liu}, \citenamefont {Xie}, \citenamefont {Huang}, \citenamefont {Normand},\ and\ \citenamefont {Xiang}}]{Liao2017}%
  \BibitemOpen
  \bibfield  {author} {\bibinfo {author} {\bibfnamefont {H.~J.}\ \bibnamefont {Liao}}, \bibinfo {author} {\bibfnamefont {Z.~Y.}\ \bibnamefont {Xie}}, \bibinfo {author} {\bibfnamefont {J.}~\bibnamefont {Chen}}, \bibinfo {author} {\bibfnamefont {Z.~Y.}\ \bibnamefont {Liu}}, \bibinfo {author} {\bibfnamefont {H.~D.}\ \bibnamefont {Xie}}, \bibinfo {author} {\bibfnamefont {R.~Z.}\ \bibnamefont {Huang}}, \bibinfo {author} {\bibfnamefont {B.}~\bibnamefont {Normand}},\ and\ \bibinfo {author} {\bibfnamefont {T.}~\bibnamefont {Xiang}},\ }\bibfield  {title} {\bibinfo {title} {Gapless Spin-Liquid Ground State in the $S=1/2$ Kagome Antiferromagnet},\ }\href {https://doi.org/10.1103/PhysRevLett.118.137202} {\bibfield  {journal} {\bibinfo  {journal} {Phys. Rev. Lett.}\ }\textbf {\bibinfo {volume} {118}},\ \bibinfo {pages} {137202} (\bibinfo {year} {2017})}\BibitemShut {NoStop}%
\bibitem [{\citenamefont {Mekata}(2003)}]{Mekata2003}%
  \BibitemOpen
  \bibfield  {author} {\bibinfo {author} {\bibfnamefont {M.}~\bibnamefont {Mekata}},\ }\bibfield  {title} {\bibinfo {title} {Kagome: The Story of the Basketweave Lattice},\ }\href {https://doi.org/10.1063/1.1564329} {\bibfield  {journal} {\bibinfo  {journal} {Physics Today}\ }\textbf {\bibinfo {volume} {56}},\ \bibinfo {pages} {12} (\bibinfo {year} {2003})}\BibitemShut {NoStop}%
\bibitem [{\citenamefont {He}\ \emph {et~al.}(2017)\citenamefont {He}, \citenamefont {Zaletel}, \citenamefont {Oshikawa},\ and\ \citenamefont {Pollmann}}]{He2017}%
  \BibitemOpen
  \bibfield  {author} {\bibinfo {author} {\bibfnamefont {Y.-C.}\ \bibnamefont {He}}, \bibinfo {author} {\bibfnamefont {M.~P.}\ \bibnamefont {Zaletel}}, \bibinfo {author} {\bibfnamefont {M.}~\bibnamefont {Oshikawa}},\ and\ \bibinfo {author} {\bibfnamefont {F.}~\bibnamefont {Pollmann}},\ }\bibfield  {title} {\bibinfo {title} {Signatures of Dirac Cones in a DMRG Study of the Kagome Heisenberg Model},\ }\href {https://doi.org/10.1103/PhysRevX.7.031020} {\bibfield  {journal} {\bibinfo  {journal} {Phys. Rev. X}\ }\textbf {\bibinfo {volume} {7}},\ \bibinfo {pages} {031020} (\bibinfo {year} {2017})}\BibitemShut {NoStop}%
\bibitem [{\citenamefont {Iqbal}\ \emph {et~al.}(2011)\citenamefont {Iqbal}, \citenamefont {Becca},\ and\ \citenamefont {Poilblanc}}]{Iqbal2011}%
  \BibitemOpen
  \bibfield  {author} {\bibinfo {author} {\bibfnamefont {Y.}~\bibnamefont {Iqbal}}, \bibinfo {author} {\bibfnamefont {F.}~\bibnamefont {Becca}},\ and\ \bibinfo {author} {\bibfnamefont {D.}~\bibnamefont {Poilblanc}},\ }\bibfield  {title} {\bibinfo {title} {Projected wave function study of ${\mathbb{Z}}_{2}$ spin liquids on the kagome lattice for the spin-$\frac{1}{2}$ quantum Heisenberg antiferromagnet},\ }\href {https://doi.org/10.1103/PhysRevB.84.020407} {\bibfield  {journal} {\bibinfo  {journal} {Phys. Rev. B}\ }\textbf {\bibinfo {volume} {84}},\ \bibinfo {pages} {020407} (\bibinfo {year} {2011})}\BibitemShut {NoStop}%
\bibitem [{\citenamefont {Clark}\ \emph {et~al.}(2013)\citenamefont {Clark}, \citenamefont {Kinder}, \citenamefont {Neuscamman}, \citenamefont {Chan},\ and\ \citenamefont {Lawler}}]{Clark2013}%
  \BibitemOpen
  \bibfield  {author} {\bibinfo {author} {\bibfnamefont {B.~K.}\ \bibnamefont {Clark}}, \bibinfo {author} {\bibfnamefont {J.~M.}\ \bibnamefont {Kinder}}, \bibinfo {author} {\bibfnamefont {E.}~\bibnamefont {Neuscamman}}, \bibinfo {author} {\bibfnamefont {G.~K.-L.}\ \bibnamefont {Chan}},\ and\ \bibinfo {author} {\bibfnamefont {M.~J.}\ \bibnamefont {Lawler}},\ }\bibfield  {title} {\bibinfo {title} {Striped Spin Liquid Crystal Ground State Instability of Kagome Antiferromagnets},\ }\href {https://doi.org/10.1103/PhysRevLett.111.187205} {\bibfield  {journal} {\bibinfo  {journal} {Phys. Rev. Lett.}\ }\textbf {\bibinfo {volume} {111}},\ \bibinfo {pages} {187205} (\bibinfo {year} {2013})}\BibitemShut {NoStop}%
\bibitem [{\citenamefont {Leung}\ and\ \citenamefont {Elser}(1993)}]{Leung1993}%
  \BibitemOpen
  \bibfield  {author} {\bibinfo {author} {\bibfnamefont {P.~W.}\ \bibnamefont {Leung}}\ and\ \bibinfo {author} {\bibfnamefont {V.}~\bibnamefont {Elser}},\ }\bibfield  {title} {\bibinfo {title} {Numerical studies of a 36-site kagome antiferromagnet},\ }\href {https://doi.org/10.1103/PhysRevB.47.5459} {\bibfield  {journal} {\bibinfo  {journal} {Phys. Rev. B}\ }\textbf {\bibinfo {volume} {47}},\ \bibinfo {pages} {5459} (\bibinfo {year} {1993})}\BibitemShut {NoStop}%
\bibitem [{\citenamefont {Lecheminant}\ \emph {et~al.}(1997)\citenamefont {Lecheminant}, \citenamefont {Bernu}, \citenamefont {Lhuillier}, \citenamefont {Pierre},\ and\ \citenamefont {Sindzingre}}]{Lecheminant1997}%
  \BibitemOpen
  \bibfield  {author} {\bibinfo {author} {\bibfnamefont {P.}~\bibnamefont {Lecheminant}}, \bibinfo {author} {\bibfnamefont {B.}~\bibnamefont {Bernu}}, \bibinfo {author} {\bibfnamefont {C.}~\bibnamefont {Lhuillier}}, \bibinfo {author} {\bibfnamefont {L.}~\bibnamefont {Pierre}},\ and\ \bibinfo {author} {\bibfnamefont {P.}~\bibnamefont {Sindzingre}},\ }\bibfield  {title} {\bibinfo {title} {Order versus disorder in the quantum Heisenberg antiferromagnet on the kagom\'e lattice using exact spectra analysis},\ }\href {https://doi.org/10.1103/PhysRevB.56.2521} {\bibfield  {journal} {\bibinfo  {journal} {Phys. Rev. B}\ }\textbf {\bibinfo {volume} {56}},\ \bibinfo {pages} {2521} (\bibinfo {year} {1997})}\BibitemShut {NoStop}%
\bibitem [{\citenamefont {L\"auchli}\ \emph {et~al.}(2011)\citenamefont {L\"auchli}, \citenamefont {Sudan},\ and\ \citenamefont {S\o{}rensen}}]{Lauchli2011}%
  \BibitemOpen
  \bibfield  {author} {\bibinfo {author} {\bibfnamefont {A.~M.}\ \bibnamefont {L\"auchli}}, \bibinfo {author} {\bibfnamefont {J.}~\bibnamefont {Sudan}},\ and\ \bibinfo {author} {\bibfnamefont {E.~S.}\ \bibnamefont {S\o{}rensen}},\ }\bibfield  {title} {\bibinfo {title} {Ground-state energy and spin gap of spin-$\frac{1}{2}$ Kagom\'e-Heisenberg antiferromagnetic clusters: Large-scale exact diagonalization results},\ }\href {https://doi.org/10.1103/PhysRevB.83.212401} {\bibfield  {journal} {\bibinfo  {journal} {Phys. Rev. B}\ }\textbf {\bibinfo {volume} {83}},\ \bibinfo {pages} {212401} (\bibinfo {year} {2011})}\BibitemShut {NoStop}%
\bibitem [{\citenamefont {L\"auchli}\ \emph {et~al.}(2019)\citenamefont {L\"auchli}, \citenamefont {Sudan},\ and\ \citenamefont {Moessner}}]{Andreas2019}%
  \BibitemOpen
  \bibfield  {author} {\bibinfo {author} {\bibfnamefont {A.~M.}\ \bibnamefont {L\"auchli}}, \bibinfo {author} {\bibfnamefont {J.}~\bibnamefont {Sudan}},\ and\ \bibinfo {author} {\bibfnamefont {R.}~\bibnamefont {Moessner}},\ }\bibfield  {title} {\bibinfo {title} {$S=\frac{1}{2}$ kagome Heisenberg antiferromagnet revisited},\ }\href {https://doi.org/10.1103/PhysRevB.100.155142} {\bibfield  {journal} {\bibinfo  {journal} {Phys. Rev. B}\ }\textbf {\bibinfo {volume} {100}},\ \bibinfo {pages} {155142} (\bibinfo {year} {2019})}\BibitemShut {NoStop}%
\bibitem [{\citenamefont {Iqbal}\ \emph {et~al.}(2021)\citenamefont {Iqbal}, \citenamefont {Ferrari}, \citenamefont {Chauhan}, \citenamefont {Parola}, \citenamefont {Poilblanc},\ and\ \citenamefont {Becca}}]{Iqbal2021}%
  \BibitemOpen
  \bibfield  {author} {\bibinfo {author} {\bibfnamefont {Y.}~\bibnamefont {Iqbal}}, \bibinfo {author} {\bibfnamefont {F.}~\bibnamefont {Ferrari}}, \bibinfo {author} {\bibfnamefont {A.}~\bibnamefont {Chauhan}}, \bibinfo {author} {\bibfnamefont {A.}~\bibnamefont {Parola}}, \bibinfo {author} {\bibfnamefont {D.}~\bibnamefont {Poilblanc}},\ and\ \bibinfo {author} {\bibfnamefont {F.}~\bibnamefont {Becca}},\ }\bibfield  {title} {\bibinfo {title} {Gutzwiller projected states for the ${J}_{1}\ensuremath{-}{J}_{2}$ Heisenberg model on the Kagome lattice: Achievements and pitfalls},\ }\href {https://doi.org/10.1103/PhysRevB.104.144406} {\bibfield  {journal} {\bibinfo  {journal} {Phys. Rev. B}\ }\textbf {\bibinfo {volume} {104}},\ \bibinfo {pages} {144406} (\bibinfo {year} {2021})}\BibitemShut {NoStop}%
\bibitem [{\citenamefont {Bauer}\ \emph {et~al.}(2014)\citenamefont {Bauer}, \citenamefont {Cincio}, \citenamefont {Keller}, \citenamefont {Dolfi}, \citenamefont {Vidal}, \citenamefont {Trebst},\ and\ \citenamefont {Ludwig}}]{Bauer2014}%
  \BibitemOpen
  \bibfield  {author} {\bibinfo {author} {\bibfnamefont {B.}~\bibnamefont {Bauer}}, \bibinfo {author} {\bibfnamefont {L.}~\bibnamefont {Cincio}}, \bibinfo {author} {\bibfnamefont {B.~P.}\ \bibnamefont {Keller}}, \bibinfo {author} {\bibfnamefont {M.}~\bibnamefont {Dolfi}}, \bibinfo {author} {\bibfnamefont {G.}~\bibnamefont {Vidal}}, \bibinfo {author} {\bibfnamefont {S.}~\bibnamefont {Trebst}},\ and\ \bibinfo {author} {\bibfnamefont {A.~W.~W.}\ \bibnamefont {Ludwig}},\ }\bibfield  {title} {\bibinfo {title} {Chiral spin liquid and emergent anyons in a Kagome lattice Mott insulator},\ }\href {https://doi.org/10.1038/ncomms6137} {\bibfield  {journal} {\bibinfo  {journal} {Nature Communications}\ }\textbf {\bibinfo {volume} {5}},\ \bibinfo {pages} {5137} (\bibinfo {year} {2014})}\BibitemShut {NoStop}%
\bibitem [{\citenamefont {Anderson}(1987)}]{rvb}%
  \BibitemOpen
  \bibfield  {author} {\bibinfo {author} {\bibfnamefont {P.~W.}\ \bibnamefont {Anderson}},\ }\bibfield  {title} {\bibinfo {title} {The Resonating Valence Bond State in $\mathrm{La_2CuO_4}$ and Superconductivity},\ }\href {https://doi.org/10.1126/science.235.4793.1196} {\bibfield  {journal} {\bibinfo  {journal} {Science}\ }\textbf {\bibinfo {volume} {235}},\ \bibinfo {pages} {1196} (\bibinfo {year} {1987})}\BibitemShut {NoStop}%
\bibitem [{\citenamefont {Schuch}\ \emph {et~al.}(2012)\citenamefont {Schuch}, \citenamefont {Poilblanc}, \citenamefont {Cirac},\ and\ \citenamefont {P\'erez-Garc\'{\i}a}}]{Schuch2012}%
  \BibitemOpen
  \bibfield  {author} {\bibinfo {author} {\bibfnamefont {N.}~\bibnamefont {Schuch}}, \bibinfo {author} {\bibfnamefont {D.}~\bibnamefont {Poilblanc}}, \bibinfo {author} {\bibfnamefont {J.~I.}\ \bibnamefont {Cirac}},\ and\ \bibinfo {author} {\bibfnamefont {D.}~\bibnamefont {P\'erez-Garc\'{\i}a}},\ }\bibfield  {title} {\bibinfo {title} {Resonating valence bond states in the PEPS formalism},\ }\href {https://doi.org/10.1103/PhysRevB.86.115108} {\bibfield  {journal} {\bibinfo  {journal} {Phys. Rev. B}\ }\textbf {\bibinfo {volume} {86}},\ \bibinfo {pages} {115108} (\bibinfo {year} {2012})}\BibitemShut {NoStop}%
\bibitem [{\citenamefont {Yang}\ and\ \citenamefont {Yao}(2012)}]{Yang2012}%
  \BibitemOpen
  \bibfield  {author} {\bibinfo {author} {\bibfnamefont {F.}~\bibnamefont {Yang}}\ and\ \bibinfo {author} {\bibfnamefont {H.}~\bibnamefont {Yao}},\ }\bibfield  {title} {\bibinfo {title} {Frustrated Resonating Valence Bond States in Two Dimensions: Classification and Short-Range Correlations},\ }\href {https://doi.org/10.1103/PhysRevLett.109.147209} {\bibfield  {journal} {\bibinfo  {journal} {Phys. Rev. Lett.}\ }\textbf {\bibinfo {volume} {109}},\ \bibinfo {pages} {147209} (\bibinfo {year} {2012})}\BibitemShut {NoStop}%
\bibitem [{\citenamefont {Parez}\ \emph {et~al.}(2023)\citenamefont {Parez}, \citenamefont {Berthiere},\ and\ \citenamefont {Witczak-Krempa}}]{Gilles2023}%
  \BibitemOpen
  \bibfield  {author} {\bibinfo {author} {\bibfnamefont {G.}~\bibnamefont {Parez}}, \bibinfo {author} {\bibfnamefont {C.}~\bibnamefont {Berthiere}},\ and\ \bibinfo {author} {\bibfnamefont {W.}~\bibnamefont {Witczak-Krempa}},\ }\bibfield  {title} {\bibinfo {title} {{Separability and entanglement of resonating valence-bond states}},\ }\href {https://doi.org/10.21468/SciPostPhys.15.2.066} {\bibfield  {journal} {\bibinfo  {journal} {SciPost Phys.}\ }\textbf {\bibinfo {volume} {15}},\ \bibinfo {pages} {066} (\bibinfo {year} {2023})}\BibitemShut {NoStop}%
\bibitem [{\citenamefont {Rokhsar}\ and\ \citenamefont {Kivelson}(1988)}]{RK1988}%
  \BibitemOpen
  \bibfield  {author} {\bibinfo {author} {\bibfnamefont {D.~S.}\ \bibnamefont {Rokhsar}}\ and\ \bibinfo {author} {\bibfnamefont {S.~A.}\ \bibnamefont {Kivelson}},\ }\bibfield  {title} {\bibinfo {title} {Superconductivity and the Quantum Hard-Core Dimer Gas},\ }\href {https://doi.org/10.1103/PhysRevLett.61.2376} {\bibfield  {journal} {\bibinfo  {journal} {Phys. Rev. Lett.}\ }\textbf {\bibinfo {volume} {61}},\ \bibinfo {pages} {2376} (\bibinfo {year} {1988})}\BibitemShut {NoStop}%
\bibitem [{\citenamefont {Tang}\ \emph {et~al.}(2011)\citenamefont {Tang}, \citenamefont {Sandvik},\ and\ \citenamefont {Henley}}]{Tang2011}%
  \BibitemOpen
  \bibfield  {author} {\bibinfo {author} {\bibfnamefont {Y.}~\bibnamefont {Tang}}, \bibinfo {author} {\bibfnamefont {A.~W.}\ \bibnamefont {Sandvik}},\ and\ \bibinfo {author} {\bibfnamefont {C.~L.}\ \bibnamefont {Henley}},\ }\bibfield  {title} {\bibinfo {title} {Properties of resonating-valence-bond spin liquids and critical dimer models},\ }\href {https://doi.org/10.1103/PhysRevB.84.174427} {\bibfield  {journal} {\bibinfo  {journal} {Phys. Rev. B}\ }\textbf {\bibinfo {volume} {84}},\ \bibinfo {pages} {174427} (\bibinfo {year} {2011})}\BibitemShut {NoStop}%
\bibitem [{\citenamefont {Albuquerque}\ and\ \citenamefont {Alet}(2010)}]{Albuquerque2010}%
  \BibitemOpen
  \bibfield  {author} {\bibinfo {author} {\bibfnamefont {A.~F.}\ \bibnamefont {Albuquerque}}\ and\ \bibinfo {author} {\bibfnamefont {F.}~\bibnamefont {Alet}},\ }\bibfield  {title} {\bibinfo {title} {Critical correlations for short-range valence-bond wave functions on the square lattice},\ }\href {https://doi.org/10.1103/PhysRevB.82.180408} {\bibfield  {journal} {\bibinfo  {journal} {Phys. Rev. B}\ }\textbf {\bibinfo {volume} {82}},\ \bibinfo {pages} {180408} (\bibinfo {year} {2010})}\BibitemShut {NoStop}%
\bibitem [{\citenamefont {{Parez}}\ and\ \citenamefont {{Witczak-Krempa}}(2024)}]{parez2024fateentanglement}%
  \BibitemOpen
  \bibfield  {author} {\bibinfo {author} {\bibfnamefont {G.}~\bibnamefont {{Parez}}}\ and\ \bibinfo {author} {\bibfnamefont {W.}~\bibnamefont {{Witczak-Krempa}}},\ }\bibfield  {title} {\bibinfo {title} {{The Fate of Entanglement}},\ }\href {https://doi.org/10.48550/arXiv.2402.06677} {\bibfield  {journal} {\bibinfo  {journal} {arXiv e-prints}\ ,\ \bibinfo {eid} {arXiv:2402.06677}} (\bibinfo {year} {2024})}\BibitemShut {NoStop}%
\bibitem [{\citenamefont {Levin}\ and\ \citenamefont {Wen}(2005)}]{Levin2005}%
  \BibitemOpen
  \bibfield  {author} {\bibinfo {author} {\bibfnamefont {M.~A.}\ \bibnamefont {Levin}}\ and\ \bibinfo {author} {\bibfnamefont {X.-G.}\ \bibnamefont {Wen}},\ }\bibfield  {title} {\bibinfo {title} {String-net condensation: A physical mechanism for topological phases},\ }\href {https://doi.org/10.1103/PhysRevB.71.045110} {\bibfield  {journal} {\bibinfo  {journal} {Phys. Rev. B}\ }\textbf {\bibinfo {volume} {71}},\ \bibinfo {pages} {045110} (\bibinfo {year} {2005})}\BibitemShut {NoStop}%
\bibitem [{\citenamefont {Cappelli}\ \emph {et~al.}(2002)\citenamefont {Cappelli}, \citenamefont {Huerta},\ and\ \citenamefont {Zemba}}]{Cappelli2002}%
  \BibitemOpen
  \bibfield  {author} {\bibinfo {author} {\bibfnamefont {A.}~\bibnamefont {Cappelli}}, \bibinfo {author} {\bibfnamefont {M.}~\bibnamefont {Huerta}},\ and\ \bibinfo {author} {\bibfnamefont {G.~R.}\ \bibnamefont {Zemba}},\ }\bibfield  {title} {\bibinfo {title} {Thermal transport in chiral conformal theories and hierarchical quantum Hall states},\ }\href {https://doi.org/https://doi.org/10.1016/S0550-3213(02)00340-1} {\bibfield  {journal} {\bibinfo  {journal} {Nuclear Physics B}\ }\textbf {\bibinfo {volume} {636}},\ \bibinfo {pages} {568} (\bibinfo {year} {2002})}\BibitemShut {NoStop}%
\bibitem [{\citenamefont {Li}\ and\ \citenamefont {Haldane}(2008)}]{Li_Haldane_2008}%
  \BibitemOpen
  \bibfield  {author} {\bibinfo {author} {\bibfnamefont {H.}~\bibnamefont {Li}}\ and\ \bibinfo {author} {\bibfnamefont {F.~D.~M.}\ \bibnamefont {Haldane}},\ }\bibfield  {title} {\bibinfo {title} {Entanglement Spectrum as a Generalization of Entanglement Entropy: Identification of Topological Order in Non-Abelian Fractional Quantum Hall Effect States},\ }\href {https://doi.org/10.1103/PhysRevLett.101.010504} {\bibfield  {journal} {\bibinfo  {journal} {Phys. Rev. Lett.}\ }\textbf {\bibinfo {volume} {101}},\ \bibinfo {pages} {010504} (\bibinfo {year} {2008})}\BibitemShut {NoStop}%
\bibitem [{\citenamefont {Calabrese}\ \emph {et~al.}(2013)\citenamefont {Calabrese}, \citenamefont {Cardy},\ and\ \citenamefont {Tonni}}]{Calabrese2013}%
  \BibitemOpen
  \bibfield  {author} {\bibinfo {author} {\bibfnamefont {P.}~\bibnamefont {Calabrese}}, \bibinfo {author} {\bibfnamefont {J.}~\bibnamefont {Cardy}},\ and\ \bibinfo {author} {\bibfnamefont {E.}~\bibnamefont {Tonni}},\ }\bibfield  {title} {\bibinfo {title} {Entanglement negativity in extended systems: a field theoretical approach},\ }\href {https://doi.org/10.1088/1742-5468/2013/02/P02008} {\bibfield  {journal} {\bibinfo  {journal} {Journal of Statistical Mechanics: Theory and Experiment}\ }\textbf {\bibinfo {volume} {2013}},\ \bibinfo {pages} {P02008} (\bibinfo {year} {2013})}\BibitemShut {NoStop}%
\bibitem [{\citenamefont {Chen}\ \emph {et~al.}(2011)\citenamefont {Chen}, \citenamefont {Liu},\ and\ \citenamefont {Wen}}]{Chen_2011}%
  \BibitemOpen
  \bibfield  {author} {\bibinfo {author} {\bibfnamefont {X.}~\bibnamefont {Chen}}, \bibinfo {author} {\bibfnamefont {Z.-X.}\ \bibnamefont {Liu}},\ and\ \bibinfo {author} {\bibfnamefont {X.-G.}\ \bibnamefont {Wen}},\ }\bibfield  {title} {\bibinfo {title} {Two-dimensional symmetry-protected topological orders and their protected gapless edge excitations},\ }\bibfield  {journal} {\bibinfo  {journal} {Physical Review B}\ }\textbf {\bibinfo {volume} {84}},\ \href {https://doi.org/10.1103/physrevb.84.235141} {10.1103/physrevb.84.235141} (\bibinfo {year} {2011})\BibitemShut {NoStop}%
\bibitem [{\citenamefont {Horodecki}\ \emph {et~al.}(1996)\citenamefont {Horodecki}, \citenamefont {Horodecki},\ and\ \citenamefont {Horodecki}}]{Horodecki1996PPT}%
  \BibitemOpen
  \bibfield  {author} {\bibinfo {author} {\bibfnamefont {M.}~\bibnamefont {Horodecki}}, \bibinfo {author} {\bibfnamefont {P.}~\bibnamefont {Horodecki}},\ and\ \bibinfo {author} {\bibfnamefont {R.}~\bibnamefont {Horodecki}},\ }\bibfield  {title} {\bibinfo {title} {Separability of mixed states: necessary and sufficient conditions},\ }\href {https://doi.org/https://doi.org/10.1016/S0375-9601(96)00706-2} {\bibfield  {journal} {\bibinfo  {journal} {Physics Letters A}\ }\textbf {\bibinfo {volume} {223}},\ \bibinfo {pages} {1} (\bibinfo {year} {1996})}\BibitemShut {NoStop}%
\bibitem [{\citenamefont {Hofmann}\ \emph {et~al.}(2014)\citenamefont {Hofmann}, \citenamefont {Moroder},\ and\ \citenamefont {Gühne}}]{Hofmann_2014}%
  \BibitemOpen
  \bibfield  {author} {\bibinfo {author} {\bibfnamefont {M.}~\bibnamefont {Hofmann}}, \bibinfo {author} {\bibfnamefont {T.}~\bibnamefont {Moroder}},\ and\ \bibinfo {author} {\bibfnamefont {O.}~\bibnamefont {Gühne}},\ }\bibfield  {title} {\bibinfo {title} {Analytical characterization of the genuine multiparticle negativity},\ }\href {https://doi.org/10.1088/1751-8113/47/15/155301} {\bibfield  {journal} {\bibinfo  {journal} {Journal of Physics A: Mathematical and Theoretical}\ }\textbf {\bibinfo {volume} {47}},\ \bibinfo {pages} {155301} (\bibinfo {year} {2014})}\BibitemShut {NoStop}%
\bibitem [{\citenamefont {Ohst}\ \emph {et~al.}(2024)\citenamefont {Ohst}, \citenamefont {Yu}, \citenamefont {Gühne},\ and\ \citenamefont {Nguyen}}]{Ties2024}%
  \BibitemOpen
  \bibfield  {author} {\bibinfo {author} {\bibfnamefont {T.-A.}\ \bibnamefont {Ohst}}, \bibinfo {author} {\bibfnamefont {X.-D.}\ \bibnamefont {Yu}}, \bibinfo {author} {\bibfnamefont {O.}~\bibnamefont {Gühne}},\ and\ \bibinfo {author} {\bibfnamefont {H.~C.}\ \bibnamefont {Nguyen}},\ }\bibfield  {title} {\bibinfo {title} {{Certifying quantum separability with adaptive polytopes}},\ }\href {https://doi.org/10.21468/SciPostPhys.16.3.063} {\bibfield  {journal} {\bibinfo  {journal} {SciPost Phys.}\ }\textbf {\bibinfo {volume} {16}},\ \bibinfo {pages} {063} (\bibinfo {year} {2024})}\BibitemShut {NoStop}%
\bibitem [{\citenamefont {L{\"{o}}fberg}(2004)}]{YALMIP2004}%
  \BibitemOpen
  \bibfield  {author} {\bibinfo {author} {\bibfnamefont {J.}~\bibnamefont {L{\"{o}}fberg}},\ }in\ \href@noop {} {\emph {\bibinfo {booktitle} {In Proceedings of the CACSD Conference}}}\ (\bibinfo {address} {Taipei, Taiwan},\ \bibinfo {year} {2004})\BibitemShut {NoStop}%
\bibitem [{mos(2024)}]{mosek}%
  \BibitemOpen
  \href {https://docs.mosek.com/10.2/toolbox/index.html} {\emph {\bibinfo {title} {MOSEK Optimization Toolbox for MATLAB 10.2.17}}},\ \bibinfo {organization} {MOSEK ApS} (\bibinfo {year} {2024})\BibitemShut {NoStop}%
\bibitem [{\citenamefont {Lewenstein}\ \emph {et~al.}(2000)\citenamefont {Lewenstein}, \citenamefont {Kraus}, \citenamefont {Cirac},\ and\ \citenamefont {Horodecki}}]{Lewenstein2000}%
  \BibitemOpen
  \bibfield  {author} {\bibinfo {author} {\bibfnamefont {M.}~\bibnamefont {Lewenstein}}, \bibinfo {author} {\bibfnamefont {B.}~\bibnamefont {Kraus}}, \bibinfo {author} {\bibfnamefont {J.~I.}\ \bibnamefont {Cirac}},\ and\ \bibinfo {author} {\bibfnamefont {P.}~\bibnamefont {Horodecki}},\ }\bibfield  {title} {\bibinfo {title} {Optimization of entanglement witnesses},\ }\href {https://doi.org/10.1103/PhysRevA.62.052310} {\bibfield  {journal} {\bibinfo  {journal} {Phys. Rev. A}\ }\textbf {\bibinfo {volume} {62}},\ \bibinfo {pages} {052310} (\bibinfo {year} {2000})}\BibitemShut {NoStop}%
\bibitem [{\citenamefont {Lieb}(1994)}]{Lieb1994}%
  \BibitemOpen
  \bibfield  {author} {\bibinfo {author} {\bibfnamefont {E.~H.}\ \bibnamefont {Lieb}},\ }\bibfield  {title} {\bibinfo {title} {Flux Phase of the Half-Filled Band},\ }\href {https://doi.org/10.1103/PhysRevLett.73.2158} {\bibfield  {journal} {\bibinfo  {journal} {Phys. Rev. Lett.}\ }\textbf {\bibinfo {volume} {73}},\ \bibinfo {pages} {2158} (\bibinfo {year} {1994})}\BibitemShut {NoStop}%
\bibitem [{\citenamefont {Zschocke}\ and\ \citenamefont {Vojta}(2015)}]{Zschocke2015}%
  \BibitemOpen
  \bibfield  {author} {\bibinfo {author} {\bibfnamefont {F.}~\bibnamefont {Zschocke}}\ and\ \bibinfo {author} {\bibfnamefont {M.}~\bibnamefont {Vojta}},\ }\bibfield  {title} {\bibinfo {title} {Physical states and finite-size effects in Kitaev's honeycomb model: Bond disorder, spin excitations, and NMR line shape},\ }\href {https://doi.org/10.1103/PhysRevB.92.014403} {\bibfield  {journal} {\bibinfo  {journal} {Phys. Rev. B}\ }\textbf {\bibinfo {volume} {92}},\ \bibinfo {pages} {014403} (\bibinfo {year} {2015})}\BibitemShut {NoStop}%
\bibitem [{\citenamefont {Yao}\ and\ \citenamefont {Qi}(2010)}]{Yao2010}%
  \BibitemOpen
  \bibfield  {author} {\bibinfo {author} {\bibfnamefont {H.}~\bibnamefont {Yao}}\ and\ \bibinfo {author} {\bibfnamefont {X.-L.}\ \bibnamefont {Qi}},\ }\bibfield  {title} {\bibinfo {title} {Entanglement Entropy and Entanglement Spectrum of the Kitaev Model},\ }\href {https://doi.org/10.1103/PhysRevLett.105.080501} {\bibfield  {journal} {\bibinfo  {journal} {Phys. Rev. Lett.}\ }\textbf {\bibinfo {volume} {105}},\ \bibinfo {pages} {080501} (\bibinfo {year} {2010})}\BibitemShut {NoStop}%
\bibitem [{\citenamefont {Baskaran}\ \emph {et~al.}(2007)\citenamefont {Baskaran}, \citenamefont {Mandal},\ and\ \citenamefont {Shankar}}]{Baskaran2007}%
  \BibitemOpen
  \bibfield  {author} {\bibinfo {author} {\bibfnamefont {G.}~\bibnamefont {Baskaran}}, \bibinfo {author} {\bibfnamefont {S.}~\bibnamefont {Mandal}},\ and\ \bibinfo {author} {\bibfnamefont {R.}~\bibnamefont {Shankar}},\ }\bibfield  {title} {\bibinfo {title} {Exact Results for Spin Dynamics and Fractionalization in the Kitaev Model},\ }\href {https://doi.org/10.1103/PhysRevLett.98.247201} {\bibfield  {journal} {\bibinfo  {journal} {Phys. Rev. Lett.}\ }\textbf {\bibinfo {volume} {98}},\ \bibinfo {pages} {247201} (\bibinfo {year} {2007})}\BibitemShut {NoStop}%
\bibitem [{\citenamefont {Hamma}\ \emph {et~al.}(2005)\citenamefont {Hamma}, \citenamefont {Ionicioiu},\ and\ \citenamefont {Zanardi}}]{Hamma2005Diagonal}%
  \BibitemOpen
  \bibfield  {author} {\bibinfo {author} {\bibfnamefont {A.}~\bibnamefont {Hamma}}, \bibinfo {author} {\bibfnamefont {R.}~\bibnamefont {Ionicioiu}},\ and\ \bibinfo {author} {\bibfnamefont {P.}~\bibnamefont {Zanardi}},\ }\bibfield  {title} {\bibinfo {title} {Bipartite entanglement and entropic boundary law in lattice spin systems},\ }\href {https://doi.org/10.1103/PhysRevA.71.022315} {\bibfield  {journal} {\bibinfo  {journal} {Phys. Rev. A}\ }\textbf {\bibinfo {volume} {71}},\ \bibinfo {pages} {022315} (\bibinfo {year} {2005})}\BibitemShut {NoStop}%
\bibitem [{\citenamefont {Ha}\ \emph {et~al.}(2018)\citenamefont {Ha}, \citenamefont {Han},\ and\ \citenamefont {Kye}}]{Ha2018}%
  \BibitemOpen
  \bibfield  {author} {\bibinfo {author} {\bibfnamefont {K.-C.}\ \bibnamefont {Ha}}, \bibinfo {author} {\bibfnamefont {K.~H.}\ \bibnamefont {Han}},\ and\ \bibinfo {author} {\bibfnamefont {S.-H.}\ \bibnamefont {Kye}},\ }\bibfield  {title} {\bibinfo {title} {Separability of multi-qubit states in terms of diagonal and anti-diagonal entries},\ }\href {https://doi.org/10.1007/s11128-018-2145-x} {\bibfield  {journal} {\bibinfo  {journal} {Quantum Information Processing}\ }\textbf {\bibinfo {volume} {18}},\ \bibinfo {pages} {34} (\bibinfo {year} {2018})}\BibitemShut {NoStop}%
\end{thebibliography}%

\clearpage
\onecolumngrid

\centerline{ \textbf{\large  Supplemental Materials for Multipartite Entanglement in Quantum Spin Liquids } }

\title{Supplemental Materials for Multipartite Entanglement in Quantum Spin Liquids}

\author{Liuke Lyu}
\affiliation{D\'epartement de Physique, Universit\'e de Montr\'eal, Montr\'eal, QC H3C 3J7, Canada}
\affiliation{
 Institut Courtois, Universit\'e de Montr\'eal, Montr\'eal (Qu\'ebec), H2V 0B3, Canada
}
\affiliation{
 Centre de Recherches Math\'ematiques, Universit\'e de Montr\'eal, Montr\'eal, QC, Canada, HC3 3J7
}

\author{Deeksha Chandorkar}
\author{Samarth Kapoor}
\author{So Takei\,\orcidlink{0000-0001-9177-1895}}
\affiliation{Department of Physics, Queens College of the City University of New York, Queens, New York 11367, USA}
\affiliation{Physics Doctoral Program, Graduate Center of the City University of New York, New York, NY 10016, USA}


\author{Erik S. S{\o}rensen\,\orcidlink{0000-0002-5956-1190}}
\email{sorensen@mcmaster.ca}
\affiliation{Department of Physics and Astronomy, McMaster University, Hamilton, Ontario L8S 4M1, Canada}

\author{William Witczak-Krempa}
\email{w.witczak-krempa@umontreal.ca}
\affiliation{D\'epartement de Physique, Universit\'e de Montr\'eal, Montr\'eal, QC H3C 3J7, Canada}
\affiliation{
 Institut Courtois, Universit\'e de Montr\'eal, Montr\'eal (Qu\'ebec), H2V 0B3, Canada
}
\affiliation{
 Centre de Recherches Math\'ematiques, Universit\'e de Montr\'eal, Montr\'eal, QC, Canada, HC3 3J7
}

\date{\today}
\maketitle
\tableofcontents

\section{Genuine Multipartite Negativity as a semidefinite program}\label{sec:GMN_SDP}

Genuine Multipartite Negativity, as defined in Eq.~(\ref{eq:CRE_GMN}), can be formulated as the following semidefinite program~\cite{Guhne2011, Hofmann_2014}
\begin{equation}
\begin{aligned}
\mathcal{N}(\rho)&=-\min \operatorname{tr} (\rho W) \\
\text { subject to } & W=P_m+Q_m^{T_m} \\
& 0 \leqslant P_m  \\
& 0 \leqslant Q_m \leqslant I \text {  for all bipartitions } m \mid \bar{m}
\end{aligned}    
\end{equation}
where $W$ is an entanglement witness that is fully decomposable with respect to all bipartitions. 
For two operators $A$ and $B$, the majorization $A \succeq B$ implies that $A-B$ is positive semidefinite. $T_m$ refers to a partial transpose with respect to either part of the bipartition $m |\bar{m}$. 

To compute the GMN, we use a modified version of the MATLAB package \texttt{pptmixer}\cite{Guhne2011}, where the constraint $P_m \leqslant I$ is removed, as suggested in the later work~\cite{Hofmann_2014} to provide a more physically transparent measure—specifically, a mixed convex roof of the minimum bipartite negativity. The semidefinite programs are modelled in MATLAB with the YALMIP toolbox~\cite{YALMIP2004} as parser, and solved using the MOSEK solver~\cite{mosek}.

To motivate such an optimization, consider first a bipartite system. The negativity can be formulated as a semidefinite program
\begin{equation}
\begin{aligned}
N(\rho) &= -\min \; \operatorname{tr}(\rho\, W) \\
\text{subject to }\quad & W = Q^{T_1}, \\
& 0 \leq Q \leq I.
\end{aligned}
\end{equation}
Thus, the negativity is associated with an optimal PPT entanglement witness \(W = Q^{T_1}\).
Expanding the witness space to fully decomposable witnesses of the form
\begin{equation}
    W' = P + Q^{T_1},
\end{equation}
with \(P \geq 0\), does not enhance the detection capability in the bipartite case. In fact, if \(P\) is positive definite, the witness \(W = Q^{T_1}\) is strictly finer than \(W = P+Q^{T_1}\)~\cite{Lewenstein2000}. However, in the multipartite case, the objective is to identify a witness that simultaneously detects entanglement across all bipartitions. In this setting, the optimal witness tailored for each bipartition may be suboptimal when a global criterion is required. Therefore, the use of fully decomposable witnesses becomes essential for quantifying genuine multipartite entanglement.

From the perspective of biseparable states, every biseparable state can be expressed as a convex combination of states that are separable with respect to the three bipartitions:
\begin{equation}
\rho^{\mathrm{bs}} = p_1\,\rho_{A\mid BC}^{\mathrm{sep}} + p_2\,\rho_{B\mid AC}^{\mathrm{sep}} + p_3\,\rho_{C\mid AB}^{\mathrm{sep}},
\end{equation}
with \(p_1+p_2+p_3=1\) and \(p_i\ge0\).
In contrast, the GMN semidefinite program detects states lying outside the larger set
\begin{equation}
\rho^{\mathrm{pmix}} = p_1\,\rho_{A\mid BC}^{\mathrm{PPT}} + p_2\,\rho_{B\mid AC}^{\mathrm{PPT}} + p_3\,\rho_{C\mid AB}^{\mathrm{PPT}},
\end{equation}
where each component \(\rho_{X\mid YZ}^{\mathrm{PPT}}\) is PPT with respect to the corresponding bipartition. 

\section{Additional analysis for the Kitaev Honeycomb Model}

\subsection{Effective Hamiltonian for the small field Kitaev Spin Liquid}~\label{sec:effHamiltonian-kitaev}

For the Kitaev model under a [111] field (main text equation~\ref{eq:Kitaev111}), a perturbation expansion around $h=0$ produces a 3-spin interaction which breaks time-reversal symmetry. The effective Hamiltonian at small field is
\begin{equation}
H_{\text{eff}} = \sum_{\langle i, j \rangle_\gamma} K_\gamma S_i^\gamma S_j^\gamma - \kappa \sum_{\langle\langle i, j, k\rangle\rangle} S_i^x S_j^y S_k^z, \label{eq:KitaevChiral}
\end{equation}
where \(\langle\langle i, j, k\rangle\rangle\) denotes ordered neighbouring sites and $\kappa= 3^{-2/3} h^3$.
The chiral term stabilizes a gapped topological phase with non-Abelian anyons~\cite{Pollmann2018}. 
This model can be solved exactly in the Majorana fermion representation~\cite{Kitaev2006}, and a computation of the hexagonal plaquette RDM can be found in section ~\ref{appendix:RDM_exact_methods}.

\subsection{Evolution of the MMES in the Kitaev model.}~\label{sec:stepwise_evo_MMES}

\begin{figure*}[h!tbp]
\centering
\includegraphics[width=0.95\textwidth]{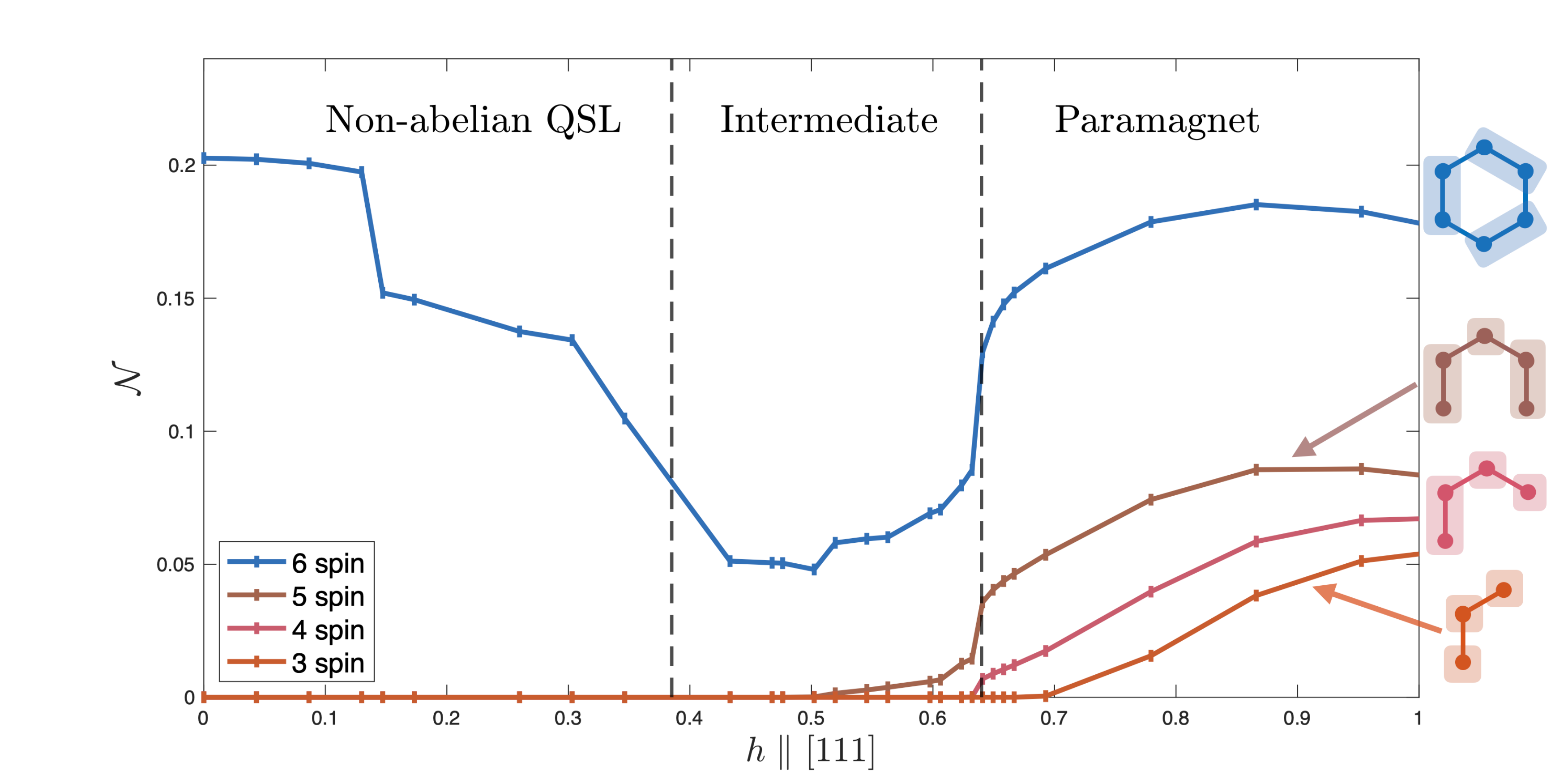}
\caption{\textbf{Evolution of the MMES in the Kitaev model.}  
Tripartite GMN \(\mathcal{N}_3\) is shown for the hexagonal plaquette (blue) and its nested five-, four-, and three-spin subregions (shades of red) as a function of the \([111]\) field \(h\) on the \(N=32\) cluster.  Vertical dashed lines mark the non-Abelian QSL (\(h<h_{c1}\)), intermediate (\(h_{c1}<h<h_{c2}\)), and paramagnetic (\(h>h_{c2}\)) regimes.  As frustration decreases with increasing field, the minimal multipartite-entangled subregion (MMES) shrinks in discrete steps: hexagon \(\rightarrow\) five-spin \(\rightarrow\) four-spin \(\rightarrow\) three-spin, illustrating the hierarchy discussed in the main text.}
\label{fig:kitaev_hex_phase_diagram}
\end{figure*}

To visualize the evolution of GME with frustration, Fig.~\ref{fig:kitaev_hex_phase_diagram} plots the tripartite GMN of the hexagonal plaquette and all its nested subregions for the Kitaev model in a [111] field. Frustration is maximum at $h=0$, and decreases with the field. The minimal multipartite-entangled subregion (MMES) changes in a stepwise fashion: at zero field the MMES is the full hexagon; it remains so up to $h\!\approx\!0.50$ in the intermediate phase, then shrinks to the five-spin cluster (hexagon minus one site); near $h= h_{c2}\approx0.64$ the four-spin cluster takes over; and beyond $h\!\approx\!0.70$ the MMES reduces to the three adjacent spins of the high-field paramagnet.  This hierarchy corroborates the MMES picture introduced in the main text.

\subsection{Certifying Separability with Adaptive Polytope Algorithm}\label{sec:certify_SEP}

Table~\ref{tab:certify_SEP} summarizes our separability analysis for key reduced density matrices.
For each subregion we bracket the critical value of the tuning parameter—field $h$ in the Kitaev model or chirality $\lambda$ in the Kagome model—between a lower bound obtained with the adaptive-polytope algorithm~\cite{Ties2024} and an upper bound extracted either from GMN (for biseparability) or the PPT criterion (for full separability).
Whenever the interval $[{\rm LB},{\rm UB}]$ lies entirely below the parameter range of interest, the state is rigorously certified biseparable; when it lies entirely above, GME is confirmed.  In several cases (e.g.\ the three-spin ring around a hexagon) the bounds overlap, leaving a narrow, inconclusive window, but the qualitative picture is unambiguous: small, non-loopy clusters are biseparable throughout broad parameter ranges, whereas GME appears only when the cluster contains a closed loop.

Because the biseparable set is convex and of full measure, a state that lies strictly inside it remains biseparable under any sufficiently small perturbation—be it a change of coupling, a weak thermal admixture, or coupling to an environment.  The intervals reported in Table \ref{tab:certify_SEP} therefore demonstrate that the loopy-only pattern of GME identified in the main text is robust: non-loopy subregions remain GME-free across finite ranges of field or interaction, while loops retain finite GMN over the same ranges.  In other words, the “entanglement frustration’’ that confines multipartite correlations to loops is not a fine-tuned artifact but a stable feature of these quantum spin liquids. 

\begin{table}[htbp]
    \centering
    \begin{tabular}{l|c|c|c}
    \hline
    State & Parameter & LB (Polytope) & UB (GMN)  \\
    \hline
    \multirow{2}{*}{Kitaev $\rho_{3-\text{hex}}$} & $h_{\text{s}}$ & $0.346$ & $0.433$ \\
                                                   & $h_{\text{bs}}$ & $0.675$  & $0.693$  \\ 
     \hline
     Kitaev $\rho_{4-\text{hex}}$ & $h_{\text{bs}}$ & $0.632$   & $0.641$   \\
     \hline
     Kitaev $\rho_{5-\text{hex}}$ & $h_{\text{bs}}$ & $0.468$   & $0.520$   \\
    \hline
    Kitaev $\rho_{6-\text{fork}}$ & $h_{\text{bs}}$ & $0.130$   & $0.147$   \\
    \hline
    Kagome $\rho_{3-\text{triangle}}$ & $\lambda_{\text{bs}}$ & $0.250$  & $0.300$\\
    \hline
    \end{tabular}
\caption{\textbf{Bounds on separability thresholds obtained from the adaptive–polytope method.}
For each type of reduced density matrices, we list a lower bound (LB) and an upper bound (UB) on the critical value of the tuning parameter at which the state changes from separable to entangled: subscripts $h_{\mathrm s}$ (or $\lambda_{\mathrm s}$) refer to \emph{full separability}, while $h_{\mathrm{bs}}$ (or $\lambda_{\mathrm{bs}}$) refer to \emph{biseparability}.
LB is obtained with the adaptive polytope algorithm of Ref.~\cite{Ties2024};  
UB is set either by the genuine multipartite negativity (for biseparability) or by the PPT criterion (for full separability), using the data in Figs.~\ref{fig:kit-phasediagram}, \ref{fig:kitaev_hex_phase_diagram}, and \ref{fig:kag-lam}.  
“3-hex”, “4-hex”, and “5-hex’’ denote three, four, and five consecutive spins around a hexagonal plaquette; “6-fork’’ is the six-spin fork subregion; “3-triangle’’ is the triangular three-spin cluster on the Kagome lattice.  
For example, \(\rho_{\text{3-hex}}\) is certified fully separable for \(h\le0.346\) and confirmed entangled for \(h\ge0.433\); the interval in between remains inconclusive.}
\label{tab:certify_SEP}
\end{table}

\subsection{GMN for N=24 Nauru Lattice}\label{sec:GMN_N24}

\begin{figure*}[h!tbp]
\centering
\includegraphics[width=0.95\textwidth]{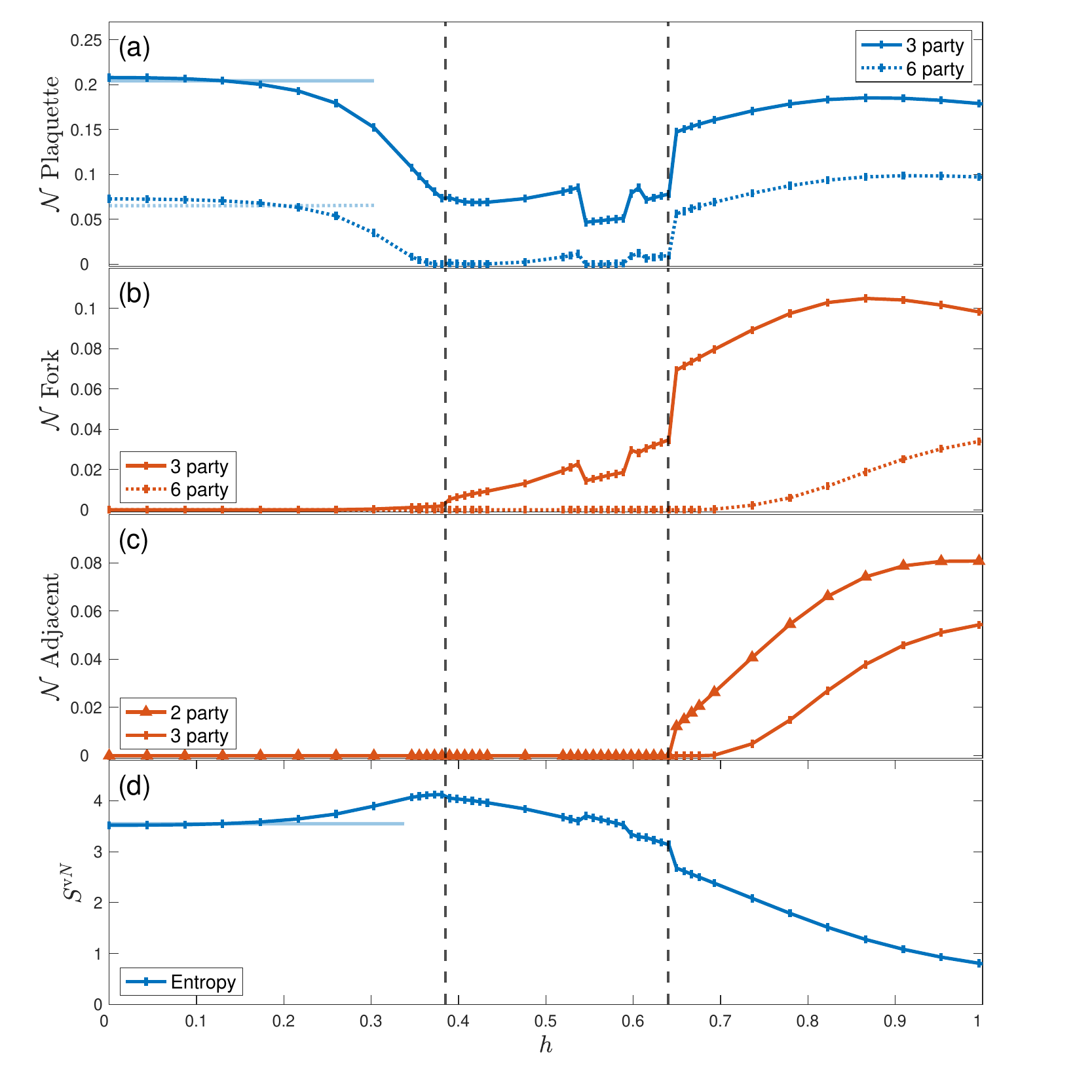}\llap{\shortstack{%
        \includegraphics[scale=.15]{Figures/hexagon-222.pdf}\\
        \rule{0ex}{5.57in}%
      }
  \rule{3.4in}{0ex}}\llap{\shortstack{%
        \includegraphics[scale=.15]{Figures/hexagon-111111.pdf}\\
        \rule{0ex}{5.45in}%
      }
  \rule{5.4in}{0ex}}\llap{\shortstack{%
        \includegraphics[scale=.15]{Figures/fork222.pdf}\\
        \rule{0ex}{4.0in}%
      }
  \rule{2.85in}{0ex}}\llap{\shortstack{%
        \includegraphics[scale=.15]{Figures/fork-111111.pdf}\\
        \rule{0ex}{3.75in}%
      }
  \rule{1.5in}{0ex}}\llap{\shortstack{%
        \includegraphics[scale=.15]{Figures/adj2.pdf}\\
        \rule{0ex}{2.7in}%
      }
  \rule{2.3in}{0ex}}\llap{\shortstack{%
        \includegraphics[scale=.15]{Figures/adj3.pdf}\\
        \rule{0ex}{2.2in}%
      }
  \rule{1.2in}{0ex}}\llap{\shortstack{%
        \includegraphics[scale=.15]{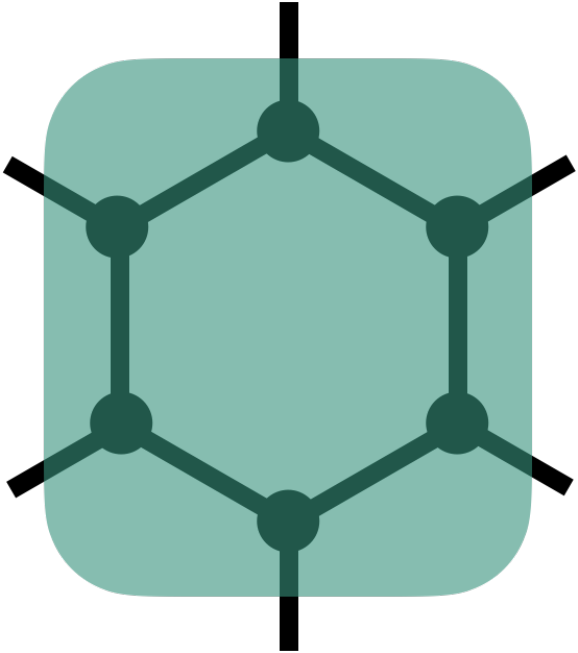}\\
        \rule{0ex}{1.1in}%
      }
  \rule{1.5in}{0ex}}
\caption{\textbf{GMN vs $h$ for the Kitaev Honeycomb Model on the $N=24$ Nauru cluster.} 
Panels mirror those in Fig.~\ref{fig:kit-phasediagram}: (a) tripartite (\(\mathcal{N}_3\), solid) and six-partite (\(\mathcal{N}_6\), dashed) GMN of the hexagon, (b) the same quantities for the fork subregion, and (c) bipartite (\(\mathcal{N}_2\)) and tripartite GMN for two- and three-spin clusters.  Panel (d) adds the von Neumann entropy of the hexagon.  The overall behaviour reproduces the \(N=32\) results: loop-confined GME at low field, the shift of the MMES from the hexagon to the three-spin cluster as the field increases, and phase boundaries at \(h_{c1}=0.38\) and \(h_{c2}=0.64\) (dashed lines).  A minor difference is the appearance of small oscillations in (a) and (b) between \(h=0.50\) and \(0.60\), attributable to finite-size effects specific to the Nauru graph.}
\label{fig:kitaev_nauru}
\end{figure*}

To further support the conclusions presented in the main text, we compute the GMN for the Kitaev Honeycomb Model on a smaller system, the \(N=24\) Nauru graph, and compare it with the results from the \(N=32\) Dyck graph. Figure~\ref{fig:kitaev_nauru} presents GMN as a function of the [111] magnetic field strength \(h\), with the same structure as Fig.~\ref{fig:kit-phasediagram} in the main text.

The results for \(N=24\) largely mirror those for $N=32$, confirming that GMN follows the same qualitative trends across system sizes. In the small-field regime, GMN is present in the hexagonal plaquette but absent in non-loopy or smaller subregions. Upon entering the intermediate phase, six-party GMN drops to zero, while GMN in the fork subregion begins to increase. In the paramagnetic phase, GMN grows across multiple subregions, regardless of their size or loop structure. The main difference observed in the \(N=24\) case is the presence of small oscillations in \(\mathcal{N}_3\) and \(\mathcal{N}_6\) between $h=0.5$ and $h=0.6$, which are absent in the \(N=32\) system. These oscillations likely stem from finite-size effects inherent to the smaller lattice geometry, rather than fundamental differences in the entanglement structure.

Overall, the consistency between the \(N=24\) and \(N=32\) results reinforces the validity of our conclusions regarding the role of loop-like multipartite entanglement in the Kitaev Honeycomb Model.

\subsection{Minimum Negativity vs. [111] Field}\label{sec:minNeg}

\begin{figure*}[h!tbp]
\centering
\includegraphics[width=0.95\textwidth]{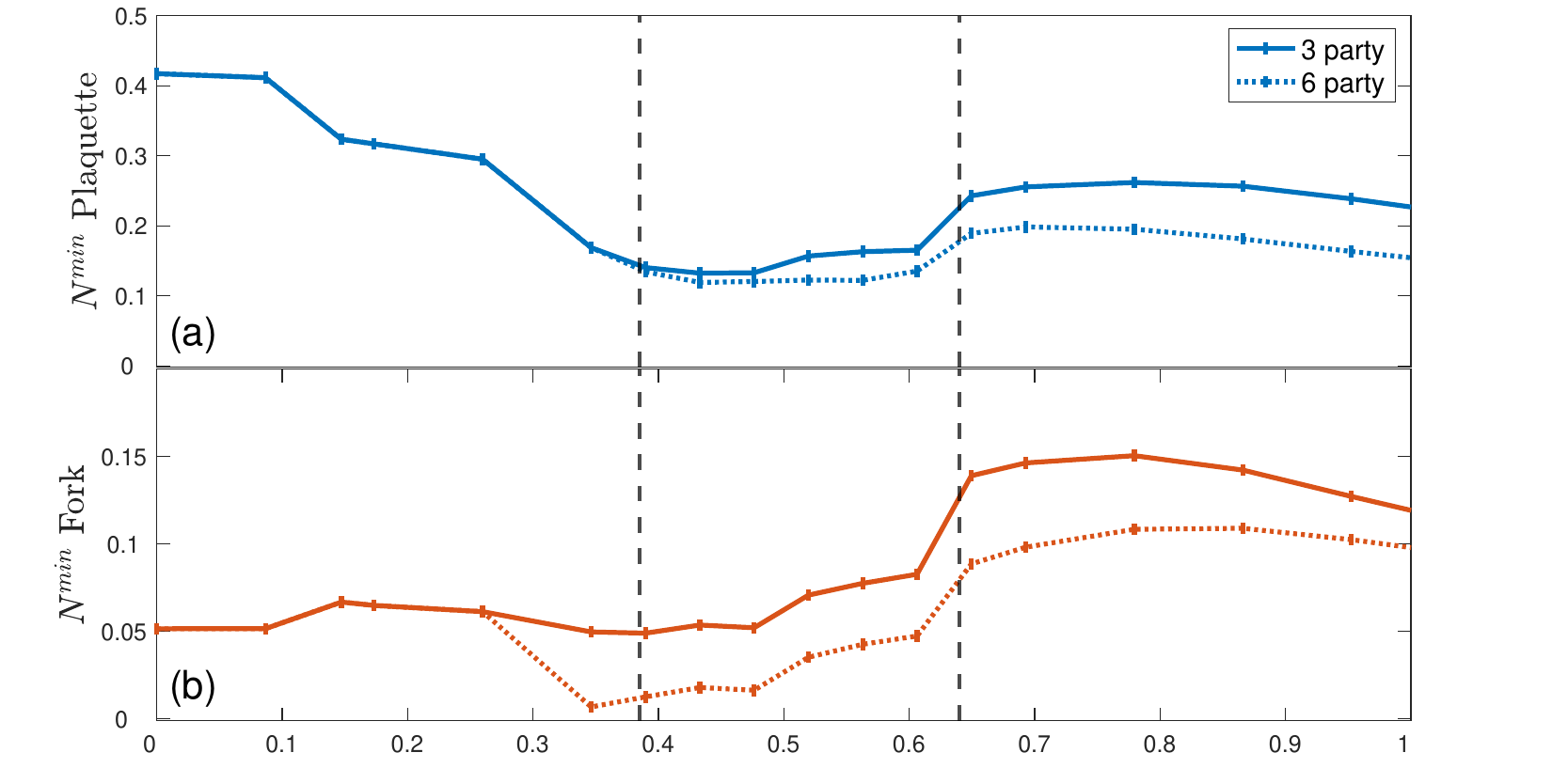}\llap{\shortstack{%
        \includegraphics[scale=.15]{Figures/hexagon-222.pdf}\\
        \rule{0ex}{2.6in}%
      }
  \rule{2.0in}{0ex}}\llap{\shortstack{%
        \includegraphics[scale=.15]{Figures/hexagon-111111.pdf}\\
        \rule{0ex}{1.75in}%
      }
  \rule{1.7in}{0ex}}\llap{\shortstack{%
        \includegraphics[scale=.15]{Figures/fork222.pdf}\\
        \rule{0ex}{0.9in}%
      }
  \rule{2.85in}{0ex}}\llap{\shortstack{%
        \includegraphics[scale=.15]{Figures/fork-111111.pdf}\\
        \rule{0ex}{0.3in}%
      }
  \rule{1.5in}{0ex}}
\caption{\textbf{Minimum Negativity vs. [111] Field for the Kitaev Honeycomb Model.} Panel (a) shows the tripartite (\(N^{\mathrm{min}}_3\)) and six-partite (\(N^{\mathrm{min}}_6\)) minimum negativity for the hexagonal plaquette subregion as functions of the magnetic field \(h\). Panel (b) presents the corresponding data for the fork subregion. Here, \(N^{\mathrm{min}}\) is obtained by minimizing the bipartite negativity over all bipartitions into 3 or 6 parties. Notably, the minimum negativity does not clearly differentiate between the fork and plaquette subregions, in contrast to the genuine multipartite entanglement measures shown in Figure~\ref{fig:kit-phasediagram} of the main text, which capture the distinct "loopy" characteristics.}
\label{fig:kitaev_minNeg}
\end{figure*}

To assess whether the distinctive “loopy” structure observed in our GMN analysis is already captured by bipartite measures, we compare the behaviour of the minimum negativity \(N^{\text{min}}\) with that of GMN. Since GMN is defined as a mixed convex roof extension of \(N^{\text{min}}\), we have $\mathcal{N}(\rho)\leq N^{\mathrm min}(\rho)$ for any mixed state $\rho$. Figure~\ref{fig:kitaev_minNeg} shows the tripartite and six-partite minimum negativity, \(N_3^{\text{min}}\) and \(N_6^{\text{min}}\), as functions of the magnetic field \(h\).

In contrast to the GMN results in Fig.~\ref{fig:kit-phasediagram}, which clearly differentiate between loopy (plaquette) and non-loopy (fork) subregions, the values of \(N^{\text{min}}\) do not exhibit such a distinction. Notably, in the small-field Kitaev spin liquid phase, \(N^{\text{min}}\) remains finite for the fork subregion even though the corresponding GMN vanishes. This observation demonstrates that the “loopy” entanglement structure is an intrinsically multipartite feature that cannot be captured by bipartite measures alone. Consequently, GMN is essential for revealing the collective entanglement properties inherent in quantum spin liquids.

\section{Exact plaquette RDM of Kitaev model from Majorana Fermion Methods}\label{appendix:RDM_exact_methods}
\subsection{Model}
\label{model}
\begin{figure}[h!]
    \centering
    \includegraphics[width=0.65\linewidth]{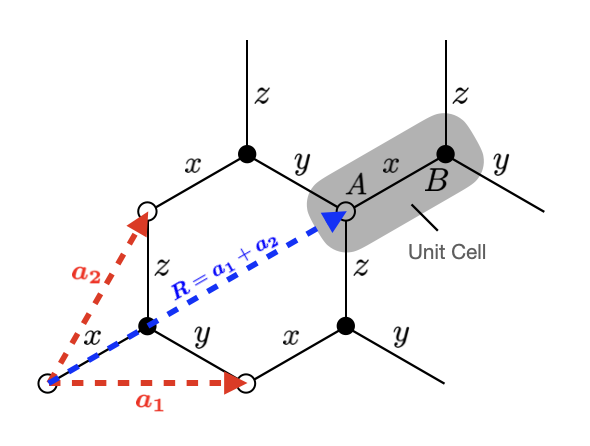}
    \caption{A \(2\times 2\) honeycomb lattice with sublattices \(A\) and \(B\). The A sites form a triangular lattice with primitive vectors \(\boldsymbol{a_1}\) and \(\boldsymbol{a_2}\). Each unit cell (denoted by \(\boldsymbol{R}\)) contains one spin from both sublattices joined by an \(x\)-link. A reference unit cell is illustrated in the figure.}
    \label{Lattice-I}
\end{figure}

The Kitaev honeycomb model is an exactly solvable model and a representative of quantum spin liquids with Majorana fermions coupled to emergent $Z_2$ gauge fields~\cite{Kitaev2006}. 
For ease of computation, we adopt a slightly different sign convention for $\hat H$ than in the main text; the Hamiltonian reads
\begin{equation}
\label{kitaev_hamiltonian}
\hat{H} = - \sum_{\{\alpha=x,y,z\} }\sum_{\braket{ij}}J^{\alpha}\hat{\sigma}^{\alpha}_i \hat{\sigma}^{\alpha}_j\,,
\end{equation}
where spin-\(1/2\) moments on the honeycomb lattice -- represented as Pauli operators -- interact only through nearest neighbour Ising exchange $J^\alpha$ between sites \(i\) and \(j\) on sublattices \(A\) and \(B\), respectively; here, $\alpha=x,y,z$ labels the different nearest neighbour bond directions, as shown in Fig.~\ref{Lattice-I}. 

Equation~\eqref{kitaev_hamiltonian} can be solved exactly via fermionization of the Pauli operators. This can be seen by first writing the spin on each site $n$ as a composite of two Majorana fermion operators: \(\hat{\sigma}_{n}^\alpha \equiv i\hat{\gamma}_n^\alpha \hat{\eta}_n\), where \(\hat{\eta}_{n}\) are typically referred to as the matter fermions and \(\hat{\gamma^{\alpha}_{n}}\) are identified as the bond fermions. These operators satisfy the canonical Majorana anticommutation relations: they square to one, and those of different types or on different sites anti-commute. Defining the bond operators as $\hat{u}^\alpha_{ij}\equiv i \hat{\gamma}^\alpha_i \hat{\gamma}^\alpha_j$, Eq.~\eqref{kitaev_hamiltonian} becomes quadratic in the matter fermions,
\begin{equation}
\label{fermion_hamiltonian}
    \hat{H} = i \sum_{\{\alpha=x,y,z\}} \sum_{\braket{ij}}J^{\alpha}\hat{u}^{\alpha}_{ij} \hat{\eta}_i \hat{\eta}_j \,.
\end{equation}

The fermionic Hilbert space introduced via the fermionization procedure above has extra states that are related by a gauge transformation. The condition \( -i \hat{\sigma}^{x}_{n} \hat{\sigma}^{y}_{n} \hat{\sigma}^{z}_{n}=1 \), which follows from the properties of the Pauli matrices, is not necessarily true in the extended Hilbert space. One may therefore define the operator $\hat{D}_{n}\equiv  -i \hat{\sigma}^{x}_{n} \hat{\sigma}^{y}_{n} \hat{\sigma}^{z}_{n}=\hat{\gamma}_{n}^x  \hat{\gamma}_{n}^y \hat{\gamma}_{n}^z \hat{\eta}_{n}$, and fix the gauge with the condition, \(\hat{D}_{n}|\Psi\rangle=|\Psi\rangle \). Any state $|\Psi\rangle$ in the extended space can then be projected into the physical space using the projector
\begin{equation}
\label{projector_eq}
\hat{P}=\bigotimes_{\forall n \in A, B} \frac{1+\hat{D}_{n}}{2}\,.
\end{equation}

\subsection{Ground State}
\label{finding_the_ground_state}
\begin{figure}[h!]
    \centering
    \includegraphics[width=0.5\linewidth]{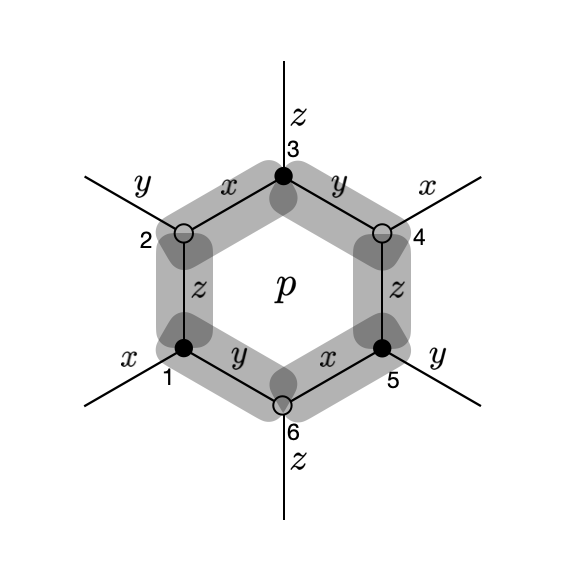}
    \caption{A plaquette \(p\) of the honeycomb lattice }
    \label{lattice-II}
\end{figure}

The flux operator \(\hat{W}_{p}=\hat{\sigma}^{x}_1 \hat{\sigma}^{y}_2 \hat{\sigma}^{z}_3 \hat{\sigma}^{x}_4 \hat{\sigma}^{y}_5 \hat{\sigma}^{z}_6\), for any plaquette $p$ on the honeycomb lattice (see Fig.~\ref{lattice-II}), is conserved. Here, the spins $1,\dots,6$  are labelled for each plaquette according to the convention in Fig.~\ref{lattice-II}.  These operators can be expressed solely in terms of the bond operators \(\hat{u}^\alpha_{ij}\). Since these bond operators commute with the Hamiltonian~\eqref{fermion_hamiltonian} and square to one, a gauge configuration $\hat{u}^\alpha_{ij} = \pm 1$ consistent with a given set of values \(W_{p}\in \{\pm 1\}\) may be found.

In the thermodynamic limit, the flux of the ground state is zero~\cite{Lieb1994}. For \(Z_2\) gauge fields, this implies that the plaquette operators \(\hat{W}_{p}\) are one for all $p$, and therefore, one can also set the bond operators to \(\hat{u}^\alpha_{ij}=1\). The resultant system is translationally invariant, rendering it suitable to Fourier methods. 

This procedure becomes more involved for the finite lattice, where the ground state is not necessarily flux-free \cite{Zschocke2015}, and may not be gauge-equivalent to the configuration \(\hat{u}^{\alpha}_{ij}=1\). Therefore, one should pick a bond configuration \(\hat{u}^{\alpha}_{ij}\), determine the ground state of the matter fermions \(\hat{\eta}_{n}\) in \eqref{fermion_hamiltonian}, and then iterate over all the independent gauge configurations.

\subsection{Fermion Ground State}
\label{fermion_ground_state}

Keeping the focus on thermodynamic systems, \(\hat{u}^\alpha_{ij}\) is set to one to find the spectrum in the flux free sector. Introducing a complex fermion $\hat{f}_{\boldsymbol{R}} =(\hat{\eta}_{\boldsymbol{R},A} + i\hat{\eta}_{\boldsymbol{R},B})/2$ for each unit cell \(\boldsymbol{R}\) (see Fig.~\ref{Lattice-I}), the Hamiltonian \eqref{fermion_hamiltonian} may be rewritten as 
\begin{equation}
   \hat{H}= \sum_{\boldsymbol{k}}\begin{pmatrix}
\hat f_{\boldsymbol{k}}^{\dagger} & \hat f_{-\boldsymbol{k}}
\end{pmatrix}  \begin{pmatrix}
\xi_{\boldsymbol{k}}  &  i \Delta_{\boldsymbol{k}} \\
- i\Delta_{\boldsymbol{k}}  &  - \xi_{\boldsymbol{k}} 
\end{pmatrix}\begin{pmatrix}
\hat f_{\boldsymbol{k}} \\
\hat f_{-\boldsymbol{k}}^{\dagger}
\end{pmatrix} \,,
\label{hf}
\end{equation}
where \(\xi_{\boldsymbol{k}}=\Re\{S_{\boldsymbol{k}}\}\), \(\Delta_{\boldsymbol{k}}=\Im\{S_{\boldsymbol{k}}\}\) and \(S_{\boldsymbol{k}} = J^x + J^y e^{i\boldsymbol{k}\cdot\boldsymbol{a}_1} + J^z e^{i\boldsymbol{k}\cdot\boldsymbol{a}_2}\). Here, ${\boldsymbol a}_1$ and ${\boldsymbol a}_2$ are the two primitive vectors of the triangular lattice formed by the unit cells, and \(\boldsymbol{k}\) labels the wave vectors within the first Brillouin zone of the triangular lattice. The values of ${\boldsymbol k}$ depend on the boundary conditions.

Equation~\eqref{hf} can be diagonalized using a Bogoliubov transformation, $\hat f_{\boldsymbol{k}} = \cos \theta_{\boldsymbol{k}} \hat{a}_{\boldsymbol{k}} - i \sin \theta_{\boldsymbol{k}} \hat{a}^\dagger_{-\boldsymbol{k}}$, where 
\begin{equation}
    \cos \theta_{\boldsymbol{k}} = \sqrt{\frac{E_{\boldsymbol{k}} +\xi_{\boldsymbol{k}}}{2E_{\boldsymbol{k}}}}\,,\ \ \sin \theta_{\boldsymbol{k}} = \text{sgn} (\Delta_{\boldsymbol{k}})\sqrt{\frac{E_{\boldsymbol{k}} -\xi_{\boldsymbol{k}}}{2E_{\boldsymbol{k}}}}\,,
\end{equation}
and \(E_{\boldsymbol{k}}=\sqrt{\xi_{\boldsymbol{k}}^2+\Delta_{\boldsymbol{k}}^2}\). All two-point correlation functions of the matter Majorana fermions can now be evaluated in the ground state and are given by
\begin{equation}
\label{fermion_correlation}
    \begin{aligned}
        \braket{\hat{\eta}_{A \boldsymbol{R}}\hat{\eta}_{B \boldsymbol{R}'}}
        &=\frac{i}{N}\sum_{\boldsymbol{k}}\frac{\xi_{\boldsymbol{k}}-i \Delta_{\boldsymbol{k}}}{E_{\boldsymbol{k}}}e^{i \boldsymbol{k}\cdot(\boldsymbol{R}-\boldsymbol{R}')}\\
        \braket{\hat{\eta}_{A \boldsymbol{R}}\hat{\eta}_{A \boldsymbol{R}'}}&=\braket{\hat{\eta}_{B \boldsymbol{R}}\hat{\eta}_{B \boldsymbol{R}'}}=\delta_{\boldsymbol{R} \boldsymbol{R}'}\,,
    \end{aligned}
\end{equation}
where \(N\) is the total number of unit cells on the lattice. 

\subsection{Reduced Density Matrix}
\label{reduced_density_matrix}
A general method to construct the reduced density matrix for the Kitaev model is presented in Ref.~\onlinecite{Yao2010}. Here, we follow a different approach that may be most useful when computing the reduced density matrix for small subregions containing a few spins. The six-spin reduced density matrix on a plaquette may be expressed generally as
\begin{equation}
\label{rdm_general_formula} 
    \hat{\rho}_6=\frac{1}{2^6}\sum_{\{\mu_i=0,x,y,z\}}\braket{\hat{\sigma}^{\mu_1}_1\hat{\sigma}^{\mu_2}_2...\hat{\sigma}^{\mu_6}_6}_{0}\hat{\sigma}^{\mu_1}_1\hat{\sigma}^{\mu_2}_2...\hat{\sigma}^{\mu_6}_6\,,
\end{equation}
where \(\braket{\hat{O}}_{0}\) represents the expectation value of \(\hat{O}\) in the ground state, \(\{\hat{\sigma}^{x}_{n},\hat{\sigma}^{y}_{n},\hat{\sigma}^{z}_{n}\}\) is the vector of Pauli matrices on site \(n\) and  \(\hat\sigma^{0}\) is the $2\times2$ identity matrix. The task is then to find all the non-zero spin expectation values to construct the density matrix. In this subsection, the results are calculated for $J^x=J^y=J^z=-1$. 

Since the flux operators \(\hat W_p\) are conserved, the correlation function for any operator that does not conserve the flux must be zero. This renders all one-spin correlation functions zero, i.e., \(\braket{\hat{\sigma}^{\mu}_{n}}=0\). In the absence of a magnetic field, all correlation functions with odd number of spins are zero by time-reversal symmetry. The only non-zero two-spin correlation functions are those that are joined by a bond with spin components matching the bond type, e.g., \(\braket{\hat{\sigma}^{z}_1\hat{\sigma}^{z}_2}\). Table~\ref{twopointcfs} lists all non-zero two-spin correlations functions on a plaquette and their values in the thermodynamic limit. 

\begin{figure}[h!]
    \centering
    \includegraphics[width=0.7\linewidth]{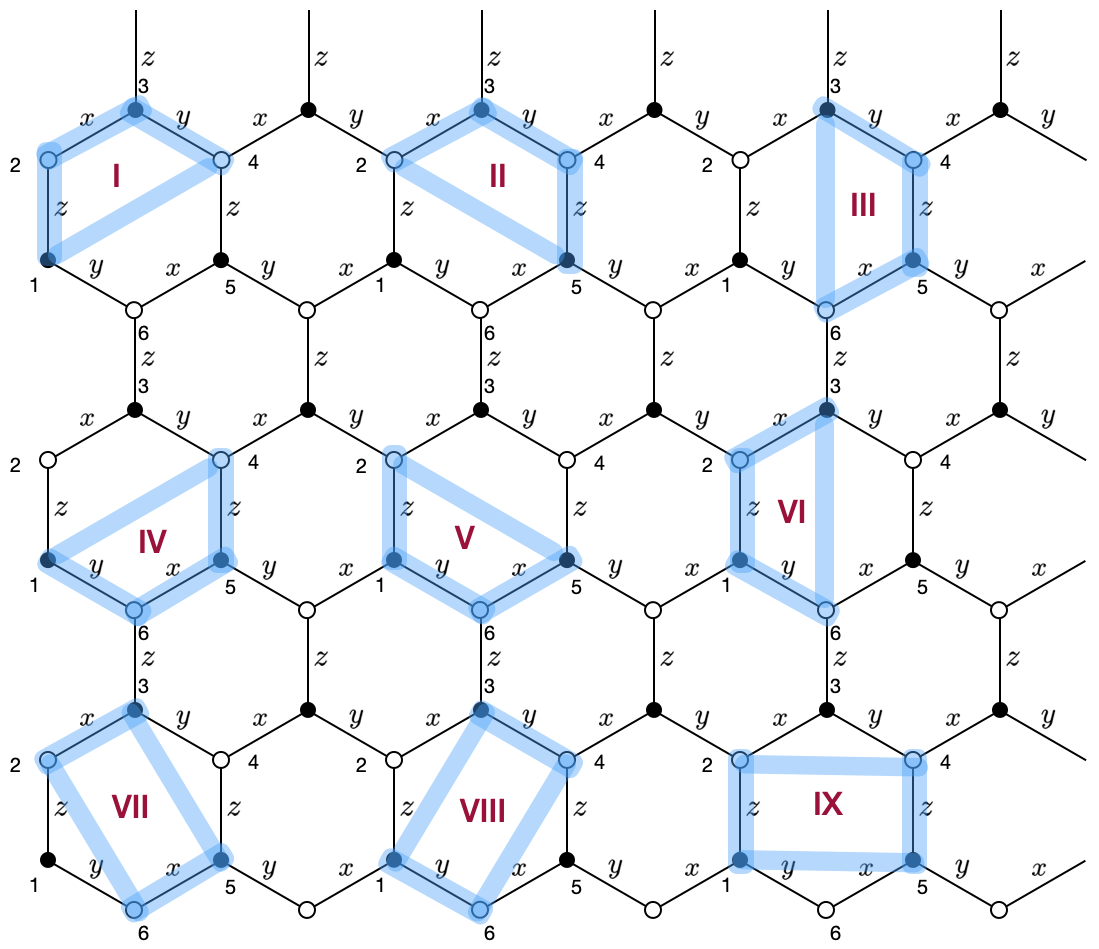}
    \caption{Site groupings contributing to four-spin correlation functions. }
    \label{four_site_grouping}
\end{figure}

Figure \ref{four_site_grouping} shows the groupings corresponding to non-zero four-spin correlation functions on a plaquette. Each grouping can give rise to more than one non-zero correlation function, and these are all listed in Table~\ref{fourpointcfs}. As shown in the third column, each correlation function can be re-expressed solely in terms of the matter Majorana fermion operators after setting the bond operators \(\hat{u}^\alpha_{ij}\) to one \cite{Baskaran2007}. Using Wick's theorem and the two-point correlation functions Eq.~\eqref{fermion_correlation}, these correlation functions can then be evaluated, and their numerical values in the thermodynamic limit are presented in the fourth column.

Similarly, all non-zero six-spin correlation functions, their corresponding matter Majorana fermion representation, and their numerical values in the thermodynamic limit are presented in Table~\ref{sixpointcfs}.

\begin{table}[H]
\centering
\caption{Two point functions for \(J^x=J^y=J^z =-1\)}
    \begin{tabular}  {  
  | p{2cm}  
  | >{\raggedright}p{2.5cm}  
  | p{2cm} | }
  \hline
     Correlation function & Matter fermion representation & Expectation value  \\
    \hline
     \( \braket{\hat{\sigma}_{1}^{z} \hat{\sigma}_{2}^{z}} \) & \(+i\braket{\hat{\eta}_1\hat{\eta}_2}\) &\(-0.5249\)\\
    \( \braket{\hat{\sigma}_{2}^{x} \hat{\sigma}_{3}^{x}} \) & \(-i\braket{\hat{\eta}_2\hat{\eta}_3}\) & \(-0.5249\)\\
     \( \braket{\hat{\sigma}_{3}^{y} \hat{\sigma}_{4}^{y}} \) & \(+i\braket{\hat{\eta}_3\hat{\eta}_4}\) &\(-0.5249\)\\
     \( \braket{\hat{\sigma}_{4}^{z} \hat{\sigma}_{5}^{z}} \) & \(-i\braket{\hat{\eta}_4\hat{\eta}_5}\) &\(-0.5249\)\\
    \( \braket{\hat{\sigma}_{5}^{x} \hat{\sigma}_{6}^{x}} \) & \(+i\braket{\hat{\eta}_5\hat{\eta}_6}\) &\(-0.5249\)\\
    \( \braket{\hat{\sigma}_{1}^{y} \hat{\sigma}_{6}^{y}} \) &  \(+i\braket{\hat{\eta}_1\hat{\eta}_6}\)&\(-0.5249\)\\
    \hline
    \end{tabular}
    \label{twopointcfs}
\end{table}

\begin{table}[H]
\centering
\caption{Four point functions for each grouping in Figure \ref{four_site_grouping} and \(J^x=J^y=J^z=-1\)}
\begin{tabular} { 
  | p{1cm} 
  | p{2cm}  
  | >{\raggedright}p{2.5cm}  
  | p{2cm} | }  
  \hline
     Group & Correlation function & Matter fermion representation & Expectation value  \\
    \hline
     I& \(  \braket{\hat{\sigma}_{1}^{z} \hat{\sigma}_{2}^{y} \hat{\sigma}_{3}^{z} \hat{\sigma}_{4}^{y}} \) & \(-i\braket{\hat{\eta}_1 \hat{\eta}_4}\) & \(-0.1858\)\\
      &\(\braket{\hat{\sigma}_{1}^{z} \hat{\sigma}_{2}^{z} \hat{\sigma}_{3}^{y} \hat{\sigma}_{4}^{y}} \) &  \(-\braket{\hat{\eta}_1\hat{\eta}_2\hat{\eta}_3\hat{\eta}_4}\)& \(+0.3730\)\\
    II &\( \braket{\hat{\sigma}_{2}^{x} \hat{\sigma}_{3}^{z} \hat{\sigma}_{4}^{x} \hat{\sigma}_{5}^{z}} \)   & \(+i\braket{\hat{\eta}_2 \hat{\eta}_5}\) & \(-0.1858\)\\
     &\( \braket{\hat{\sigma}_{2}^{x} \hat{\sigma}_{3}^{x} \hat{\sigma}_{4}^{z} \hat{\sigma}_{5}^{z}} \) &   \(-\braket{\hat{\eta}_2\hat{\eta}_3\hat{\eta}_4\hat{\eta}_5}\) & \(+0.3730\)\\
    III&\( \braket{\hat{\sigma}_{3}^{y} \hat{\sigma}_{4}^{x} \hat{\sigma}_{5}^{y} \hat{\sigma}_{6}^{x}}\)  &  \(-i\braket{\hat{\eta}_3 \hat{\eta}_6}\) & \(-0.1858\)\\
     & \(\braket{\hat{\sigma}_{3}^{y} \hat{\sigma}_{4}^{y} \hat{\sigma}_{5}^{x} \hat{\sigma}_{6}^{x}}\) &  \(-\braket{\hat{\eta}_3\hat{\eta}_4\hat{\eta}_5\hat{\eta}_6}\) & \(+0.3730\)\\
    IV &\(\braket{\hat{\sigma}_{1}^{y} \hat{\sigma}_{4}^{z} \hat{\sigma}_{5}^{y} \hat{\sigma}_{6}^{z}}\)  &  \(-i\braket{\hat{\eta}_1 \hat{\eta}_4}\) & \(-0.1858\)\\
     &  \(\braket{\hat{\sigma}_{1}^{y}  \hat{\sigma}_{4}^{z} \hat{\sigma}_{5}^{z} \hat{\sigma}_{6}^{y}}\) & \(+\braket{\hat{\eta}_1\hat{\eta}_4\hat{\eta}_5\hat{\eta}_6}\) & \(+0.3730\)\\
    V&\(\braket{\hat{\sigma}_{1}^{x} \hat{\sigma}_{2}^{z}  \hat{\sigma}_{5}^{x} \hat{\sigma}_{6}^{z}}\)  & \(+i\braket{\hat{\eta}_2 \hat{\eta}_5}\) & \(-0.1858\)\\
    &\(\braket{\hat{\sigma}_{1}^{z} \hat{\sigma}_{2}^{z}  \hat{\sigma}_{5}^{x} \hat{\sigma}_{6}^{x}}\) & \(-\braket{\hat{\eta}_1\hat{\eta}_2\hat{\eta}_5\hat{\eta}_6}\) & \(+0.3730\)\\
    VI &\(\braket{\hat{\sigma}_{1}^{x} \hat{\sigma}_{2}^{y} \hat{\sigma}_{3}^{x} \hat{\sigma}_{6}^{y}}\) &  \(-i\braket{\hat{\eta}_3 \hat{\eta}_6}\) & \(-0.1858\)\\
     &\(\braket{\hat{\sigma}_{1}^{y} \hat{\sigma}_{2}^{x} \hat{\sigma}_{3}^{x}  \hat{\sigma}_{6}^{y}}\) & \(+\braket{\hat{\eta}_1\hat{\eta}_2\hat{\eta}_3\hat{\eta}_6}\) & \(+0.3730\)\\
    VII&\(\braket{\hat{\sigma}_{2}^{x} \hat{\sigma}_{3}^{x} \hat{\sigma}_{5}^{x} \hat{\sigma}_{6}^{x}}\) &  \(+\braket{ \hat{\eta}_2 \hat{\eta}_3 \hat{\eta}_5 \hat{\eta}_6 }\) & \(+0.2410\)\\
    VIII&\(\braket{\hat{\sigma}_{1}^{y} \hat{\sigma}_{3}^{y} \hat{\sigma}_{4}^{y}  \hat{\sigma}_{6}^{y}}\)&  \(-\braket{  \hat{\eta}_1\hat{\eta}_3 \hat{\eta}_4 \hat{\eta}_6  }\)  & \(+0.2410\)\\
     IX &\(\braket{\hat{\sigma}_{1}^{z} \hat{\sigma}_{2}^{z} \hat{\sigma}_{4}^{z} \hat{\sigma}_{5}^{z}} \) & \(+\braket{  \hat{\eta}_1\hat{\eta}_2 \hat{\eta}_4 \hat{\eta}_5  }\)& \(+0.2410\)\\
    \hline
\end{tabular}
    \label{fourpointcfs}
\end{table}

\begin{table}[H]
\centering
\caption{Six point functions for \(J^x=J^y=J^z=-1\)}
    \begin{tabular}  {  
  | p{3.0cm}  
  | >{\raggedright}p{3.0cm}  
  | p{2cm} | }
  \hline
     Correlation function & Matter fermion representation & Expectation value  \\
    \hline
    \(\braket{\hat{\sigma}_{1}^{y} \hat{\sigma}_{2}^{x} \hat{\sigma}_{3}^{z} \hat{\sigma}_{4}^{x} \hat{\sigma}_{5}^{y} \hat{\sigma}_{6}^{z}}\) & \(+i\braket{\hat{\eta}_1\hat{\eta}_2}\)& \(-0.5249\)\\
    \(\braket{\hat{\sigma}_{1}^{x} \hat{\sigma}_{2}^{z} \hat{\sigma}_{3}^{y} \hat{\sigma}_{4}^{x} \hat{\sigma}_{5}^{y} \hat{\sigma}_{6}^{z}}\) & \(-i\braket{\hat{\eta}_2 \hat{\eta}_3}\)& \(-0.5249\)\\
    \(\braket{\hat{\sigma}_{1}^{x} \hat{\sigma}_{2}^{y} \hat{\sigma}_{3}^{x} \hat{\sigma}_{4}^{z} \hat{\sigma}_{5}^{y} \hat{\sigma}_{6}^{z}}\) & \(+i\braket{\hat{\eta}_3 \hat{\eta}_4}\)& \(-0.5249\)\\
    \(\braket{\hat{\sigma}_{1}^{x} \hat{\sigma}_{2}^{y} \hat{\sigma}_{3}^{z} \hat{\sigma}_{4}^{y} \hat{\sigma}_{5}^{x} \hat{\sigma}_{6}^{z}}\) & \(-i\braket{\hat{\eta}_4\hat{\eta}_5}\)& \(-0.5249\)\\
    \(\braket{\hat{\sigma}_{1}^{x} \hat{\sigma}_{2}^{y} \hat{\sigma}_{3}^{z} \hat{\sigma}_{4}^{x} \hat{\sigma}_{5}^{z} \hat{\sigma}_{6}^{y}}\) &  \(+i\braket{\hat{\eta}_5\hat{\eta}_6}\) & \(-0.5249\)\\
    \(\braket{\hat{\sigma}_{1}^{z} \hat{\sigma}_{2}^{y} \hat{\sigma}_{3}^{z} \hat{\sigma}_{4}^{x} \hat{\sigma}_{5}^{y} \hat{\sigma}_{6}^{x}}\) & \(+i\braket{\hat{\eta}_1 \hat{\eta}_6}\) & \(-0.5249\)\\
    \(\braket{\hat{\sigma}_{1}^{y} \hat{\sigma}_{2}^{x} \hat{\sigma}_{3}^{x} \hat{\sigma}_{4}^{z} \hat{\sigma}_{5}^{y} \hat{\sigma}_{6}^{z}}\)  &  \(-\braket{\hat{\eta}_1\hat{\eta}_2\hat{\eta}_3\hat{\eta}_4}\)& \(+0.3730\)\\
    \(\braket{\hat{\sigma}_{1}^{x} \hat{\sigma}_{2}^{z} \hat{\sigma}_{3}^{y} \hat{\sigma}_{4}^{y} \hat{\sigma}_{5}^{x} \hat{\sigma}_{6}^{z}}\)  &   \(-\braket{\hat{\eta}_2\hat{\eta}_3\hat{\eta}_4\hat{\eta}_5}\) & \(+0.3730\)\\
    \(\braket{\hat{\sigma}_{1}^{x} \hat{\sigma}_{2}^{y} \hat{\sigma}_{3}^{x} \hat{\sigma}_{4}^{z} \hat{\sigma}_{5}^{z} \hat{\sigma}_{6}^{y}}\)  &  \(-\braket{\hat{\eta}_3\hat{\eta}_4\hat{\eta}_5\hat{\eta}_6}\) & \(+0.3730\)\\
    \(\braket{\hat{\sigma}_{1}^{z} \hat{\sigma}_{2}^{y} \hat{\sigma}_{3}^{z} \hat{\sigma}_{4}^{y} \hat{\sigma}_{5}^{x} \hat{\sigma}_{6}^{x}}\)  & \(+\braket{\hat{\eta}_1\hat{\eta}_4\hat{\eta}_5\hat{\eta}_6}\) & \(+0.3730\)\\
    \(\braket{\hat{\sigma}_{1}^{y} \hat{\sigma}_{2}^{x} \hat{\sigma}_{3}^{z} \hat{\sigma}_{4}^{x} \hat{\sigma}_{5}^{z} \hat{\sigma}_{6}^{y}}\)  & \(-\braket{\hat{\eta}_1\hat{\eta}_2\hat{\eta}_5\hat{\eta}_6}\) & \(+0.3730\)\\
    \(\braket{\hat{\sigma}_{1}^{z} \hat{\sigma}_{2}^{z} \hat{\sigma}_{3}^{y} \hat{\sigma}_{4}^{x} \hat{\sigma}_{5}^{y} \hat{\sigma}_{6}^{x}}\) & \(+\braket{\hat{\eta}_1\hat{\eta}_2\hat{\eta}_3\hat{\eta}_6}\) & \(+0.3730\)\\
    \(\braket{\hat{\sigma}_{1}^{x} \hat{\sigma}_{2}^{z} \hat{\sigma}_{3}^{y} \hat{\sigma}_{4}^{x} \hat{\sigma}_{5}^{z} \hat{\sigma}_{6}^{y}}\) & \(+\braket{ \hat{\eta}_2 \hat{\eta}_3 \hat{\eta}_5 \hat{\eta}_6 }\)& \(+0.2410\) \\
    \(\braket{\hat{\sigma}_{1}^{z} \hat{\sigma}_{2}^{y} \hat{\sigma}_{3}^{x} \hat{\sigma}_{4}^{z} \hat{\sigma}_{5}^{y} \hat{\sigma}_{6}^{x}}\) & \(-\braket{  \hat{\eta}_1\hat{\eta}_3 \hat{\eta}_4 \hat{\eta}_6  }\)& \(+0.2410\) \\
    \(\braket{\hat{\sigma}_{1}^{y} \hat{\sigma}_{2}^{x} \hat{\sigma}_{3}^{z} \hat{\sigma}_{4}^{y} \hat{\sigma}_{5}^{x} \hat{\sigma}_{6}^{z}}\) &\(+\braket{  \hat{\eta}_1\hat{\eta}_2 \hat{\eta}_4 \hat{\eta}_5  }\) & \(+0.2410\) \\
    \(\braket{\hat{\sigma}_{1}^{z} \hat{\sigma}_{2}^{z} \hat{\sigma}_{3}^{y} \hat{\sigma}_{4}^{y} \hat{\sigma}_{5}^{x} \hat{\sigma}_{6}^{x}}\) &\(-i\braket{\hat{\eta}_1\hat{\eta}_2 \hat{\eta}_3\hat{\eta}_4  \hat{\eta}_5\hat{\eta}_6 }\) & \(-0.4363\)\\
    \(\braket{\hat{\sigma}_{1}^{y} \hat{\sigma}_{2}^{x} \hat{\sigma}_{3}^{x} \hat{\sigma}_{4}^{z} \hat{\sigma}_{5}^{z} \hat{\sigma}_{6}^{y}}\) & \(-i\braket{ \hat{\eta}_{1}\hat{\eta}_{2} \hat{\eta}_{3}\hat{\eta}_{4} \hat{\eta}_{5}\hat{\eta}_{6}}\) & \(-0.4363\)\\
    \(\braket{\hat{\sigma}_{1}^{x}\hat{\sigma}_{2}^{y} \hat{\sigma}_3^{z}\hat{\sigma}_{4}^{x} \hat{\sigma}_{5}^{y} \hat{\sigma}_{6}^{z}}\) & +1& \(+1\)\\
    \hline
    \end{tabular}
    \label{sixpointcfs}
\end{table}

\subsection{Under a [111] magnetic field}\label{sec:3-spin-chirality}
\begin{figure}[h!]
\begin{center}
    \includegraphics[width=0.7\linewidth]{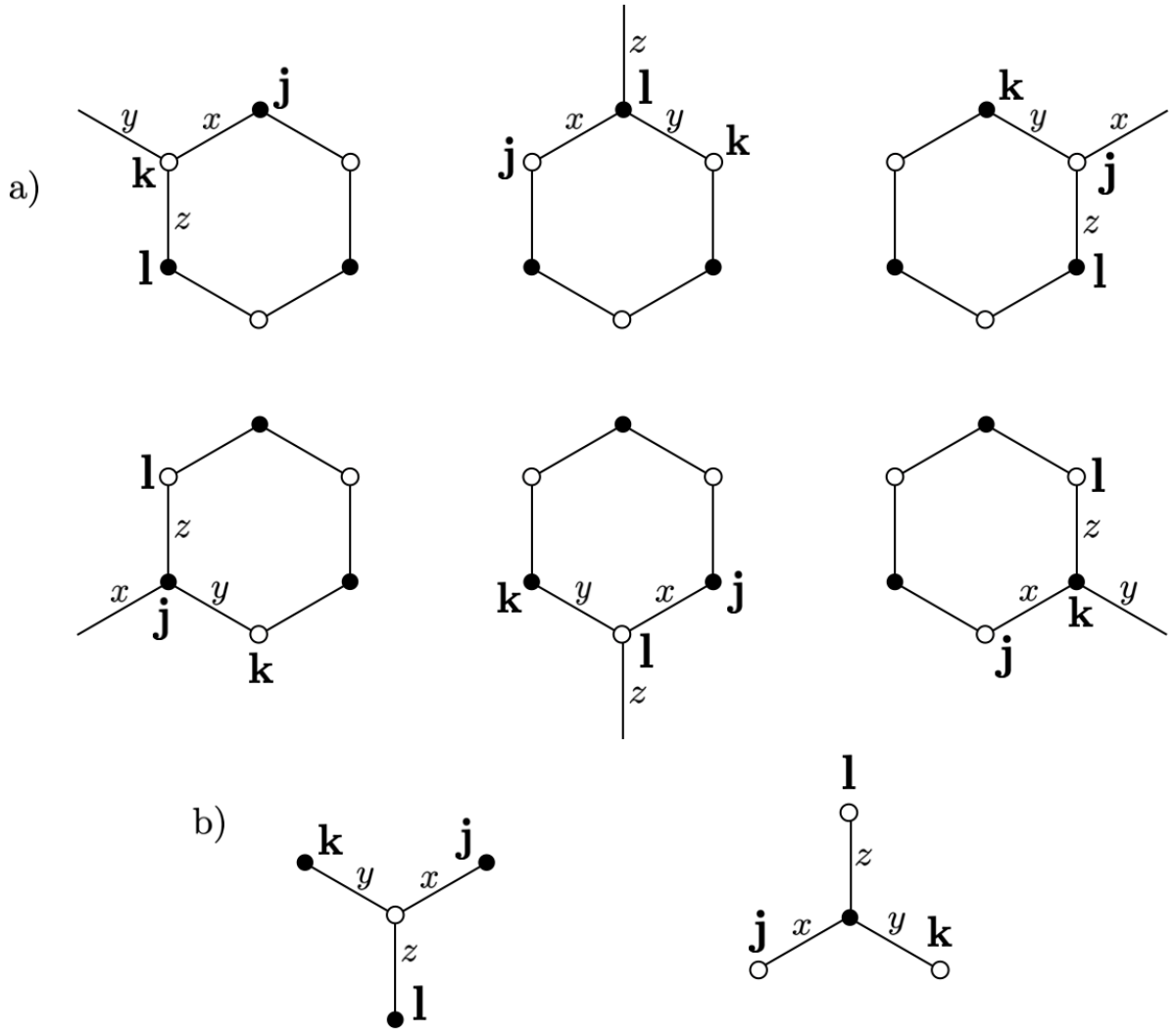}
    \end{center}
    \caption{Spin triple arrangements contributing to the projected Hamiltonian. (a) Spin triples on a plaquette. These correspond to operators quadratic in the matter fermions. (b) These configurations are associated with terms quartic in the matter fermions.}
    \label{spin_triples}
\end{figure}

Now we consider the addition of a magnetic field term,
\begin{equation}
    H_b = -\sum_{i}\sum_{\{\alpha=x,y,z\}}h_{\alpha}\hat{\sigma}^{\alpha}_{i}\,,
\end{equation}
where $h_\alpha$ are the components of the field. In the case of a weak magnetic field, an effective Hamiltonian may be found in the subspace of the flux-free states using perturbation theory.~\cite{Kitaev2006} In the isotropic case (\(J^x=J^y=J^z=J\)), the additional magnetic field term can then be expressed as
\begin{equation}
 H_b \approx-\kappa \sum_{\llangle jkl \rrangle }\hat{\sigma}_{j}^{x}\hat{\sigma}_{k}^{y}\hat{\sigma}_{l}^{z}\,,   
\end{equation}
where \(\kappa=h_x h_y h_z/J^2\) in this section, and \(\llangle jkl \rrangle \) represents the sum over all spin triples that conserve the flux operators \(\hat{W}_p\), as listed in Fig.~\ref{spin_triples}.~\cite{Kitaev2006}
\\\\
Fermionizing the spin operators as in Sec.~\ref{model}, the Hamiltonian can be expressed as a sum of terms containing either two or four matter fermions. Following Ref.~\onlinecite{Kitaev2006}, we drop the four-fermion contributions in the weak magnetic field limit. Using the gauge choice \(\hat{u}_{ij}^{\alpha}=1\) and employing the same notation as Sec.~\ref{fermion_ground_state}, the Hamiltonian is given by,  
\begin{equation}
   \hat{H}= \sum_{\boldsymbol{k}}\begin{pmatrix}
\hat f_{\boldsymbol{k}}^{\dagger} & \hat f_{-\boldsymbol{k}}
\end{pmatrix}  
\begin{pmatrix}
\xi_{\boldsymbol{k}}  &  i \Delta_{\boldsymbol{k}} +\delta_{\boldsymbol{k}}\\
- i\Delta_{\boldsymbol{k}}+\delta_{\boldsymbol{k}}  &  - \xi_{\boldsymbol{k}} 
\end{pmatrix}\begin{pmatrix}
\hat f_{\boldsymbol{k}} \\
\hat f_{-\boldsymbol{k}}^{\dagger}
\end{pmatrix} \,,
\end{equation}
where \(\delta_{\boldsymbol{k}}=-2\kappa[\sin \boldsymbol{k}\cdot\boldsymbol{a}_1-\sin \boldsymbol{k}\cdot\left(\boldsymbol{a}_1-\boldsymbol{a}_2\right)-\sin \boldsymbol{k}\cdot\boldsymbol{a}_2]\). This is diagonalized by the Boguliubov transformation,  $\hat f_{\boldsymbol{k}} = \cos \theta_{\boldsymbol{k}} \hat{a}_{\boldsymbol{k}} - i \sin \theta_{\boldsymbol{k}} \hat{a}^\dagger_{-\boldsymbol{k}}$,  where
\begin{equation}
    \cos \theta_{\boldsymbol{k}} = \sqrt{\frac{E_{\boldsymbol{k}} +\xi_{\boldsymbol{k}}}{2E_{\boldsymbol{k}}}}\,, \quad \sin \theta_{\boldsymbol{k}} =  \frac{\Delta_{\boldsymbol{k}}-i \delta_{\boldsymbol{k}}}{\sqrt{\Delta_{\boldsymbol{k}}^2+\delta_{\boldsymbol{k}}^2 }}\sqrt{\frac{E_{\boldsymbol{k}} -\xi_{\boldsymbol{k}}}{2E_{\boldsymbol{k}}}}\,,
\end{equation}
and
\(E_{\boldsymbol{k}}=\sqrt{\xi_{\boldsymbol{k}}^2+\Delta_{\boldsymbol{k}}^2+\delta_{\boldsymbol{k}}^2}\). The two-point correlation functions for the matter fermions can then be found, 
\begin{equation}
    \begin{aligned}
        &\braket{\hat{\eta}_{A \boldsymbol{R}}\hat{\eta}_{B \boldsymbol{R}'}}
        =\frac{i}{N}\sum_{\boldsymbol{k}}\frac{\xi_{\boldsymbol{k}}-i \Delta_{\boldsymbol{k}}}{E_{\boldsymbol{k}}}e^{i \boldsymbol{k}\cdot(\boldsymbol{R}-\boldsymbol{R}')}\\
        &\braket{\hat{\eta}_{A \boldsymbol{R}}\hat{\eta}_{A \boldsymbol{R}'}}=\delta_{\boldsymbol{R} \boldsymbol{R}'}+\frac{1}{N}\sum_{\boldsymbol{k}}\frac{\delta_{\boldsymbol{k}}}{E_{\boldsymbol{k}}}e^{i \boldsymbol{k}\cdot(\boldsymbol{R}-\boldsymbol{R}')}\\
        &\braket{\hat{\eta}_{B \boldsymbol{R}}\hat{\eta}_{B \boldsymbol{R}'}}=\delta_{\boldsymbol{R} \boldsymbol{R}'}-\frac{1}{N}\sum_{\boldsymbol{k}}\frac{\delta_{\boldsymbol{k}}}{E_{\boldsymbol{k}}}e^{i \boldsymbol{k}\cdot(\boldsymbol{R}-\boldsymbol{R}')}\,.
    \end{aligned}
\end{equation}

The reduced density matrix for the plaquette can be calculated in terms of spin correlation functions using Eq.~\eqref{rdm_general_formula}. While the flux-conserving two-spin, four-spin, and six-spin operators contribute to the sum in Eq.~\eqref{rdm_general_formula} at zero field, the breaking of time-reversal symmetry allows the inclusion of operators involving an odd number of spins, provided that they conserve flux. In particular, the one-spin contribution to the reduced density matrix remains absent. The three-spin correlation functions are nonzero for the operators listed in Fig.~\ref{spin_triples}, and Table~\ref{three_point_correlation_finite_field} lists all nonzero three-spin correlations functions on a plaquette and their values in the thermodynamic limit.

The list of all nonzero spin correlation functions, their corresponding matter Majorana fermion representations and their numerical values for \(J =-1\) and \(\kappa=0.1\) are presented in Tables \ref{two_point_correlation_finite_field}, \ref{three_point_correlation_finite_field}, \ref{four_point_correlation_finite_field}, \ref{five_point_correlation_finite_field} and \ref{six_point_correlation_finite_field}.

\begin{table}[H]
\centering
\caption{Nonzero two point functions for \(J^x=J^y=J^z =-1\) and \(\kappa=0.1\)}
\label{two_point_correlation_finite_field}
    \begin{tabular}  {  
  | p{2cm}  
  | >{\raggedright}p{2.5cm}  
  | p{2cm} | }
  \hline
     Correlation function & Matter fermion representation & Expectation value  \\
    \hline
     \( \braket{\hat{\sigma}_{1}^{z} \hat{\sigma}_{2}^{z}} \) & \(+i\braket{\hat{\eta}_1\hat{\eta}_2}\) &\(-0.5152\)\\
    \( \braket{\hat{\sigma}_{2}^{x} \hat{\sigma}_{3}^{x}} \) & \(-i\braket{\hat{\eta}_2\hat{\eta}_3}\) & \(-0.5152\)\\
     \( \braket{\hat{\sigma}_{3}^{y} \hat{\sigma}_{4}^{y}} \) & \(+i\braket{\hat{\eta}_3\hat{\eta}_4}\) &\(-0.5152\)\\
     \( \braket{\hat{\sigma}_{4}^{z} \hat{\sigma}_{5}^{z}} \) & \(-i\braket{\hat{\eta}_4\hat{\eta}_5}\) &\(-0.5152\)\\
    \( \braket{\hat{\sigma}_{5}^{x} \hat{\sigma}_{6}^{x}} \) & \(+i\braket{\hat{\eta}_5\hat{\eta}_6}\) &\(-0.5152\)\\
    \( \braket{\hat{\sigma}_{1}^{y} \hat{\sigma}_{6}^{y}} \) &  \(+i\braket{\hat{\eta}_1\hat{\eta}_6}\)&\(-0.5152\)\\
    \hline
    \end{tabular}
\end{table}

\begin{table}[H]
\centering
\caption{Non-zero three point functions for \(J^x=J^y=J^z=-1\) and \(\kappa=0.1\)}
\label{three_point_correlation_finite_field}
\begin{tabular} {  
  | p{2cm}  
  | >{\raggedright}p{2.5cm}  
  | p{2cm} | }  
  \hline
     Correlation function & Matter fermion representation & Expectation value  \\
    \hline
     \(  \braket{\hat{\sigma}_{1}^{z} \hat{\sigma}_{2}^{y} \hat{\sigma}_{3}^{x}} \) & \(-i \braket{\hat{\eta}_1\hat{\eta}_3}\) & \(+0.1119\)\\
     \(  \braket{\hat{\sigma}_{2}^{x} \hat{\sigma}_{3}^{z} \hat{\sigma}_{4}^{y}} \) & \(-i \braket{\hat{\eta}_2\hat{\eta}_4}\) & \(+0.1119\)\\
     \(  \braket{\hat{\sigma}_{3}^{y} \hat{\sigma}_{4}^{x} \hat{\sigma}_{5}^{z}} \) & \(-i \braket{\hat{\eta}_3\hat{\eta}_5}\) & \(+0.1119\)\\
     \(  \braket{\hat{\sigma}_{4}^{z} \hat{\sigma}_{5}^{y} \hat{\sigma}_{6}^{x}} \) & \(-i \braket{\hat{\eta}_4\hat{\eta}_6}\) & \(+0.1119\)\\
     \(  \braket{\hat{\sigma}_{1}^{y}\hat{\sigma}_{5}^{x} \hat{\sigma}_{6}^{z} } \) & \(+i \braket{\hat{\eta}_1\hat{\eta}_5}\) & \(+0.1119\)\\
     \(  \braket{\hat{\sigma}_{1}^{x} \hat{\sigma}_{2}^{z}\hat{\sigma}_{6}^{y} } \) & \(+i \braket{\hat{\eta}_2\hat{\eta}_6}\) & \(+0.1119\)\\
     \hline
\end{tabular}
\end{table}
\begin{table}[H]
\centering
\caption{Non-zero four point functions for each grouping in Figure \ref{four_site_grouping} and \(J^x=J^y=J^z=-1\) and \(\kappa=0.1\)}
\label{four_point_correlation_finite_field}
\begin{tabular} { 
  | p{1cm} 
  | p{2cm}  
  | >{\raggedright}p{2.5cm}  
  | p{2cm} | }  
  \hline
     Group & Correlation function & Matter fermion representation & Expectation value  \\
    \hline
     I& \(  \braket{\hat{\sigma}_{1}^{z} \hat{\sigma}_{2}^{y} \hat{\sigma}_{3}^{z} \hat{\sigma}_{4}^{y}} \) & \(-i\braket{\hat{\eta}_1 \hat{\eta}_4}\) & \(-0.1694\)\\
      &\(\braket{\hat{\sigma}_{1}^{z} \hat{\sigma}_{2}^{z} \hat{\sigma}_{3}^{y} \hat{\sigma}_{4}^{y}} \) &  \(-\braket{\hat{\eta}_1\hat{\eta}_2\hat{\eta}_3\hat{\eta}_4}\)& \(+0.3402\)\\
    II &\( \braket{\hat{\sigma}_{2}^{x} \hat{\sigma}_{3}^{z} \hat{\sigma}_{4}^{x} \hat{\sigma}_{5}^{z}} \)   & \(+i\braket{\hat{\eta}_2 \hat{\eta}_5}\) & \(-0.1694\)\\
     &\( \braket{\hat{\sigma}_{2}^{x} \hat{\sigma}_{3}^{x} \hat{\sigma}_{4}^{z} \hat{\sigma}_{5}^{z}} \) &   \(-\braket{\hat{\eta}_2\hat{\eta}_3\hat{\eta}_4\hat{\eta}_5}\) & \(+0.3402\)\\
    III&\( \braket{\hat{\sigma}_{3}^{y} \hat{\sigma}_{4}^{x} \hat{\sigma}_{5}^{y} \hat{\sigma}_{6}^{x}}\)  &  \(-i\braket{\hat{\eta}_3 \hat{\eta}_6}\) & \(-0.1694\)\\
     & \(\braket{\hat{\sigma}_{3}^{y} \hat{\sigma}_{4}^{y} \hat{\sigma}_{5}^{x} \hat{\sigma}_{6}^{x}}\) &  \(-\braket{\hat{\eta}_3\hat{\eta}_4\hat{\eta}_5\hat{\eta}_6}\) & \(+0.3402\)\\
    IV &\(\braket{\hat{\sigma}_{1}^{y} \hat{\sigma}_{4}^{z} \hat{\sigma}_{5}^{y} \hat{\sigma}_{6}^{z}}\)  &  \(-i\braket{\hat{\eta}_1 \hat{\eta}_4}\) & \(-0.1694\)\\
     &  \(\braket{\hat{\sigma}_{1}^{y}  \hat{\sigma}_{4}^{z} \hat{\sigma}_{5}^{z} \hat{\sigma}_{6}^{y}}\) & \(+\braket{\hat{\eta}_1\hat{\eta}_4\hat{\eta}_5\hat{\eta}_6}\) & \(+0.3402\)\\
    V&\(\braket{\hat{\sigma}_{1}^{x} \hat{\sigma}_{2}^{z}  \hat{\sigma}_{5}^{x} \hat{\sigma}_{6}^{z}}\)  & \(+i\braket{\hat{\eta}_2 \hat{\eta}_5}\) & \(-0.1694\)\\
    &\(\braket{\hat{\sigma}_{1}^{z} \hat{\sigma}_{2}^{z}  \hat{\sigma}_{5}^{x} \hat{\sigma}_{6}^{x}}\) & \(-\braket{\hat{\eta}_1\hat{\eta}_2\hat{\eta}_5\hat{\eta}_6}\) & \(+0.3402\)\\
    VI &\(\braket{\hat{\sigma}_{1}^{x} \hat{\sigma}_{2}^{y} \hat{\sigma}_{3}^{x} \hat{\sigma}_{6}^{y}}\) &  \(-i\braket{\hat{\eta}_3 \hat{\eta}_6}\) & \(-0.1694\)\\
     &\(\braket{\hat{\sigma}_{1}^{y} \hat{\sigma}_{2}^{x} \hat{\sigma}_{3}^{x}  \hat{\sigma}_{6}^{y}}\) & \(+\braket{\hat{\eta}_1\hat{\eta}_2\hat{\eta}_3\hat{\eta}_6}\) & \(+0.3402\)\\
    VII&\(\braket{\hat{\sigma}_{2}^{x} \hat{\sigma}_{3}^{x} \hat{\sigma}_{5}^{x} \hat{\sigma}_{6}^{x}}\) &  \(+\braket{ \hat{\eta}_2 \hat{\eta}_3 \hat{\eta}_5 \hat{\eta}_6 }\) & \(+0.2493\)\\
    VIII&\(\braket{\hat{\sigma}_{1}^{y} \hat{\sigma}_{3}^{y} \hat{\sigma}_{4}^{y}  \hat{\sigma}_{6}^{y}}\)&  \(-\braket{  \hat{\eta}_1\hat{\eta}_3 \hat{\eta}_4 \hat{\eta}_6  }\)  & \(+0.2493\)\\
     IX &\(\braket{\hat{\sigma}_{1}^{z} \hat{\sigma}_{2}^{z} \hat{\sigma}_{4}^{z} \hat{\sigma}_{5}^{z}} \) & \(+\braket{  \hat{\eta}_1\hat{\eta}_2 \hat{\eta}_4 \hat{\eta}_5  }\)& \(+0.2493\)\\
    \hline
\end{tabular}
\end{table}

\begin{table}[H]
\centering
\caption{Non-zero five point functions for \(J^x=J^y=J^z=-1\) and \(\kappa=0.1\)}
\label{five_point_correlation_finite_field}
    \begin{tabular}  {  
  | p{3.0cm}  
  | >{\raggedright}p{3.0cm}  
  | p{2cm} | }
  \hline
     Correlation function & Matter fermion representation & Expectation value  \\
    \hline
    \(\braket{\hat{\sigma}_{1}^{y} \hat{\sigma}_{2}^{0} \hat{\sigma}_{3}^{y} \hat{\sigma}_{4}^{x} \hat{\sigma}_{5}^{y} \hat{\sigma}_{6}^{z}}\) & \(-i \braket{\hat{\eta}_1\hat{\eta}_3}\)& \(+0.1119\)\\
    \(\braket{\hat{\sigma}_{1}^{y} \hat{\sigma}_{2}^{0} \hat{\sigma}_{3}^{y} \hat{\sigma}_{4}^{y} \hat{\sigma}_{5}^{x} \hat{\sigma}_{6}^{z}}\) & \(- \braket{\hat{\eta}_1 \hat{\eta}_3 \hat{\eta}_4 \hat{\eta}_5}\)& \(-0.0963\)\\
    \(\braket{\hat{\sigma}_{1}^{y} \hat{\sigma}_{2}^{0} \hat{\sigma}_{3}^{y} \hat{\sigma}_{4}^{x} \hat{\sigma}_{5}^{z} \hat{\sigma}_{6}^{y}}\) & \(+\braket{\hat{\eta}_1 \hat{\eta}_3 \hat{\eta}_5 \hat{\eta}_6}\)& \(-0.0963\)\\
    \(\braket{\hat{\sigma}_{1}^{x} \hat{\sigma}_{2}^{z} \hat{\sigma}_{3}^{0} \hat{\sigma}_{4}^{z} \hat{\sigma}_{5}^{y} \hat{\sigma}_{6}^{z}}\) & \(-i \braket{\hat{\eta}_2\hat{\eta}_4}\)& \(+0.1119\)\\
    \(\braket{\hat{\sigma}_{1}^{x} \hat{\sigma}_{2}^{z} \hat{\sigma}_{3}^{0} \hat{\sigma}_{4}^{z} \hat{\sigma}_{5}^{z} \hat{\sigma}_{6}^{y}}\) & \( +\braket{\hat{\eta}_2\hat{\eta}_4\hat{\eta}_5\hat{\eta}_6}\)& \(-0.0963\)\\
    \(\braket{\hat{\sigma}_{1}^{z} \hat{\sigma}_{2}^{z} \hat{\sigma}_{3}^{0} \hat{\sigma}_{4}^{z} \hat{\sigma}_{5}^{y} \hat{\sigma}_{6}^{x}}\) & \( +\braket{\hat{\eta}_1 \hat{\eta}_2\hat{\eta}_4\hat{\eta}_6}\)& \(-0.0963\)\\
    \(\braket{\hat{\sigma}_{1}^{x} \hat{\sigma}_{2}^{y} \hat{\sigma}_{3}^{x} \hat{\sigma}_{4}^{0} \hat{\sigma}_{5}^{x} \hat{\sigma}_{6}^{z}}\) & \(-i \braket{\hat{\eta}_3\hat{\eta}_5}\) & \(+0.1119\)\\
    \(\braket{\hat{\sigma}_{1}^{y} \hat{\sigma}_{2}^{x} \hat{\sigma}_{3}^{x} \hat{\sigma}_{4}^{0} \hat{\sigma}_{5}^{x} \hat{\sigma}_{6}^{z}}\) & \(+\braket{\hat{\eta}_1 \hat{\eta}_2 \hat{\eta}_3 \hat{\eta}_5}\)& \(-0.0963\)\\
    \(\braket{\hat{\sigma}_{1}^{z} \hat{\sigma}_{2}^{y} \hat{\sigma}_{3}^{x} \hat{\sigma}_{4}^{0} \hat{\sigma}_{5}^{x} \hat{\sigma}_{6}^{x}}\) & \(+\braket{\hat{\eta}_1 \hat{\eta}_3 \hat{\eta}_5 \hat{\eta}_6}\)& \(-0.0963\)\\
    \(\braket{\hat{\sigma}_{1}^{x} \hat{\sigma}_{2}^{y} \hat{\sigma}_{3}^{z} \hat{\sigma}_{4}^{y} \hat{\sigma}_{5}^{0} \hat{\sigma}_{6}^{y}}\) &  \(-i \braket{\hat{\eta}_4\hat{\eta}_6}\) & \(+0.1119\)\\
    \(\braket{\hat{\sigma}_{1}^{x} \hat{\sigma}_{2}^{z} \hat{\sigma}_{3}^{y} \hat{\sigma}_{4}^{y} \hat{\sigma}_{5}^{0} \hat{\sigma}_{6}^{y}}\) & \( -\braket{\hat{\eta}_2\hat{\eta}_3\hat{\eta}_4\hat{\eta}_6}\)& \(-0.0963\)\\
    \(\braket{\hat{\sigma}_{1}^{y} \hat{\sigma}_{2}^{x} \hat{\sigma}_{3}^{z} \hat{\sigma}_{4}^{y} \hat{\sigma}_{5}^{0} \hat{\sigma}_{6}^{y}}\) & \( +\braket{\hat{\eta}_1 \hat{\eta}_2\hat{\eta}_4\hat{\eta}_6}\)& \(-0.0963\)\\
    \(\braket{\hat{\sigma}_{1}^{z} \hat{\sigma}_{2}^{y} \hat{\sigma}_{3}^{z} \hat{\sigma}_{4}^{x} \hat{\sigma}_{5}^{z} \hat{\sigma}_{6}^{0}}\) & \(+i \braket{\hat{\eta}_1\hat{\eta}_5}\)& \(+0.1119\)\\
    \(\braket{\hat{\sigma}_{1}^{z} \hat{\sigma}_{2}^{z} \hat{\sigma}_{3}^{y} \hat{\sigma}_{4}^{x} \hat{\sigma}_{5}^{z} \hat{\sigma}_{6}^{0}}\) & \(+\braket{\hat{\eta}_1 \hat{\eta}_2 \hat{\eta}_3 \hat{\eta}_5}\)& \(-0.0963\)\\
    \(\braket{\hat{\sigma}_{1}^{z} \hat{\sigma}_{2}^{y} \hat{\sigma}_{3}^{x} \hat{\sigma}_{4}^{z} \hat{\sigma}_{5}^{z} \hat{\sigma}_{6}^{0}}\) & \(- \braket{\hat{\eta}_1 \hat{\eta}_3 \hat{\eta}_4 \hat{\eta}_5}\)& \(-0.0963\)\\
    \(\braket{\hat{\sigma}_{1}^{0} \hat{\sigma}_{2}^{x} \hat{\sigma}_{3}^{z} \hat{\sigma}_{4}^{x} \hat{\sigma}_{5}^{y} \hat{\sigma}_{6}^{x}}\) & \(+i \braket{\hat{\eta}_2\hat{\eta}_6}\)& \(+0.1119\)\\
    \(\braket{\hat{\sigma}_{1}^{0} \hat{\sigma}_{2}^{x} \hat{\sigma}_{3}^{z} \hat{\sigma}_{4}^{y} \hat{\sigma}_{5}^{x} \hat{\sigma}_{6}^{x}}\) & \( +\braket{\hat{\eta}_2\hat{\eta}_4\hat{\eta}_5\hat{\eta}_6}\)& \(-0.0963\)\\
    \(\braket{\hat{\sigma}_{1}^{0} \hat{\sigma}_{2}^{x} \hat{\sigma}_{3}^{x} \hat{\sigma}_{4}^{z} \hat{\sigma}_{5}^{y} \hat{\sigma}_{6}^{x}}\) & \( -\braket{\hat{\eta}_2 \hat{\eta}_3\hat{\eta}_4\hat{\eta}_6}\)& \(-0.0963\)\\
    \hline
    \end{tabular}
\end{table}

\begin{table}[H]
\centering
\caption{Non-zero six point functions for \(J^x=J^y=J^z=-1\) and \(\kappa=0.1\)}
\label{six_point_correlation_finite_field}
    \begin{tabular}  {  
  | p{3.0cm}  
  | >{\raggedright}p{3.0cm}  
  | p{2cm} | }
  \hline
     Correlation function & Matter fermion representation & Expectation value  \\
    \hline
    \(\braket{\hat{\sigma}_{1}^{y} \hat{\sigma}_{2}^{x} \hat{\sigma}_{3}^{z} \hat{\sigma}_{4}^{x} \hat{\sigma}_{5}^{y} \hat{\sigma}_{6}^{z}}\) & \(+i\braket{\hat{\eta}_1\hat{\eta}_2}\)& \(-0.5152\)\\
    \(\braket{\hat{\sigma}_{1}^{x} \hat{\sigma}_{2}^{z} \hat{\sigma}_{3}^{y} \hat{\sigma}_{4}^{x} \hat{\sigma}_{5}^{y} \hat{\sigma}_{6}^{z}}\) & \(-i\braket{\hat{\eta}_2 \hat{\eta}_3}\)& \(-0.5152\)\\
    \(\braket{\hat{\sigma}_{1}^{x} \hat{\sigma}_{2}^{y} \hat{\sigma}_{3}^{x} \hat{\sigma}_{4}^{z} \hat{\sigma}_{5}^{y} \hat{\sigma}_{6}^{z}}\) & \(+i\braket{\hat{\eta}_3 \hat{\eta}_4}\)& \(-0.5152\)\\
    \(\braket{\hat{\sigma}_{1}^{x} \hat{\sigma}_{2}^{y} \hat{\sigma}_{3}^{z} \hat{\sigma}_{4}^{y} \hat{\sigma}_{5}^{x} \hat{\sigma}_{6}^{z}}\) & \(-i\braket{\hat{\eta}_4\hat{\eta}_5}\)& \(-0.5152\)\\
    \(\braket{\hat{\sigma}_{1}^{x} \hat{\sigma}_{2}^{y} \hat{\sigma}_{3}^{z} \hat{\sigma}_{4}^{x} \hat{\sigma}_{5}^{z} \hat{\sigma}_{6}^{y}}\) &  \(+i\braket{\hat{\eta}_5\hat{\eta}_6}\) & \(-0.5152\)\\
    \(\braket{\hat{\sigma}_{1}^{z} \hat{\sigma}_{2}^{y} \hat{\sigma}_{3}^{z} \hat{\sigma}_{4}^{x} \hat{\sigma}_{5}^{y} \hat{\sigma}_{6}^{x}}\) & \(+i\braket{\hat{\eta}_1 \hat{\eta}_6}\) & \(-0.5152\)\\
    \(\braket{\hat{\sigma}_{1}^{y} \hat{\sigma}_{2}^{x} \hat{\sigma}_{3}^{x} \hat{\sigma}_{4}^{z} \hat{\sigma}_{5}^{y} \hat{\sigma}_{6}^{z}}\)  &  \(-\braket{\hat{\eta}_1\hat{\eta}_2\hat{\eta}_3\hat{\eta}_4}\)& \(+0.3402\)\\
    \(\braket{\hat{\sigma}_{1}^{x} \hat{\sigma}_{2}^{z} \hat{\sigma}_{3}^{y} \hat{\sigma}_{4}^{y} \hat{\sigma}_{5}^{x} \hat{\sigma}_{6}^{z}}\)  &   \(-\braket{\hat{\eta}_2\hat{\eta}_3\hat{\eta}_4\hat{\eta}_5}\) & \(+0.3402\)\\
    \(\braket{\hat{\sigma}_{1}^{x} \hat{\sigma}_{2}^{y} \hat{\sigma}_{3}^{x} \hat{\sigma}_{4}^{z} \hat{\sigma}_{5}^{z} \hat{\sigma}_{6}^{y}}\)  &  \(-\braket{\hat{\eta}_3\hat{\eta}_4\hat{\eta}_5\hat{\eta}_6}\) & \(+0.3402\)\\
    \(\braket{\hat{\sigma}_{1}^{z} \hat{\sigma}_{2}^{y} \hat{\sigma}_{3}^{z} \hat{\sigma}_{4}^{y} \hat{\sigma}_{5}^{x} \hat{\sigma}_{6}^{x}}\)  & \(+\braket{\hat{\eta}_1\hat{\eta}_4\hat{\eta}_5\hat{\eta}_6}\) & \(+0.3402\)\\
    \(\braket{\hat{\sigma}_{1}^{y} \hat{\sigma}_{2}^{x} \hat{\sigma}_{3}^{z} \hat{\sigma}_{4}^{x} \hat{\sigma}_{5}^{z} \hat{\sigma}_{6}^{y}}\)  & \(-\braket{\hat{\eta}_1\hat{\eta}_2\hat{\eta}_5\hat{\eta}_6}\) & \(+0.3402\)\\
    \(\braket{\hat{\sigma}_{1}^{z} \hat{\sigma}_{2}^{z} \hat{\sigma}_{3}^{y} \hat{\sigma}_{4}^{x} \hat{\sigma}_{5}^{y} \hat{\sigma}_{6}^{x}}\) & \(+\braket{\hat{\eta}_1\hat{\eta}_2\hat{\eta}_3\hat{\eta}_6}\) & \(+0.3402\)\\
    \(\braket{\hat{\sigma}_{1}^{x} \hat{\sigma}_{2}^{z} \hat{\sigma}_{3}^{y} \hat{\sigma}_{4}^{x} \hat{\sigma}_{5}^{z} \hat{\sigma}_{6}^{y}}\) & \(+\braket{ \hat{\eta}_2 \hat{\eta}_3 \hat{\eta}_5 \hat{\eta}_6 }\)& \(+0.2493\) \\
    \(\braket{\hat{\sigma}_{1}^{z} \hat{\sigma}_{2}^{y} \hat{\sigma}_{3}^{x} \hat{\sigma}_{4}^{z} \hat{\sigma}_{5}^{y} \hat{\sigma}_{6}^{x}}\) & \(-\braket{  \hat{\eta}_1\hat{\eta}_3 \hat{\eta}_4 \hat{\eta}_6  }\)& \(+0.2493\) \\
    \(\braket{\hat{\sigma}_{1}^{y} \hat{\sigma}_{2}^{x} \hat{\sigma}_{3}^{z} \hat{\sigma}_{4}^{y} \hat{\sigma}_{5}^{x} \hat{\sigma}_{6}^{z}}\) &\(+\braket{  \hat{\eta}_1\hat{\eta}_2 \hat{\eta}_4 \hat{\eta}_5  }\) & \(+0.2493\) \\
    \(\braket{\hat{\sigma}_{1}^{z} \hat{\sigma}_{2}^{z} \hat{\sigma}_{3}^{y} \hat{\sigma}_{4}^{y} \hat{\sigma}_{5}^{x} \hat{\sigma}_{6}^{x}}\) &\(-i\braket{\hat{\eta}_1\hat{\eta}_2 \hat{\eta}_3\hat{\eta}_4  \hat{\eta}_5\hat{\eta}_6 }\) & \(-0.3712\)\\
    \(\braket{\hat{\sigma}_{1}^{y} \hat{\sigma}_{2}^{x} \hat{\sigma}_{3}^{x} \hat{\sigma}_{4}^{z} \hat{\sigma}_{5}^{z} \hat{\sigma}_{6}^{y}}\) & \(-i\braket{ \hat{\eta}_{1}\hat{\eta}_{2} \hat{\eta}_{3}\hat{\eta}_{4} \hat{\eta}_{5}\hat{\eta}_{6}}\) & \(-0.3712\)\\
    \(\braket{\hat{\sigma}_{1}^{x}\hat{\sigma}_{2}^{y} \hat{\sigma}_3^{z}\hat{\sigma}_{4}^{x} \hat{\sigma}_{5}^{y} \hat{\sigma}_{6}^{z}}\) & +1& \(+1\)\\
    \hline
    \end{tabular}
\end{table}

\section{Finite Size Analysis for the Kagomé Heisenberg Antiferromagnetic}

\begin{figure}[h!tbp]
\includegraphics[width=\columnwidth]{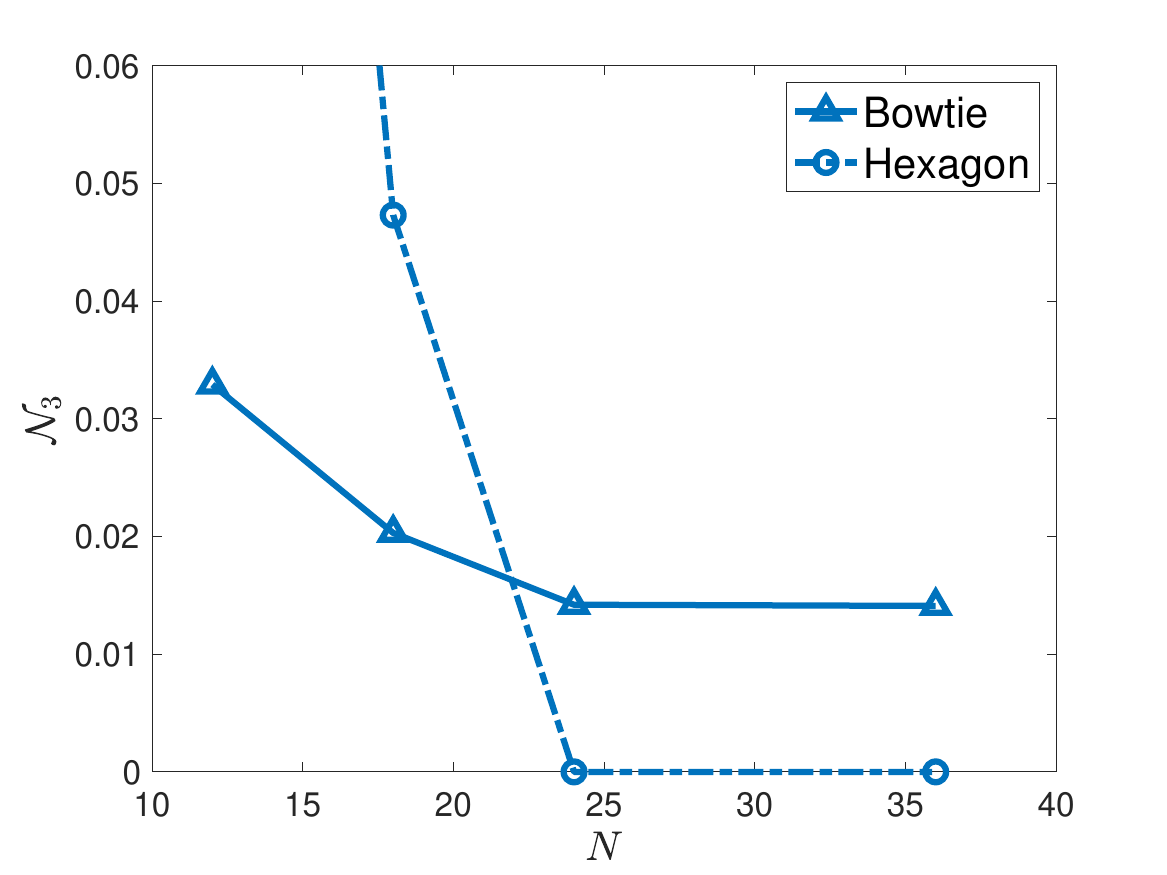}\llap{\shortstack{%
        \includegraphics[scale=.25]{Figures/Bowtie-212.pdf}\\
        \rule{0ex}{1.7in}%
      }
  \rule{1.7in}{0ex}}\llap{\shortstack{%
        \includegraphics[scale=.25]{Figures/hexagon-222.pdf}\\
        \rule{0ex}{3.2in}%
      }
  \rule{3.5in}{0ex}}
\caption{\textbf{Finite-size scaling of GMN in the Kagome Heisenberg Model.} 
Tripartite GMN (\(\mathcal{N}_3\)) for the Bowtie and Hexagon subregions as a function of system size \(N\), computed for the Kagome model with \(J_2 = \lambda =  0\). Results are shown for \(N = 12, 18, 24,\) and \(36\). The values of \(\mathcal{N}_3\) appear to have converged by \(N=36\), with the Bowtie subregion stabilizing at \(0.0141\) and the Hexagon subregion at \(0\). These results suggest that the thermodynamic limit is well-approximated by the largest system sizes considered.}
\label{fig:kag-finitesize}
\end{figure}

To assess the robustness of our ED results in the thermodynamic limit, we consider the model at $J_2=\lambda=0$ and compute the tripartite GMN \(\mathcal{N}_3\) for the Bowtie and Hexagon subregions across system sizes \(N = 12, 18, 24,\) and \(36\) (see Fig.~\ref{fig:kag-finitesize}). The values of \(\mathcal{N}_3\) appear to have converged by \(N=36\), with the Bowtie subregion stabilizing at \(0.0141\) and the Hexagon subregion at \(0\). This suggests that the absence of GMN in the Hexagon and its persistence in the Bowtie are intrinsic properties of the Kagome Heisenberg Antiferromagnet rather than finite-size effects. These findings reinforce the distinction between loopy and non-loopy subregions in hosting multipartite entanglement. 

\section{Multiparty entanglement in topological string-net models}~\label{sec:GME_string-net}

\begin{figure}[htp!]
\centering
\includegraphics[width=0.5\columnwidth]{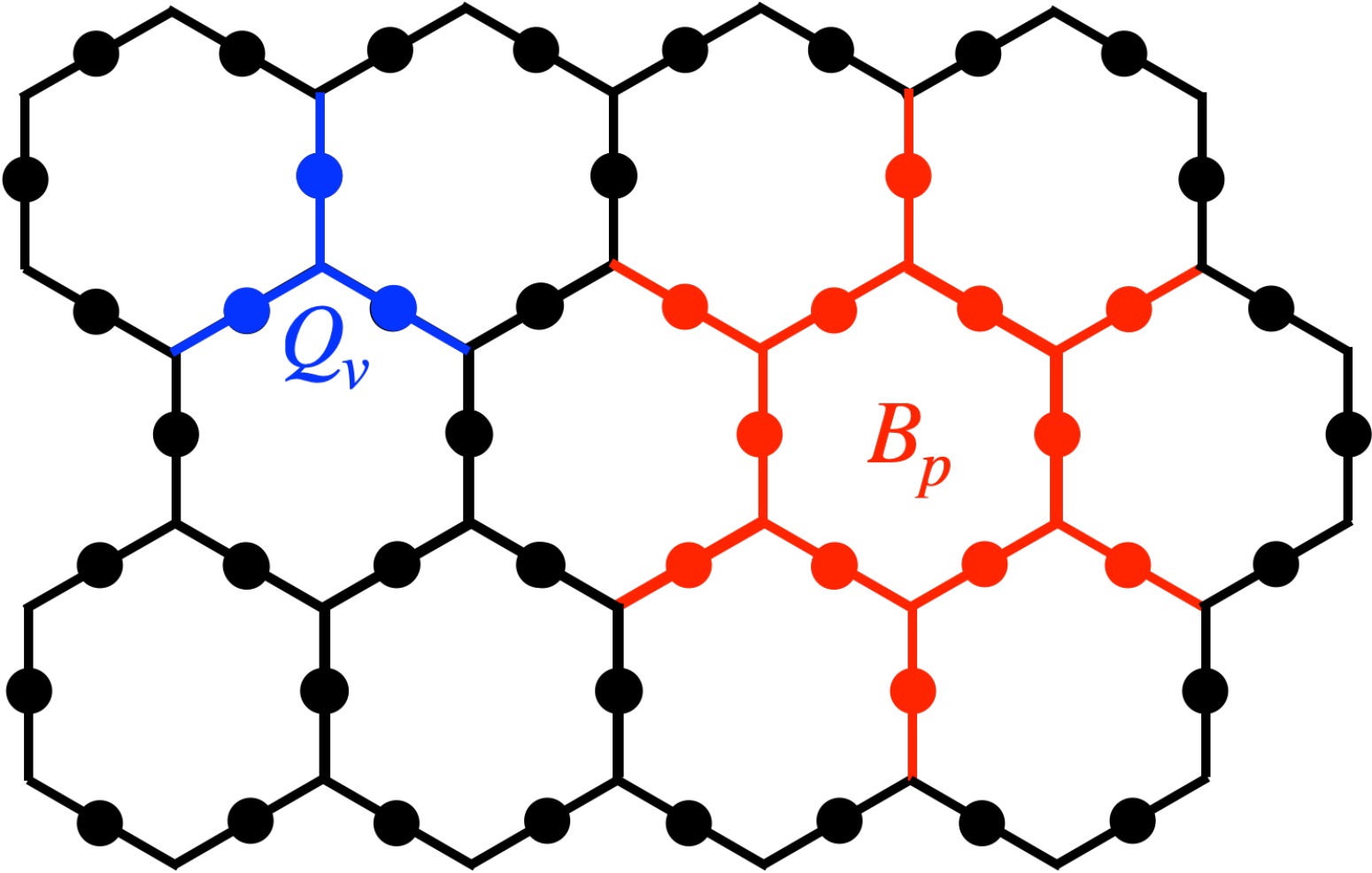}
\caption{\textbf{Hexagonal lattice for the string-net model.} The diagram illustrates the honeycomb lattice used in the string-net model of Levin and Wen~\cite{Levin2005}, where spins reside on the edges. The vertex operator \(Q_v\) (a three-spin interaction) enforces local constraints at each trivalent vertex, while the plaquette operator \(B_p\) (a 12-spin interaction) measures the magnetic flux through each plaquette and drives the dynamics of the string-net configurations.}
\label{fig:heavyhex}
\end{figure}

The string-net model proposed by Levin and Wen~\cite{Levin2005} describes a large class of topologically ordered states, including quantum spin liquids. It is defined on a honeycomb lattice with spin degrees of freedom residing on the edges (see Fig.~\ref{fig:heavyhex}). The Hamiltonian takes the form
\begin{equation}
H = -\sum_v Q_v - \sum_p B_p,
\end{equation}
where \( Q_v \) is a vertex operator enforcing local constraints at each trivalent vertex, and \( B_p \) is a plaquette operator that measures the flux through a plaquette and provides the dynamics of string-net configurations. The ground states of this model satisfy \(\langle Q_v \rangle = \langle B_p \rangle = 1\), ensuring local gauge invariance. 
A key feature of the string-net construction is that its fixed-point ground state is \emph{fully specified} by a pair of local tensors, namely the six-index tensor \(F_{kln}^{ijm}\) and the one-index tensor \(d_i\). 
In particular, we can derive the branching rules from the $F$ tensor, which provides a local gauge constraint for any three edges which share a vertex. Consider the subspace $S$ of string configurations which satisfy all local constraints, the fixed point wavefunction can be written as $\ket{\Phi}=\sum_{s\in S} a(s) \ket{s}$, and the matrix elements of the RDM of a subregion $A$ can be written as
\begin{equation}
    [\rho_A]_{s_A s_{A^\prime}} = \sum_{s_B} a(s_A s_B) a^*(s_{A^\prime} s_B)
\end{equation}
where $(s_A,s_B),(s_{A^\prime},s_B)\in S$. If we cannot find any subregion $s_B$ which is compatible with two distinct configurations $s_A$ and $s_{A^\prime}$, then all off-diagonal elements vanish and thus the RDM is fully separable. 
A similar argument was made for a general spin-1/2 system where the valid spin flips form a group~\cite{Hamma2005Diagonal}, which can be applied to show that non-loopy subregions of the $\mathbb{Z}_2$ gauge theory are fully separable. We generalize this to arbitrary Abelian string-nets:
\begin{center}
    \textit{Reduced state $\rho_A$ of any non-loopy subregion $A$ is fully separable in Abelian string-nets.}
\end{center}
To see this, notice that in an Abelian string-net, the branching rules at each vertex have no multiplicity: once two of the three incident edge labels are fixed, the third is determined uniquely. Now, if the region $A$ is a tree (i.e. has no loops), then fixing a global configuration $s_B$ on the complement immediately fixes the labels on the edges of $A$ that touch $B$. From there, we can "peel" the tree inward: at each trivalent vertex in $A$, two of its edges have already been determined, so the third edge label follows uniquely from the branching rule. By iterating this process, we recover at most one valid $s_A$ for each $s_B$. Hence, no two distinct $s_A$ can coexist with the same $s_B$, all off-diagonal matrix elements $\rho_{s_A, s_{A^{\prime}}}$ vanish, and $\rho_A$ is fully separable.

For a general string-net, the reduced density matrix of a simply connected region $A$ takes a diagonal form \cite{Levin2006TEE}
\begin{equation}
    \rho_A = \sum_{\{q,m\}} p_{\{q,m\}} \ket{\{q,m\}} \bra{\{q,m\}}
\end{equation}
where $\{q,m\}$ labels the string configuration on the tree-like boundary of $A$.  Each boundary configuration specifies a unique interior wavefunction $\ket{\{q,m\}}$ via successive F-moves, and its weight factorizes as $p_{\{q,m\}}=\prod_m d_{q_m}$, depending only on the external legs that link A to its complement.
Using this expression, we evaluate $\rho_A$ for the 12-site region that encloses a single hexagon—the support of the $B_p$ operator (red area in Fig.~\ref{fig:heavyhex}).  From this parent RDM, we can trace out selected spins to obtain smaller subregions and then compute their GMN, allowing a systematic study of multiparty entanglement at and below the hexagon scale.
Because the Hilbert space dimension grows rapidly with the number of string types, exact GMN can be computed only for modest cluster sizes: up to six sites when the local dimension is $d=2$, and up to four sites for $d=3$.  For some larger clusters, we instead determine rigorous lower bounds.
Armed with these tools, we now survey GME across a range of Abelian and non-Abelian string-net models.

\noindent\textbf{$\mathbb{Z}_2$ gauge theory and the Double Semion Model}\\
The simplest case is an Abelian string-net with one string type of no branching rules, the model is specified by
\begin{align}
d_0 &= 1,\quad d_1 = F_{110}^{110} = \pm 1,\nonumber\\[1mm]
F_{000}^{000} &= F_{101}^{101} = F_{011}^{011} = 1,\nonumber\\[1mm]
F_{111}^{000} &= F_{001}^{110} = F_{010}^{101} = F_{100}^{011} = 1,
\end{align}
with all other elements of \(F\) vanishing. The corresponding fixed-point wave functions are given by
\begin{equation}
\Phi_{\pm}(X) = (\pm 1)^{X_c},
\end{equation}
where \(X_c\) is the number of disconnected closed string components in the configuration \(X\). 
\(\Phi_{+}\) is associated with a \(Z_2\) gauge theory, while \(\Phi_{-}\) corresponds to a double semion model.

To investigate the local entanglement properties of these fixed-point wave functions \(\Phi_{\pm}\), we study the RDMs for various local subregions. 
For subregions that do not enclose any loop, the RDM is diagonal, so the state is fully separable.
For the hexagonal plaquette the RDM takes the form of an \emph{X state}—a state whose only non-zero matrix elements lie on the diagonal and anti-diagonal. More precisely, an \(n\)-qubit X state is defined by
\begin{equation}
[X(a,c)]_{ij} = \begin{cases}
a_i, & \text{if } i = j,\\[1mm]
c_i, & \text{if } j = \overline{i},\\[1mm]
0,   & \text{otherwise,}
\end{cases}  
\end{equation}
where \(I_{[n]}\) is the set of all \(n\)-bit indices (e.g., \(I_{[6]}\) for a hexagon), \(a_i\) and \(c_i\) are real numbers corresponding to the diagonal and anti-diagonal elements, and \(\overline{i}\) denotes the bitwise complement of \(i\).

In our case, the state \(\rho_+\) corresponding to \(\Phi_+\) has uniform diagonal elements \(a_i = 1/64\) and uniform anti-diagonal elements \(c_i = 1/64\) for all \(i\in I_{[6]}\). In contrast, while \(\rho_-\) has the same diagonal entries \(a_i = 1/64\), its anti-diagonal elements \(c_i\) depend on the number of groups of consecutive 1’s (under periodic boundary conditions): if the number of groups is even, then \(c_i = -1/64\); if odd, \(c_i = 1/64\). For instance, configurations like \(000000\), \(000111\), \(001101\), and \(101010\) (which correspond to 0, 1, 2, and 3 groups respectively) acquire the appropriate sign for \(c_i\).
Using Theorem 5.3 in~\cite{Ha2018}, it can be shown that \(\rho_+\) is fully separable. Although \(\rho_-\) is not fully separable, it has zero GMN along the tripartition (12)(34)(56) around the plaquette and can be further shown biseparable using the separability algorithm.


\noindent\textbf{Fibonacci String-Net Model}\\
The Fibonacci string-net is a non-abelian string-net with one type of string, specified by 
\begin{align}
d_0 &= 1,\quad d_1 = \gamma, \nonumber\\[1mm]
F^{110}_{110} &= \gamma^{-1},\quad F^{110}_{111} = \gamma^{-1/2},\nonumber\\[1mm]
F^{111}_{110} &= \gamma^{-1/2},\quad F^{111}_{111} = -\gamma^{-1}.
\end{align}
In the above, the nontrivial entries correspond to the case \(i=j=k=l=1\) when using the notation \(F^{ijm}_{kln}\).
The remaining elements of \(F\) are defined by
\begin{equation}
F^{ijm}_{kln} = \delta_{m,i,j}\,\delta_{m,k,l}\,\delta_{n,j,k}\,\delta_{n,i,l},
\label{eq:Ftensor-trivial}
\end{equation}
where
\begin{equation}
\delta_{ijk}=\begin{cases}
1, & \text{if } \{i,j,k\} \text{ is allowed},\\[1mm]
0, & \text{otherwise.}
\end{cases}    
\end{equation}

All contiguous subregions up to six spins exhibit no detectable GME: every three- and four-site cluster has a diagonal reduced density matrix and is therefore fully separable (hence zero bipartite negativity and zero GMN), while five-site clusters are fully PPT along all bipartitions—and thus no GMN—and six-site clusters, although some contain bipartite negativity, still carry zero GMN.
The hexagonal plaquette likewise has no GMN.
However, we find a finite tripartite GMN for the 12-site region that encloses a single hexagon by evaluating a lower bound: $\mathcal N_3\geq 0.0607$.

\noindent\textbf{$\mathbb{Z}_3$ gauge theory}\\
The next case is an Abelian model with two types of strings, $1$ and $2$ where $2=1^{*}$, so they correspond to two orientations of the string. The model is specified by the branching rules $\{\{1,1,1\},\{2,2,2\}\}$ and the F tensor
\begin{align}
d_0 &= d_1 = d_2=1, \nonumber\\[1mm]
F^{ijm}_{kln} &=\delta_{m,i,j}\,\delta_{m,k,l}\,\delta_{n,j,k}\,\delta_{n,i,l}, 
\end{align}
and the fixed point wavefunction $\Phi(X)=1$ for every configuration satisfying the branching rules.

All non-loopy subregions are diagonal and thus fully separable, as proved by the general result for Abelian string-nets. For the hexagonal plaquette, the RDM for the $\mathbb{Z}_3$ gauge theory is fully PPT, and thus has no GMN.

\noindent\textbf{$S_3$ gauge theory}\\
We now consider the first non-abelian string-net model, with two string types and branching rules $\{\{1,2,2\},\{2,2,2\}\}$. The strings are unoriented, so $1^*=1,2^*=2$. The $d$ and $F$ tensors are 
\begin{align}
    d_0=d_1=1, \quad d_2=2 \\
    F^{110}_{110} = F^{110}_{222} = F^{212}_{212} = 1\\
    F^{212}_{222} = -1
\end{align}
and $F^{22m}_{22n}$ is given by the matrix 
\begin{equation}
    F_{22 n}^{22 m}=\left(\begin{array}{ccc}
\frac{1}{2} & \frac{1}{2} & \frac{1}{\sqrt{2}} \\
\frac{1}{2} & \frac{1}{2} & -\frac{1}{\sqrt{2}} \\
\frac{1}{\sqrt{2}} & -\frac{1}{\sqrt{2}} & 0
\end{array}\right) .
\end{equation}

All three- and four-spin clusters have diagonal reduced density matrices and are therefore fully separable (hence zero bipartite negativity and zero GMN). Five-spin patches, although non-diagonal, remain fully PPT and thus exhibit neither bipartite negativity nor GMN. 
The six-spin hexagonal plaquette does display bipartite entanglement, but our lower-bound does not detect a finite GMN. 
Similarly, no GMN was detected for the 12-site subregion enclosing a hexagon.

\noindent\textbf{Ising String-Net Model}\\ 
Finally, we consider the Ising string-net model, a non-abelian model with local dimension 3 that accommodates two distinct string types. In this model, following Kitaev's notation (See Table.1 in Ref.~\cite{Kitaev2006}), we define
\[
1\ (\text{vacuum}),\quad \varepsilon\ (\text{fermion}),\quad \sigma\ (\text{vortex}),
\]
The quantum dimensions are $d_1 = d_{\varepsilon} = 1, d_{\sigma} = \sqrt{2}$,  and the fusion rules are given by
\[
\varepsilon \times \varepsilon = 1,\quad \varepsilon \times \sigma = \sigma,\quad \sigma \times \sigma = 1 + \varepsilon.
\]
The nontrivial elements of the \(F\) tensor are specified as follows:
\begin{align}
F^{\sigma\sigma 1}_{\sigma\sigma 1} &= \frac{1}{\sqrt{2}},\quad  F^{\sigma\sigma 1}_{\sigma\sigma \varepsilon} = \frac{1}{\sqrt{2}},\nonumber\\
F^{\sigma\sigma \varepsilon}_{\sigma\sigma 1} &= \frac{1}{\sqrt{2}}, \quad F^{\sigma\sigma \varepsilon}_{\sigma\sigma \varepsilon} = -\frac{1}{\sqrt{2}},\nonumber\\
F^{\varepsilon\sigma\sigma}_{\varepsilon\sigma\sigma} &= -1, \quad F^{\sigma\varepsilon\sigma}_{\sigma\varepsilon\sigma} = -1.
\end{align}
In all other cases, the \(F\) tensor elements are determined by Eq.~(\ref{eq:Ftensor-trivial}), analogous to the Fibonacci string-net model.

All three- and four-site clusters have diagonal reduced density matrices and are therefore fully separable (zero bipartite negativity and zero GMN). Five-site clusters, though non-diagonal, remain fully PPT and thus likewise exhibit neither bipartite negativity nor GMN. In contrast,
the 6-site hexagonal plaquette retains genuine tripartite entanglement: our lower-bound calculation yields $\mathcal{N}_3 \ge 0.0112$.

\end{document}